\documentclass{article}

\usepackage{amsmath}
\usepackage{amssymb}
\usepackage[nopatch]{microtype}
\usepackage{booktabs}
\usepackage{makecell}
\usepackage{comment}
\usepackage{hyperref}
\usepackage{cleveref}
\usepackage{fancyhdr}
\usepackage{fullpage}
\usepackage{graphicx} 
\usepackage{setspace}
\usepackage{authblk}
\usepackage{caption}
\usepackage{comment}
\usepackage[authoryear,sort]{natbib}
   \setcitestyle{aysep={}} 
\usepackage{algorithm}
\usepackage{algpseudocode}
\newcommand\E{\mathbb{E}}
\newcommand\V{\mathbb{V}}
\newcommand{\K}{\mathbf{K}}
\newcommand{\X}{\mathbf{X}}

\usepackage{scalerel,stackengine}
\stackMath
\newcommand\reallywidehat[1]{%
\savestack{\tmpbox}{\stretchto{%
  \scaleto{%
    \scalerel*[\widthof{\ensuremath{#1}}]{\kern-.6pt\bigwedge\kern-.6pt}%
    {\rule[-\textheight/2]{1ex}{\textheight}}
  }{\textheight}%
}{0.5ex}}%
\stackon[1pt]{#1}{\tmpbox}%
}
\title{Inference at the data's edge: Gaussian processes for estimation and inference in the face of extrapolation uncertainty \thanks{Authorship order is alphabetical for the first two authors; Hazlett is senior co-author. All authors contributed equally. We thank two anonymous reviewers, Max Goplerud, Ian Lundberg, attendees of the 2024 Summer Polmeth meeting, and members of the UCLA Practical Causal Inference Lab learning group for their comments and suggestions.}}

\author[1]{Soonhong Cho}
\author[1]{Doeun Kim}
\author[1,2,3]{Chad Hazlett}

\affil[1]{Department of Political Science, UCLA}
\affil[2]{Department of Statistics and Data Science, UCLA}
\affil[3]{Co-director, Practical Causal Inference Lab, UCLA}
\date{\today}

\begin{document}
\maketitle

\begin{abstract}
Many inferential tasks involve fitting models to observed data and predicting outcomes at new covariate values, requiring interpolation or extrapolation. Conventional methods select a single best-fitting model, discarding fits that were similarly plausible in-sample but would yield sharply different predictions out-of-sample. Gaussian Processes (GPs) offer a principled alternative. Rather than committing to one conditional expectation function, GPs deliver a posterior distribution over outcomes at any covariate value. This posterior effectively retains the range of models consistent with the data, widening uncertainty intervals where extrapolation magnifies divergence. In this way, the GP's uncertainty estimates reflect the implications of extrapolation on our predictions, helping to tame the ``dangers of extreme counterfactuals'' \citep{king2006dangers}. The approach requires (i) specifying a covariance function linking outcome similarity to covariate similarity, and (ii) assuming Gaussian noise around the conditional expectation. We provide an accessible introduction to GPs with emphasis on this property, along with a simple, automated procedure for hyperparameter selection implemented in the R package \texttt{gpss}. We illustrate the value of GPs for capturing counterfactual uncertainty in three settings: (i) treatment effect estimation with poor overlap, (ii) interrupted time series requiring extrapolation beyond pre-intervention data, and (iii) regression discontinuity designs where estimates hinge on boundary behavior.
\end{abstract}

\textit{Keywords:} causal inference, Gaussian process regression, machine learning, regression discontinuity, interrupted time series, positivity

\section{Introduction}

Many inferential tasks rely on training models to predict outcomes from observed covariates, then extending these predictions to new covariate locations through interpolation or extrapolation. During model fitting, different parameter choices can often produce models that fit similarly well in-sample but diverge sharply in data-sparse regions or beyond the support of the data. Nevertheless, standard practice selects only a single best-fitting conditional expectation function (CEF). When extrapolating, uncertainty intervals from this single model cannot account for the different predictions that might have been produced by other plausible fits. This approach thus omits a source of uncertainty we refer to as ``extrapolation uncertainty''. The problem is especially acute in settings requiring predictions in data-sparse regions, such as covariate-adjusted comparisons lacking common support, interrupted time series (ITS), and regression discontinuity (RD).

We recommend Gaussian Process (GP) regression as one principled approach to address this concern in such settings. GPs yield a posterior distribution for outcomes at each covariate location, not just a single best-fitting CEF estimate. In data-sparse regions, this posterior effectively retains and encompasses the range of plausible fits, widening uncertainty intervals to reflect their divergence. 

Despite their appeal, GPs remain relatively unfamiliar in the social sciences, with only a few recent exceptions (e.g. \citealp{branson2019nonparametric,ornstein2022gaussian,hinne2022bayesian,prati2023gaussian,ben2023estimating}). We therefore begin with an accessible introduction. One barrier to wider adoption has been the practical challenge of tuning three highly interdependent hyperparameters. We simplify this by fixing one parameter, selecting a second based only on the covariates, and estimating the third with an automated line search. This fully automated procedure, implemented in the R package \texttt{gpss}, has proven stable in our simulations and applications.

We illustrate the practical value of GPs in three settings. First, in treatment-control comparisons, we illustrate how GPs handle failures of overlap or common support in terms of their uncertainty intervals for both average and conditional effect estimates. Second, in ITS designs, GPs provide a principled framework for extrapolating beyond pre-intervention data while making specification choices explicit.
Third, in RD designs, GPs approach edge estimation differently from standard optimal-bandwidth local polynomial methods (e.g., \citealp{calonico2014robust}), relying on the GP’s posterior distribution for the outcome at the cutoff from models trained on either side. This proves especially valuable in samples smaller than those suitable for the local polynomial approach. In particular, the GP approach reduces the influence of noisy observations near the cutoff that can drive extreme polynomial fits, improving bias and RMSE while maintaining good coverage, often with shorter intervals.

\section{The GP framework}
We briefly outline the GP framework, drawing on concepts familiar to most quantitative social scientists. For a classical treatment, readers may also wish to reference \citet{rasmussen2006gaussian}. Our discussion is motivated by---and emphasizes the implications for---how we think about uncertainty estimation for new observations that may lie at varying distances from the observed data. Throughout, we assume (training) data consisting of $n$ independent tuples \( \{Y_i, X_i\}_{i=1}^n \) drawn from a common joint distribution or data-generating process (DGP). Here, \( Y_i \in \mathcal{Y} \) denotes a scalar outcome, and \( X_i \in \mathcal{X} \subseteq \mathbb{R}^d \) denotes a vector of covariates.

\paragraph{A distributional outcome.} Consider first the outcome $Y$, a vector of $n$ observations ($i \in \{1, ..., n\}$) drawn from a multivariate normal distribution $Y \sim \mathcal{N}(\mu, \Sigma)$. We defer momentarily how $\mu$ and $\Sigma$ will be determined. 

\paragraph{Smoothness, covariance, and functional form.} Second, suppose that $Y_i$ and $Y_j$ should be more similar in value when $X_i$ and $X_j$ are similar in value. This can be restated as requiring that the covariance of $Y_i$ with $Y_j$ (across varying data generating process) must be higher when $X_i$ and $X_j$ are more similar. More formally, choose a kernel function $k(\cdot, \cdot)$ governing the relationship between the covariance of $Y_i$ with $Y_j$ and the distance between $X_i$ and $X_j$ according to $cov(Y_i,Y_j)=\sigma_f k(X_i,X_j)$. This can take a particular form, such as $k(X_i,X_j)= exp(-||X_i-X_j||^2/b)$ (the Gaussian kernel) whereby observations $i$ and $j$ will have maximal covariance when $X_i=X_j$, and covariance decreases towards zero as $X_i$ and $X_j$ become distant. Other kernel functions can be applied to various purposes we discuss below. 

Consider the kernel matrix $\K$, containing all the pairwise kernel evaluations, i.e., $\K_{i,j}=k(X_i,X_j)$. As this describes the full variance-covariance matrix of the vector $Y$, we now have the prior model
\begin{align}
        Y \mid X \sim \mathcal{N}(\mu,\sigma_f \K). \label{eq:prior}
\end{align}

\noindent where $\mu$ is a length-$n$ vector giving the prior mean at each point, and $\sigma_f $ is a scaling parameter governing the prior variance. Further, the CEF is not expected to go through each observation; rather the observations are spread around their conditional expectations with some irreducible error, $\sigma^2$. This revises our prior model to
\begin{align}
    Y \mid X \sim \mathcal{N}(\mu,\sigma_f\K+\sigma^2 I).
\end{align}

\paragraph{Simplification.} As a prior model, we emphasize that this represents only possible draws of the function before observing any data. We make two simplifications. First, we set $\mu=0$ at all points. This is because, (i) having not yet seen the data, we have no preference for any particular choice of $\mu$, and (ii) we globally demean $Y$ for modeling purposes (adding the mean back to predictions afterward). As we elaborate in Section~\ref{subsec.additionaldetails}, this does not effectively constrain the shape of the posterior CEF, which will shape itself around the data as we describe below. 

Second, we set $\sigma_f=1$. Recall that $\sigma_f$ governs the prior variance of the underlying function at any point $X=x$, before observing data. With this choice, the solution for the posterior CEF at the training points becomes $\K(\K + \sigma^2 I)^{-1}Y$ (see Expression~\ref{eq.main} below). This exactly reproduces the solution for the CEF obtained through kernel ridge regression (e.g., \citealp{hainmueller2014kernel}), which rests only on regularized loss-minimization in a function space without invoking any notion of prior variance, except implicitly through the chosen degree of regularization. Fixing $\sigma_f=1$ can thus be understood not as a restriction but as parameterization that will rescale and alter the interpretation of $\sigma^2$. Practically speaking, this greatly improves identifiability of $\sigma^2$, allowing it to be tuned efficiently via a simple line search (see Section~\ref{subsec.additionaldetails}). Together these adjustments leave us with a prior model of
\begin{align}\label{eq.prior3}
Y \mid X \sim \mathcal{N}(0,\K+\sigma^2 I).
\end{align}

\begin{figure}[hbt!]
\centering
\includegraphics[width=.7\linewidth]{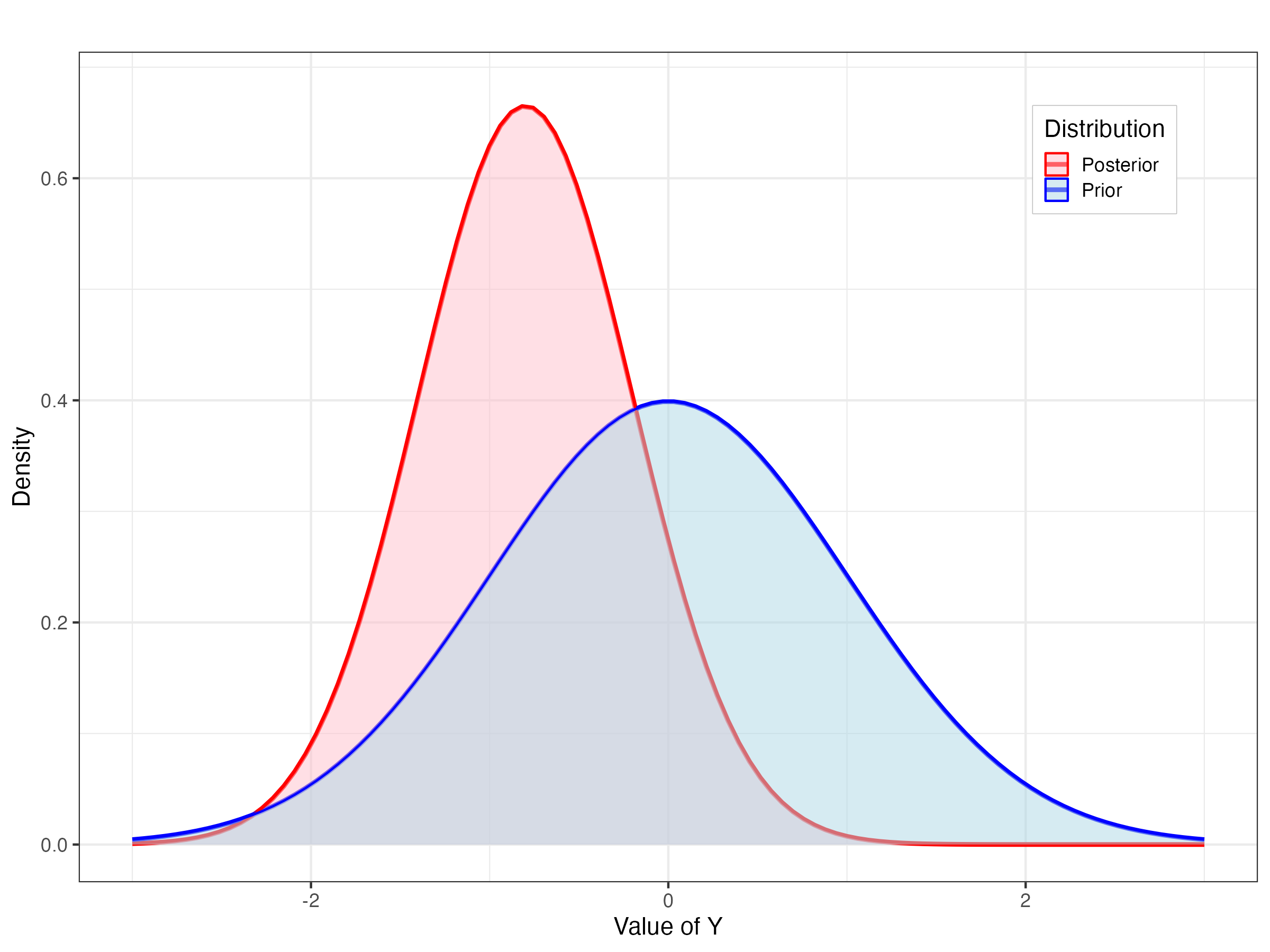}
\caption{\textit{Blue} (prior): Distributional belief for an unseen $Y^*$ knowing only that it will come from a normal distribution with mean 0 and variance $\sigma^2$. \textit{Red} (posterior): Our revised belief regarding $Y^*$ assuming $cor(Y^*,Y_{obs})=.8$, and having observed $Y_{obs}=-1$.
}
\label{fig:oneobs}
\end{figure}

\paragraph{Conditioning on the data.} If we are given $X_j$ and asked to guess $Y_j$, we can only guess $Y_j$ is distributed $N(0, \sigma^2)$; i.e., we are maximally ignorant (Figure~\ref{fig:oneobs}, blue). However, suppose we observe another unit, $\{X_i, Y_i\}$. The distance between $X_i$ and $X_j$ (or more properly, $k(X_i,X_j)$) tells us how $Y_i$ and $Y_j$ will covary. If $X_j$ is close to $X_i$, we can guess that $Y_j$ is more likely to be close in value to $Y_i$. If $Y_i$ takes a large value, for example, we can expect $Y_j$ to be larger, and this information somewhat reduces our uncertainty. Conditioning on the data in this way leads to a (posterior) distribution for the unobserved $Y_j$ (Figure~\ref{fig:oneobs}, red).  

Scaling this logic, consider $n$ (training) observations $\{X_i,Y_i\}_{i=1}^{n}$ and $n^*$ test observations with observed covariates $\{X^*\}$ and unknown outcome $Y^*$ (an $n^*$-length vector). Our belief about $Y^*$ is then given by:
    \begin{align}\label{eq.main}
    Y^* \mid X^*, Y,X \sim \mathcal{N}(\K_* (\K+\sigma^2 I_n)^{-1}Y, \; \K_{*,*}+\sigma^2 I_{n^*} - \K_* (\K+\sigma^2 I_n)^{-1}\K_*^{\top}),
    \end{align}
where $\K_{*,*}$ denotes the $(n^* \times n^*)$ kernel matrix between test points, and $\K_*$ the $(n^* \times n)$ kernel matrix between test and training points. We defer derivations to \citet{rasmussen2006gaussian} or other sources. This is a posterior distribution for $Y^*$ given the data (and the choice of kernel). Further, since the normal distribution has identical mean and mode, the mean argument in posterior ($\K_* (\K+\sigma^2 I_n)^{-1}Y$) is both the CEF and the \textit{maximum a posteriori} (MAP) estimator. 

\begin{figure}[hbt!]
    \centering
    \includegraphics[scale=.5]{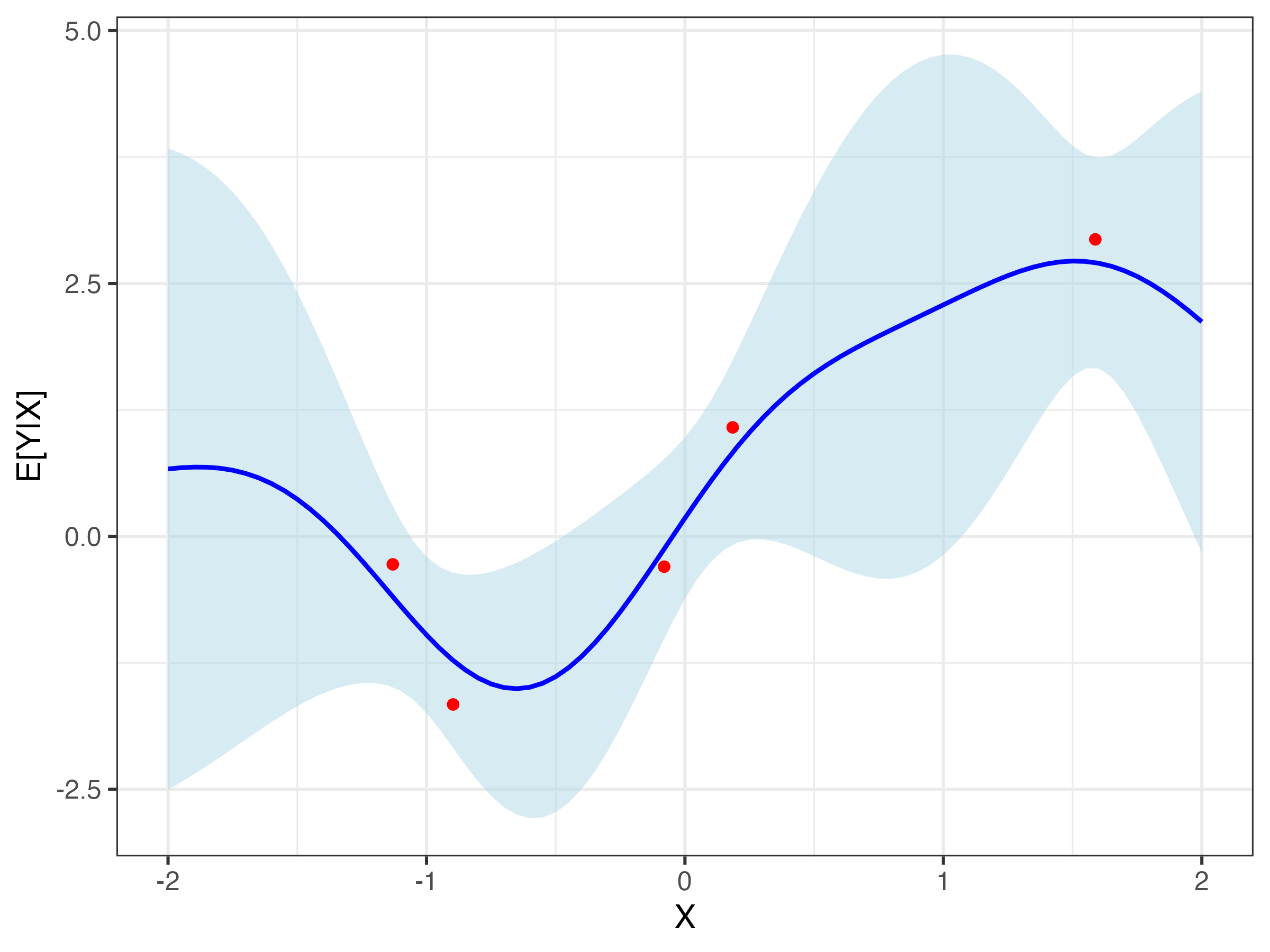}
    \caption{Posterior CEF with its 95\% CI. Inferred distribution (spread vertically) about $Y^*$ as a function of covariate $X$, after seeing five observations (red dots). The key assumption is on the covariance between points as a function of their $X$ values, here given by $cov(Y^*,Y_i) = exp(-||X_i-X^*||^2/b)$.}
    \label{fig:fiveobs}
\end{figure}

The consequences of this expression are illustrated in Figure~\ref{fig:fiveobs}. For values of $X$ closer to observations in the data, the covariance of $Y^*$ with the corresponding observed $Y$ will be higher, providing more information to update our guess of $Y^*$  and reduce our uncertainty, whereas uncertainty  balloons for predictions farther from observed data. This entire posterior distribution for any point is available in closed-form, making Markov chain sampling unnecessary. 

Note also that the variance in Expression~\ref{eq.main}, $\K_{*,*}+\sigma^2 I_{n^*} - \K_* (\K+\sigma^2 I_n)^{-1}\K_*^{\top}$, describes our (un)certainty about points $Y^*$. In contrast, when we are interested in uncertainty about the conditional expectation function itself, the relevant variance is $\K_{*,*} - \K_* (\K+\sigma^2 I_n)^{-1}\K_*^{\top}$, which excludes the irreducible noise term ($\sigma^2 I_{n^*}$) that reflects how individual $Y$ values scatter around their conditional mean.

\paragraph{Homoskedasticity.} A key assumption employed by this model choice is that $Y$ is distributed normally around its CEF/MAP, with constant variance. Violating this assumption can influence interval size in particular, as we illustrate in Section~\ref{sec:rd} and ~\ref{app.hetero}. Note however that this does not imply constant variance in the posterior uncertainty, which will widen or narrow adaptively, reflecting local data sparsity.

\paragraph{Interpreting the function space and comparison to kernel regularized least squares (KRLS).} For additional context, in ~\ref{app.krlscompare} we remark on the relationship between GP and KRLS, which have identical function spaces for the CEF, but handle uncertainty very differently (discussed in Section~\ref{subsec:reflectinguncertainty}). Both also involve regularization, but motivated and tuned differently.  

\subsection{Additional details}\label{subsec.additionaldetails}

We now consider a set of interrelated scaling and hyperparameter selection choices, including the kernel choice. It is not clear whether an optimal solution to these questions exists in a general sense, yet users must make these choices, and some choices can lead to poor performance.  We offer guidance that we believe is at least reasoned, practical, and shows good performance to the limits of our testing.

\paragraph{Demeaning and rescaling.}
We demean and scale the covariates to have unit variance. This is common for kernel-based machine learning methods, and avoids dependency on unit-of-measure decisions.
As noted, we also (globally) demean and scale $Y$ to have unit variance, as in some other kernelized approaches (see e.g., \citealp{hainmueller2014kernel}). This is also required for the interpretation of $\sigma^2$ as 1-$R^2$ noted below.

\paragraph{Bandwidth selection.} For the Gaussian kernel $k(X_i,X_j) = exp(-||X_i-X_j||^2/b)$, we must choose the bandwidth or ``length-scale'', $b$. There are numerous strategies for doing so. One reasonable approach is to set $b$ equal to (or proportional to) the number of dimensions of $X$, so that the result does not explode or go to zero as the number of dimensions changes (as in \citealp{hainmueller2014kernel}). Another recent proposal, offered in \citet{kpop}, chooses the value of $b$ that leads to the highest estimated variance across the (off-diagonal) elements of $\K$. This simply ensures that the columns of $\K$ stand to be highly informative rather than having $\K$ approach the identity matrix (if $b$ is too small) or a block of ones (if $b$ is too large). This is the approach we take.

\paragraph{Choosing $\sigma^2$.}  Having fixed $\sigma_f=1$ and selected the kernel bandwidth based directly on the distribution of $\K$, we are left with only one hyperparameter to tune. As advocated by \citet{rasmussen2006gaussian}, we tune the residual variance term, $\sigma^2$, by maximizing the log marginal likelihood, 
 \begin{equation}
    \log~p(Y \mid X) = -\frac{1}{2} Y^\top (\K + \sigma^2 I)^{-1}Y - \frac{1}{2}(\log |\K|+\sigma^2 I) - \frac{n}{2} \log 2\pi.
  \end{equation}

In principal, since one could always choose a ``wigglier CEF'' (even for a fixed kernel bandwidth) together with a smaller $\sigma^2$, this quantity might not always be well-identified on the data. While we remain concerned about this in the general case, empirically this approach appears to perform very well at least over the cases we examine. In particular, we see that for each DGP we attempt, the estimated $\sigma^2$ values are very close to 1-$R^2$, as our scaling choice should lead to. This may fail when the DGP does not have the Gaussian, homoskedastic noise on which this likelihood estimate is based.

\paragraph{Combining kernels.}A useful fact about kernels used for such kernel regression tasks is that they can be added together, with the resulting kernel remaining valid (positive semi-definite), and providing access to a space of functions that combines the underlying features of the constituent kernels (see, e.g., \citealp{schulz2017compositional}). For example, consider a kernel $\K^{poly}$ that provides access to polynomial models, i.e.  $\K^{poly}_{i,j}= 1 + \langle x_i, x_j \rangle^p$, where $p$ is the desired degree of the polynomial.  Such a kernel matrix can be added pointwise to the Gaussian kernel matrix. The resulting kernel matrix allows the GP to represent functions that are both locally smooth (from the Gaussian kernel), and exhibit polynomial growth rather than translation invariance (from the polynomial). Similarly, including a periodic kernel enables the model to capture cyclic structures such as seasonality. We demonstrate this below to explore models for ITS analysis.

\paragraph{Stationary vs. non-stationary kernels for mild vs. extreme extrapolation.} The next question is whether one is interested in making inferences nearer to observed data (e.g., interpolation, edge estimation, or very mild extrapolation), where we can rely on the covariance between observed points and target points directly, or if one wishes to make inferences well beyond the data's edge. In the former case, we can employ a ``stationary kernel'' that operates only on the relative distances between observations. In these settings, we rely on the Gaussian kernel because (i) it directly works with the logic that observations with more similar values should have greater covariance while covariance should drop to zero as observations grow far apart in the covariate space; (ii) it is commonly taken to be the work-horse kernel for many machine learning approaches involving kernels; and (iii) it is an example of a ``universal kernel'' for the continuous functions, meaning that it can represent any continuous function given sufficient data \citep{micchelli2006universal}.  

By contrast, for some applications, inferences are attempted farther from the edge of the data. Our ITS application below is a leading example, though our poor-overlap case also requires moderate to severe extrapolation. Under stationary kernels like the Gaussian, the predicted CEF will ``return to the mean'' far from the edge of the data. While this may be reasonable, the uncertainty can reach a maximum (determined by $\hat{\sigma}^2$) and stop growing. Without addressing this issue, it would appear that the GP does not extrapolate well (see e.g., \citealp{wilson2013gaussian}). However, when investigators need to extrapolate farther (relative to the scale of the kernel bandwidth), they may need to entertain functions that show (i) periodic behavior (in the case of time series data), and/or (ii) continued growth, e.g., with some continuity in the first or higher derivative, as in linear or polynomial prediction. The GP can accommodate this by combining kernels, for example, adding a Gaussian kernel with a polynomial kernel of the desired order. In these settings, the GP becomes a device for illustrating the wide range of possible functions (and our uncertainty over them). We illustrate this use in Section~\ref{sec:its} below. 

\paragraph{Prior mean functions.} As noted above, we set $\mu=0$. This is because prior to seeing any data we would have no particular reason to prefer a given functional form, direction of change, or any other feature of the CEF not already encoded by choosing the multivariate normal distribution with covariance given by the kernel function. After conditioning, the posterior CEF can approximate whatever smooth functional form the data support.\footnote{These CEFs are linear in the kernel matrix, corresponding to the empirical subspace of the reproducing kernel Hilbert space (RKHS) associated with the kernel.} This choice does not restrict the posterior fit, which adapts around the observed data as in Expression~\ref{eq.main} (see e.g. Figure~\ref{fig:fiveobs}). Regularization nevertheless occurs through $\sigma^2$, which governs the trade-off between fit and smoothness and plays the same role as the regularization parameter in kernel regression (e.g., $\lambda$ in KRLS; \citealt{hainmueller2014kernel}).

Taking RD as an example, the GP can represent CEFs that are approximately linear, polynomial, or otherwise smooth on each side of the cutoff. If extrapolation far beyond the observed data is needed, prior expectations about functional form can be encoded through a (non-stationary) kernel rather than a mean function. This distinguishes our strategy from approaches in \citet{branson2019nonparametric, rischard2021school, ornstein2022gaussian}.\footnote{We expect that adding a non-zero prior mean function will have little impact on estimation, particularly when the mean function lies within the function space implied by the kernel and inference is restricted to regions near the observed data. \citet{branson2019nonparametric} likewise report that using a zero mean instead of a linear mean function produced “largely the same” results in their simulations.}

\subsection{Comparison to conventional uncertainty estimation under extrapolation}\label{subsec:reflectinguncertainty}

For tasks involving extrapolation, a key advantage of the GP is its ability to incorporate what we refer to as ``extrapolation uncertainty'' in its reported posterior uncertainty intervals. Conventional approaches to uncertainty estimation in parametric models involve selecting a single best-fitting model and using that fitted model alone for extrapolation (as well as to compute residuals that inform uncertainty intervals). This is problematic because, within a sufficiently flexible model space, many model fits can explain the observed data similarly well but will make radically different predictions outside of the observed data. Our uncertainty over which of those fits to choose must be maintained if we wish to sustain appropriate uncertainty over the predictions they make when extrapolated. This ``extrapolation uncertainty'', however, is lost once we choose a single fitted model, discarding all others. Contrast this with the GP. While the GP does commit to a specific model space and set of assumptions described above, the process of conditioning on the data and inspecting the posterior is different from that of choosing and extrapolating only a single best-fitting model. The GP’s posterior uncertainty band instead reflects the range of CEFs that remain plausible, to varying degrees, given the observed data. The posterior uncertainty at new points effectively extends these functions beyond the data, encompassing their divergent predictions and thereby incorporating increased uncertainty as we move farther from the conditioning data.

\begin{figure}[hbt!]
\centering
\includegraphics[width=0.75\linewidth]{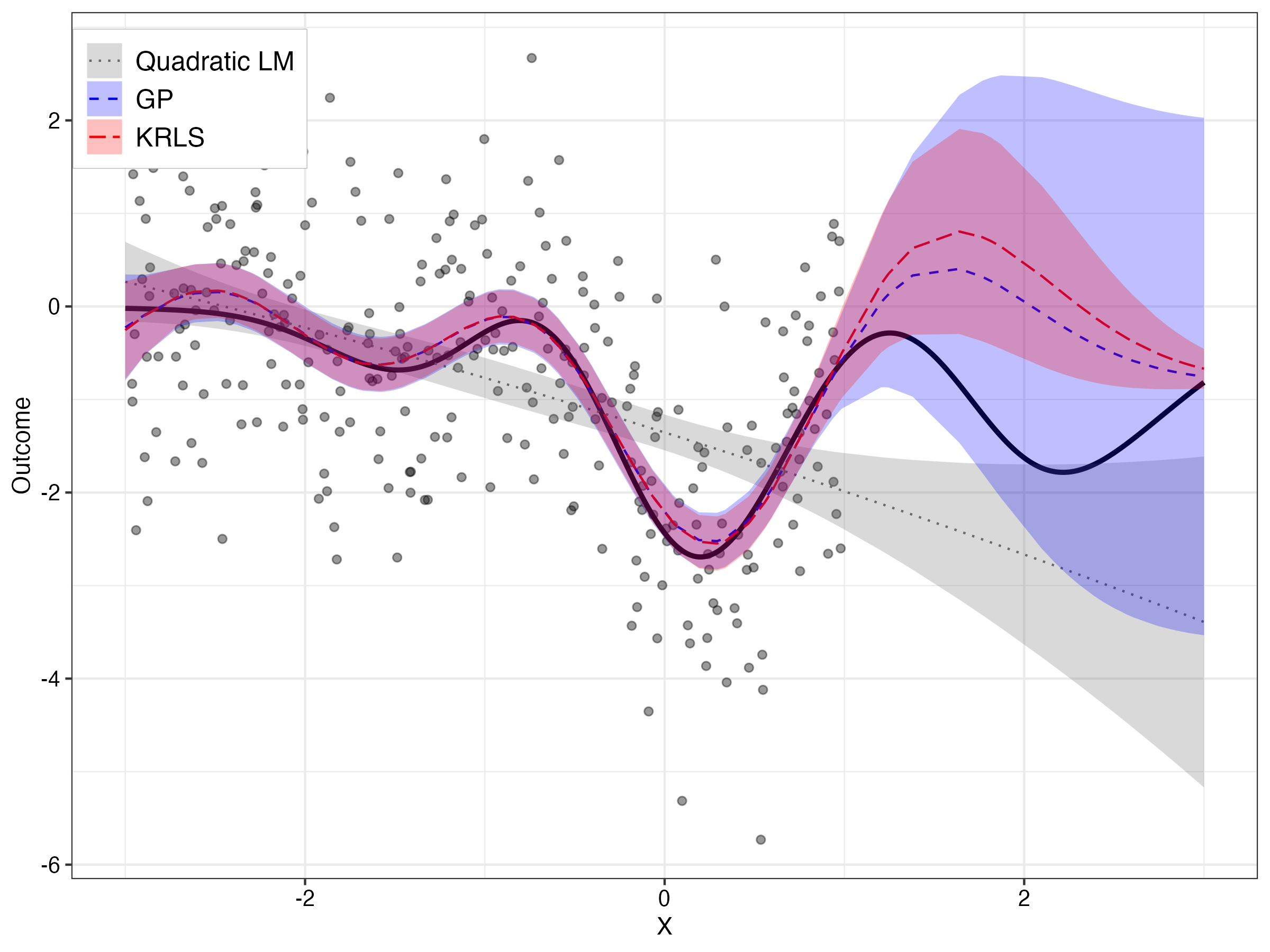}
\caption{Comparison of uncertainty for conditional expectation functions given by GP (blue), KRLS (red), and a quadratic polynomial (gray). Models are fitted on $X$ data between -3 and 1, then extrapolated over $X$ between 1 and 3. The solid line shows the (randomly drawn) true function, while dashed/dotted lines indicate each model's estimated CEF. Bands indicate 95\% CIs for the CEFs.}
\label{fig:krls_gp_quadratic}
\end{figure}

We illustrate this in Figure~\ref{fig:krls_gp_quadratic}. First, consider a linear model fitted by OLS. The classical variance estimate for the coefficient under homoskedastic and independent (spherical) errors will be $\widehat{\V}(\hat{\beta} \mid \X,Y) = (\X^{\top} \X)^{-1}\hat{\sigma}^2 I$. The value of $\hat{\sigma}^2$ is proportional to $\sum_i (Y_i - \X_i^{\top}\hat{\beta})^2$, that is, the sum of the squared \textit{fitted} residuals.\footnote{Alternative variance estimators such as heteroskedasticity robust, cluster-robust, etc. will have a more complex form rooted in a different choice of variance-covariance matrix for the residuals, $\Sigma$. However, the problem we illustrate here is not due to misspecification of the $\Sigma$ and persists even when $\Sigma$ is correctly chosen. } Consequently, the estimated variance of the predicted value of the CEF at some point $\X_j$ will be $\widehat{\V}(\hat{Y}(\X_j)) = \hat{\V}(\X_j^{\top}\hat{\beta}) = \X_j^{\top} \hat{\V}(\hat{\beta}) \X_j$. While this quantity increases for predictions at $\X_j$ points farther from the mean of the data, this is blind to a point's distance from supporting data. The \texttt{quadratic LM} line and uncertainty envelope in Figure~\ref{fig:krls_gp_quadratic} illustrate this in the case of a fitted quadratic model.

This problem is not one of relying on a restrictive model space. We can demonstrate this by comparing GP and KRLS, employing the same kernel (Figure~\ref{fig:krls_gp_quadratic}). Since these models have identical model spaces for the CEF, their CEFs match very closely, differing only due to hyperparameter tuning. The concern, however, is the uncertainty intervals. KRLS employs the conventional fitted-function variance estimation. This results in a collapse at points farther from the observed data, because the model is structured such that for any one model we can be sure the CEF it predicts returns to the mean of $Y$ as we leave the support of the data. However, this is the opposite of what we would want, which is to see increased uncertainty about the value of $Y$ as we move further from the support of the data. As Figure~\ref{fig:krls_gp_quadratic} shows, the predictive uncertainty from the GP model provides this, widening as we move towards and then beyond the edge of the data. 

\section{Use cases for GP}

In this section we illustrate the useful properties of GPs in three settings where inferences depend on understanding uncertainty that arises from model dependency as we move towards or beyond the edge of the observed data.\footnote{Replication code and data are available at \cite{our_data}.}

\subsection{Comparing groups with poor overlap}
\label{sec:pooroverlap}

In cross-sectional comparisons, treated and control groups must often be compared as if they had the same covariate distributions.  For example, we have an observed confounder $X$, and assuming the absence of unobserved confounders (i.e., under selection on observables or conditional ignorability), we wish to make comparisons between treated and control units adjusting for $X$.

Adjustment strategies of all kinds are vulnerable to extrapolation bias when there are regions of $X$ containing only treated or only control units. Here, treated units have values of $X$ between -3 and 1, whereas control units have values between -2 and 3. Thus, a model for the treatment outcome must be extrapolated in areas with few or no treated units ($X > 1$) and a model for the control outcome must be extrapolated in areas with few or no control units ($X < -2)$. Treatment effect estimates---computed by comparing these two models for each unit (i.e., g-computation, regression imputation, etc.)---consequently suffer in the regions of poor overlap insofar as these model fits fail to predict well in those regions.

\begin{figure}[hbt!]
\centering
\includegraphics[width=1\linewidth]{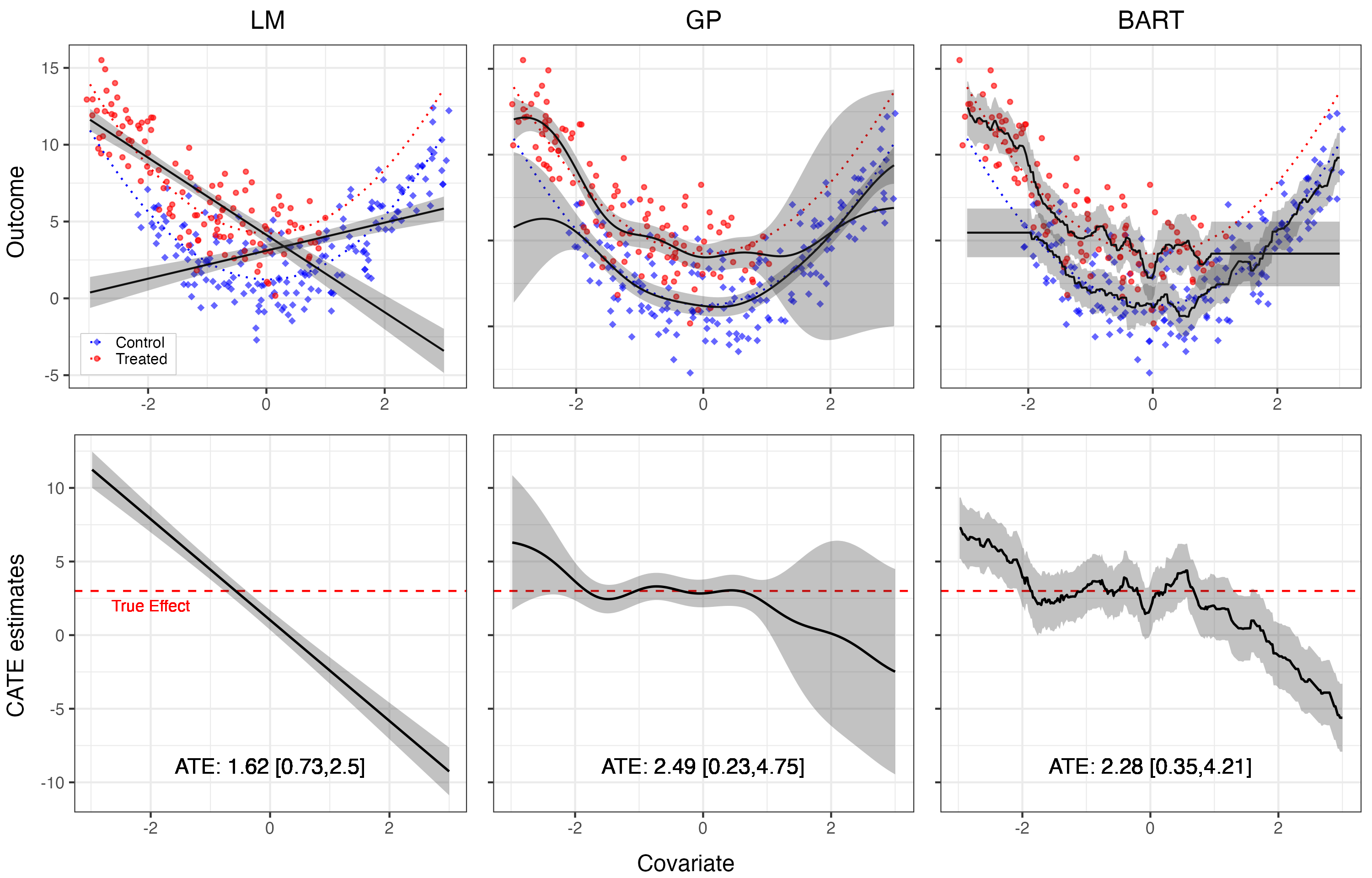}
\caption{Uncertainty quantification of CATE under varying overlap. \textit{Top row}: Illustration from one draw of the simulation setting, with a good overlap region ($-2<X<1$) and no overlap elsewhere. For treated units (red) and control units (blue), the true CEF (dotted lines) and estimated CEF (black solid lines) with corresponding uncertainty bands are drawn. As shown at both ends of the covariate value, GP allows for growing uncertainty bands adaptive to the degree of overlap. The upper bound of the confidence interval in extrapolation depends on the variability in the fitted data. \textit{Bottom row}: The CATE estimates with 95\% confidence bands. The wider bands of GP in poor-overlap regions propagate to higher uncertainty in CATE estimation. The red dashed line represents the true effect size.}
\label{fig:pooroverlap_illustration}
\end{figure}

For each simulation iteration, we draw a random function to serve as the CEF of $Y$ given $X$, with a constant treatment effect of 3. To favor non-GP methods, we draw the CEF as a quadratic function $f(x) = \alpha_0 + \alpha_1 x + \alpha_2 x^2$ with the coefficients independently drawn from a standard normal distribution. Following this, ``observed'' samples are drawn from each of these CEFs ($N=500$) with the addition of noise, distributed normally and with variance calibrated to ensure an overall true $R^2$ of 0.5.\footnote{Throughout our analyses, for realism we calibrate $R^2$ values in simulations to 0.3-0.5, except where we vary them more widely to demonstrate how performance depends on it. In this simulation we set this $R^2$ to 0.5, though in Figure~\ref{fig:pooroverlap_illustration} $R^2=0.8$ for improved visualization.}

Figure~\ref{fig:pooroverlap_illustration} presents a single example of a random function drawn from this space (dotted lines), along with the simulated data under treatment (red) and control (blue). The three models tested fit the data similarly well within the region of common support ($-2 < X < 1$). In the regions with poor overlap ($X< -2$ and $X>1$), however, the result is highly sensitive to the choice of modeling approach. Notably, the uncertainty estimates for the Linear Model (LM) and BART show insufficient uncertainty to accommodate this model dependency under poor overlap. 

The bottom row of Figure~\ref{fig:pooroverlap_illustration} shows what these estimated CEFs and uncertainty intervals imply for conditional average treatment effects (CATE) and the average treatment effect (ATE). Since the true CATE is constant, the apparently linear change in the treatment effect as a function of $X$ for the LM (left) is erroneous. This is a byproduct of the poor overlap, which led to fitted models with very different slopes for treated versus control groups. While the problem is less pronounced for BART, its uncertainty estimates still fail to reflect distance from relevant observed data. Its uncertainty over the CATE remains nearly constant across areas of good and poor overlap, producing widely insufficient coverage in the areas of weaker overlap. In contrast, GP's uncertainty estimates adaptively reflect model dependence due to data sparsity---narrower where common support is better, wider where it is weaker.

\begin{figure}[hbt!]
\centering
\includegraphics[width=.9\linewidth]{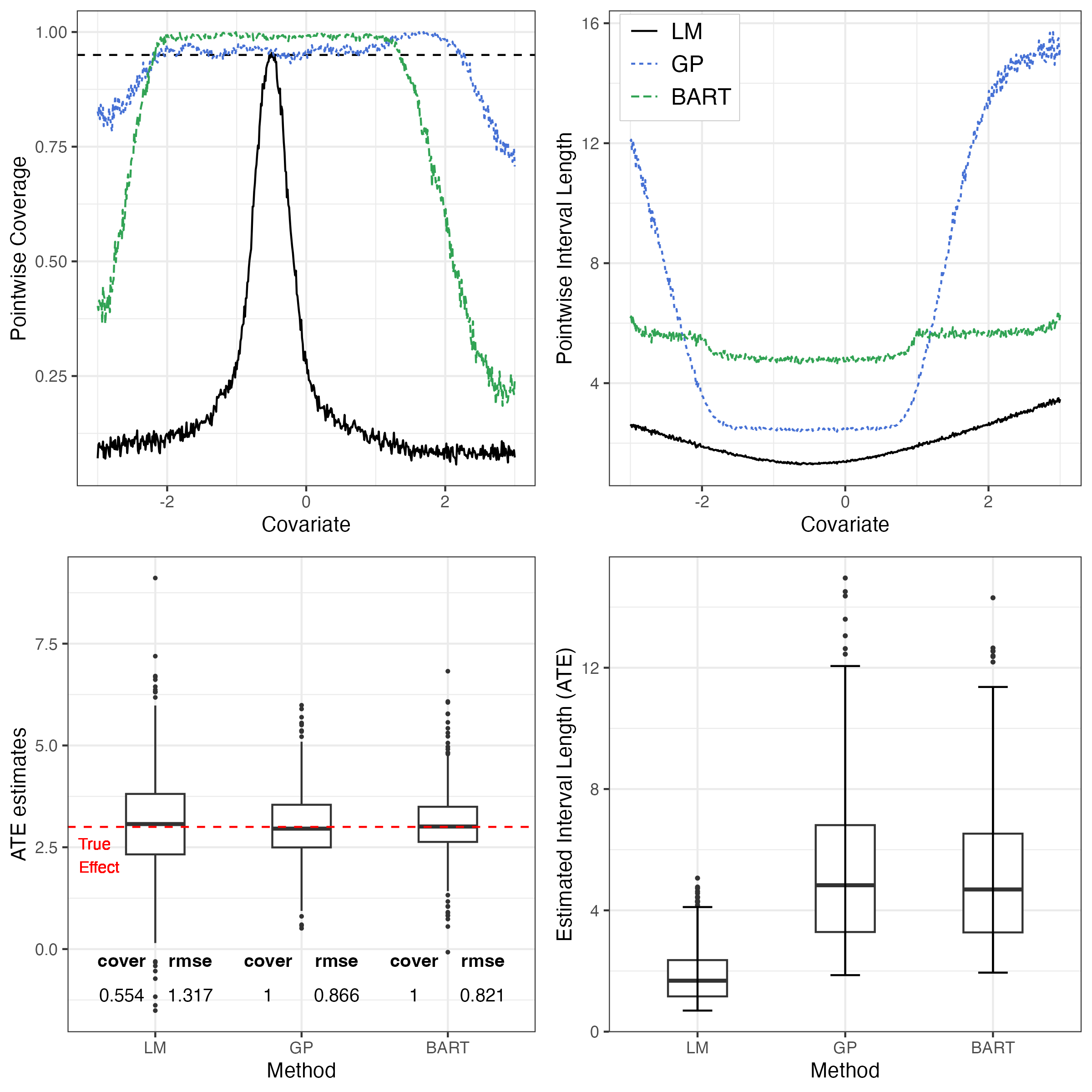}
\caption{Pointwise (top) and average (bottom) performance of GP model compared to LM and BART. \textit{Top row}: The graph on the left displays the pointwise coverage rate of the ATE estimators by the three models. On the right, the lengths of corresponding 95\% confidence intervals are shown over 500 simulations (n=500). \textit{Bottom row}: Boxplots represent the distributions of the ATE estimates (\textit{left}) and average interval lengths (\textit{right}). The true treatment effect is denoted by the red dashed line. The coverage rates and RMSEs (of ATE) for each method are also shown.}
\label{fig:pooroverlap_summary}
\end{figure}

We repeat this process across 500 iterations, drawing different random functions each time with sample sizes of 500 as described above. Figure~\ref{fig:pooroverlap_summary} illustrates the behaviors of each method across these iterations. The coverage rate for the CATE at each possible value of $X$ (top left) is problematic for the linear model at all values of $X$. BART generally over-covers in areas of common support but radically under-covers in non-overlap regions, dropping below 50\% on one side and 25\% on the other. The GP achieves approximately nominal coverage where common support is good. In non-overlap areas, it sustains good coverage initially on one side, but still transitions to undercoverage, dropping as low as roughly 75\%. While this undercoverage is much less severe than the others, it remains imperfect. This imperfection serves as a warning of the limitations of the GP approach. Specifically, the example is an adversarial one with common support failing for half of the range of $X$, while the underlying DGPs are quadratic functions that can diverge radically beyond the edges of their training data. In practice, we would recommend GP with a non-stationary kernel when this degree of extrapolation is required (see Section~\ref{sec:its}).

The interval lengths required to achieve this (top right) are also of interest. The drastic under-coverage of LM results from inappropriately short intervals. BART maintains almost constant interval lengths throughout, which are excessively long for the region of good common-support. GP exhibits adaptive behavior, maintaining narrow intervals in regions of good overlap and wider intervals in areas of poor overlap. 

Finally, while conditional uncertainty and coverage are of potential interest for many queries and demonstrate the behavior of interest here, investigators may be interested primarily in average effects. Even here, however, the poor behavior of LM under weak overlap leads to an RMSE 1.52 times larger than that of GP, together with severe undercoverage (55\%). 

In~\ref{app:pooroverlap}, we expand these findings with additional simulations across various function spaces and signal-to-noise ratios ($R^2$).

\subsubsection{Illustration with a multivariate benchmark under varying overlap} \label{subsec:lalonde}

In ~\ref{app:lalonde} we use the well-known National Supported Work Demonstration (NSW) example \citep{LaLonde1986, dehejia1999causal} to demonstrate how to apply GP for covariate-adjusted comparison with multiple covariates, with varying overlap quality. GP produces a point estimate very close to the experimental benchmark. Further, the counterfactual uncertainty estimates from GP demonstrate their expected relationship to local data density.

\subsection{Interrupted time series with GP} \label{sec:its}

\paragraph{Background.} The ITS design is used to assess the effect of events or shocks experienced universally after a specific time point \citep{box1975intervention, box1976time, bernal2017interrupted}. We observe a time series of non-treatment outcomes, $Y_t(0)$ prior to the event of interest. The event occurs at time $t=T$ and all $Y_t$ outcomes measured after that time are treatment outcomes, $Y_t(1)$. We train a model on pre-treatment data then must extrapolate it to the post-treatment era. These extrapolated predictions of $Y_t(0)$ are then compared to actual post-treatment outcomes ($Y_t(1)$) at any $t \geq T$.

The GP approach is potentially valuable here for several reasons. First, it handles autocorrelation in the outcomes naturally. Second, the covariance function (kernel) can be designed to accommodate not only the ``smoothness'' we expect but also secular trends that will continue on over time, and periodic trends such as seasonality. Third and most relevant to our discussion, its estimates of uncertainty take into account the distance (in time) to the observed data and the consequent model dependency. We note that regardless of estimation approach, making causal claims from ITS requires additional demanding identification assumptions (See~\ref{app.causalITS}). 

\begin{figure}[hbt!]
\centering
\includegraphics[width=0.9\linewidth]{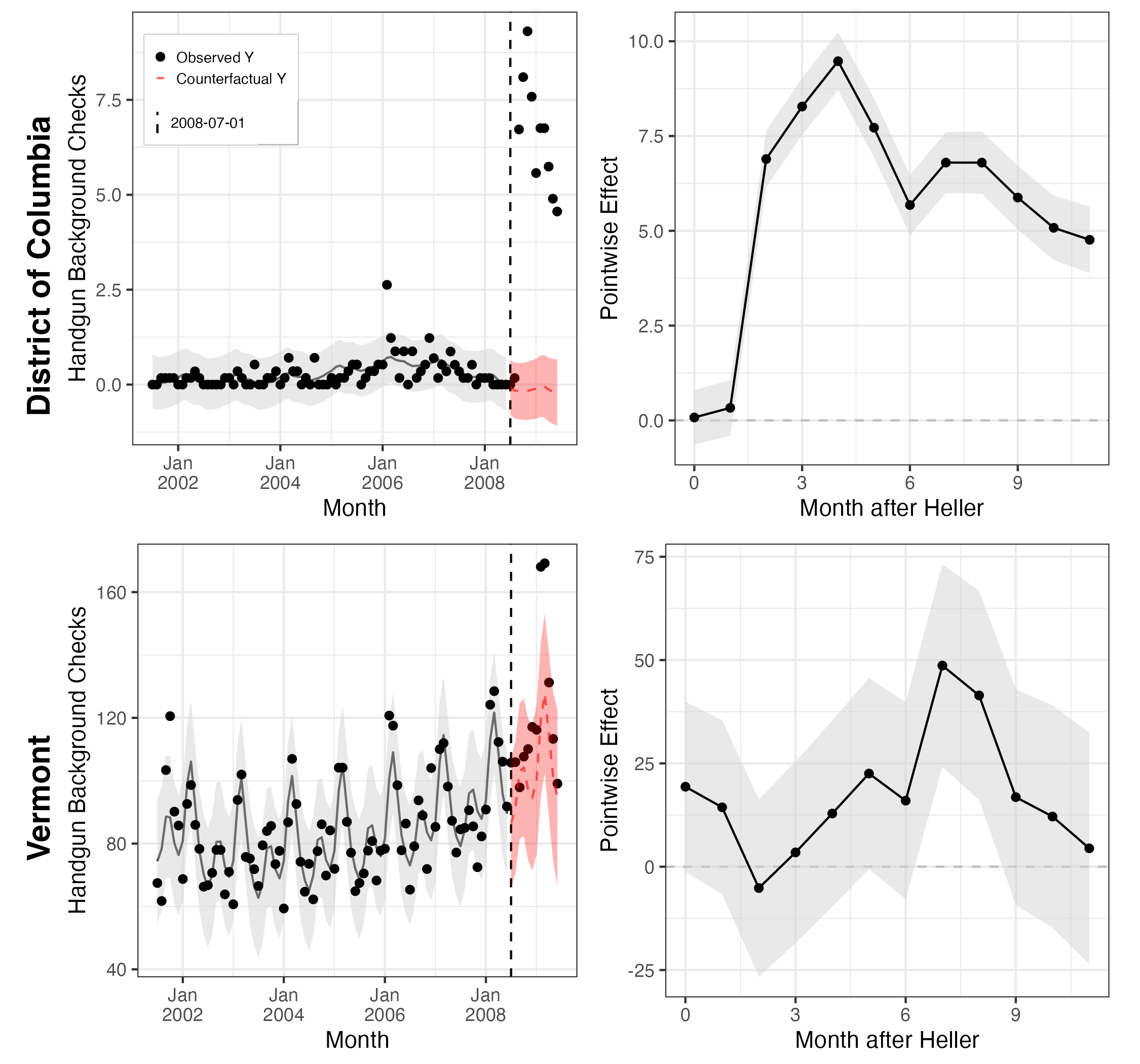}
\caption{\textit{Left column:} Monthly rate of handgun background checks per 100,000 population in D.C. (top) and Vermont (bottom), 7 years before and 1 year after the \textit{Heller} ruling. The red envelope shows the 95\% predictive interval for the GP-estimated non-treatment outcome, post-treatment. \textit{Right column:} Point estimates and confidence band for the differences between the observed outcome (treated) and the predicted (non-treatment) counterfactual at each month. The kernel choice combines (adds) the linear, periodic, and Gaussian kernels.}
\label{fig:its_heller}
\end{figure}

\paragraph{Illustrative application.} In June 2008, the Supreme Court ruling in \textit{District of Columbia v. Heller}, 554 U.S. 570 (2008) affirmed the individual's right to keep and bear arms for self-defense and other purposes, striking down DC's prior ban. We examine how the number of handgun background checks per 100,000 population (as a proxy for handgun sales) responded to this decision.\footnote{Thanks to Jack Kappelman for these data.}  Figure~\ref{fig:its_heller} shows results for DC and Vermont, with seven years of pre-treatment and one year of post-treatment data. In DC, the increase in background checks immediately after \textit{Heller} is substantial compared to the expectations suggested by the GP model, even given additional uncertainty due to extrapolation. Results are less clear in Vermont, with many of the post-treatment observations falling within the interval expected post-treatment (Figure~\ref{fig:its_heller}, \textit{bottom}). 

\begin{figure}[hbt!]
\centering
\includegraphics[width=1\linewidth]{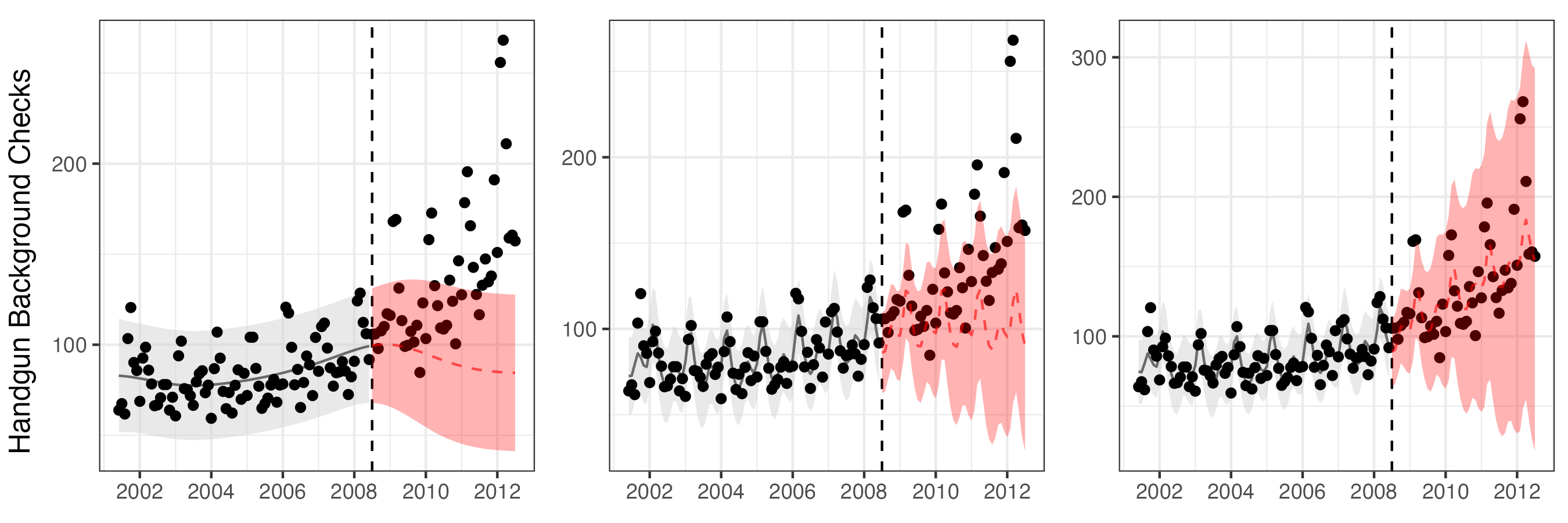}
\caption{GP results with three different kernels for Vermont. The post-treatment period is extended to four years to better illustrate the behavior under extreme extrapolation. \textit{Left.} Gaussian kernel;  \textit{Center.} ``Gaussian + periodic + linear'' kernel; \textit{Right.} ``Gaussian + periodic + quadratic'' kernel.}
\label{fig:its_heller_vermont}
\end{figure}
 
Further, the choice of kernel encodes the set of extrapolating functions the investigator deems plausible. Figure~\ref{fig:its_heller_vermont} demonstrates the impact that these choices can have. The left panel shows results with only the Gaussian kernel. As a stationary kernel, this is not suitable for conservative extrapolation as noted above. 

We then consider two additions to the kernel. First, we add a periodic kernel (with a period of one year) to capture possible annual cycles. Second, we consider polynomial functions that describe how the CEF is allowed to extend beyond the edge of the data. The \textit{center} panel of Figure~\ref{fig:its_heller_vermont} employs a Gaussian + periodic + linear kernel. A sizable fraction of post-treatment observations still fall above the predicted counterfactual level. Adding a quadratic growth component rather than linear (\textit{right} panel), however, the observed post-treatment outcomes fall almost entirely within the predicted envelope, implying no estimated effect.

We do not expect the user will typically know which choice best describes how $Y(0)$ would evolve once it becomes unobservable. Further, the observed pre-treatment data are not well-equipped to choose among these options without additional strong assumptions. The benefit of GP is not that it can somehow ``know'' the right extrapolation model, but that it provides a tool to consider capacious spaces of functions while fully characterizing uncertainty over predictions given this. The price or our inability to rule out more extreme extrapolating models (e.g. higher order polynomials) is seen in greater uncertainty over the post-treatment counterfactual, and weakened ability to make inferences. The GP makes this tradeoff explicit and clear.  This approach can also be used in the inverse, revealing what assumption (on the extrapolation function) a user would have to defend in order to support a given conclusion. Here, for instance, to claim there is a putative effect requires arguing that the extrapolating function is linear, but not quadratic.

\subsection{Regression discontinuity} \label{sec:rd}
The RD design is a widely used tool in the social sciences for cases when decisions about treatment turn on a sharp cutoff in some variable. While this approach rests on relatively credible identification assumptions, it poses estimation challenges as it relies on point estimates with uncertainty intervals precisely where the underlying model(s) are at the very edge of their support. 

When fitting models below/above the cutoff in RD, decisions must be made regarding the bias-variance tradeoff: how much should we seek to reduce variance by using observations farther from the cutoff, given that including such data can bias our estimate? Even when optimizing this choice by some criterion, the consequent bias must be dealt with when estimating uncertainty. \citet{stommes2023reliability} review varied approaches to dealing with this bias and its implications for inference. Among such approaches, \citet{calonico2014robust} provide a now standard optimal bandwidth selection technique, bias-adjustment of the point estimate, and bias-adjusted inference, which we refer to by the name of the widely used software implementation, \texttt{rdrobust} \citep{calonico2015rdrobust, calonico2017rdrobust}. 

The GP approach is attractive for RD, first, because it offers a flexible function space that makes only weak assumptions about smoothness (when using the Gaussian kernel, for example). Second, and more importantly, it provides a principled approach to uncertainty quantification for edge prediction or mild extrapolation. In cases with considerable data near the threshold, an optimally chosen parametric or semi-parametric approach, like those employed by \texttt{rdrobust}, can establish reliable point estimates and uncertainty intervals at the edge of the data, as demonstrated by \citet{calonico2014robust}. However, with smaller samples or sparse data near the cutoff, multiple models may fit the observed data similarly well while yielding different predictions at the boundary. The GP approach provides uncertainty estimates that retain this  uncertainty over the CEF, allowing inference without relying on large-sample asymptotics.  

\subsubsection{RD case 1: Total random simulation.} 
We first consider the simple ``total random case'', in which the running variable ($X$) and the outcome ($Y$) are both drawn independently at random, with zero treatment effect.\footnote{Here, we examine RD designs with a single running variable and no covariates. As shown by \cite{rd_covariates}, including covariates may be desirable in some cases, if the intention is to improve precision. To include covariates in GP, they need only be included as additional columns in the data alongside the running variable. The kernel matrix entries are computed over the running variable and covariates, just as it is in the multivariate examples above. The RD estimator itself is still taken with respect to the running variable: for each individual $i$, the predicted $Y_i$ is obtained from the model below the cutoff and from the model above the cutoff, setting $X$ to the cutoff value but leaving the covariates at observed values.  Consequently, the models account for any differences between groups on covariates above and below the cutoff, through their impact on the predictions of $Y$.} For each of 500 simulations, we draw a sample of size 500, with $X_i \sim \mathcal{N}(0, 1)$, and $Y_i \sim \mathcal{N}(0, \sigma^2)$. The value of $\sigma$ is varied from near zero up to 3, reflecting various possible scales of the outcome variable relative to the forcing variable. We use a cutoff of 0 for $X$ to determine treatment eligibility status. We employ \texttt{rdrobust} with software defaults and utilizing the bias-adjusted point estimates with the ``robust'' confidence intervals. To compute the GP estimate (\texttt{GPrd}) we first use the GP function from \texttt{gpss} to estimate a model for $Y$ given $X$ for the treated units ($X > 0$), and another for the control units ($X \leq 0$). From this we obtain the point estimate and standard error for the prediction of the expected $Y$ at precisely $X=0$ from each model. The treatment effect is estimated as the difference between these point estimates, with standard error given by the square root of the sum of the variances of the two predicted outcomes.

\begin{figure}[hbt!]
\centering \textit{(a)} True effect of zero, $n=500$\\
\includegraphics[width=\linewidth]{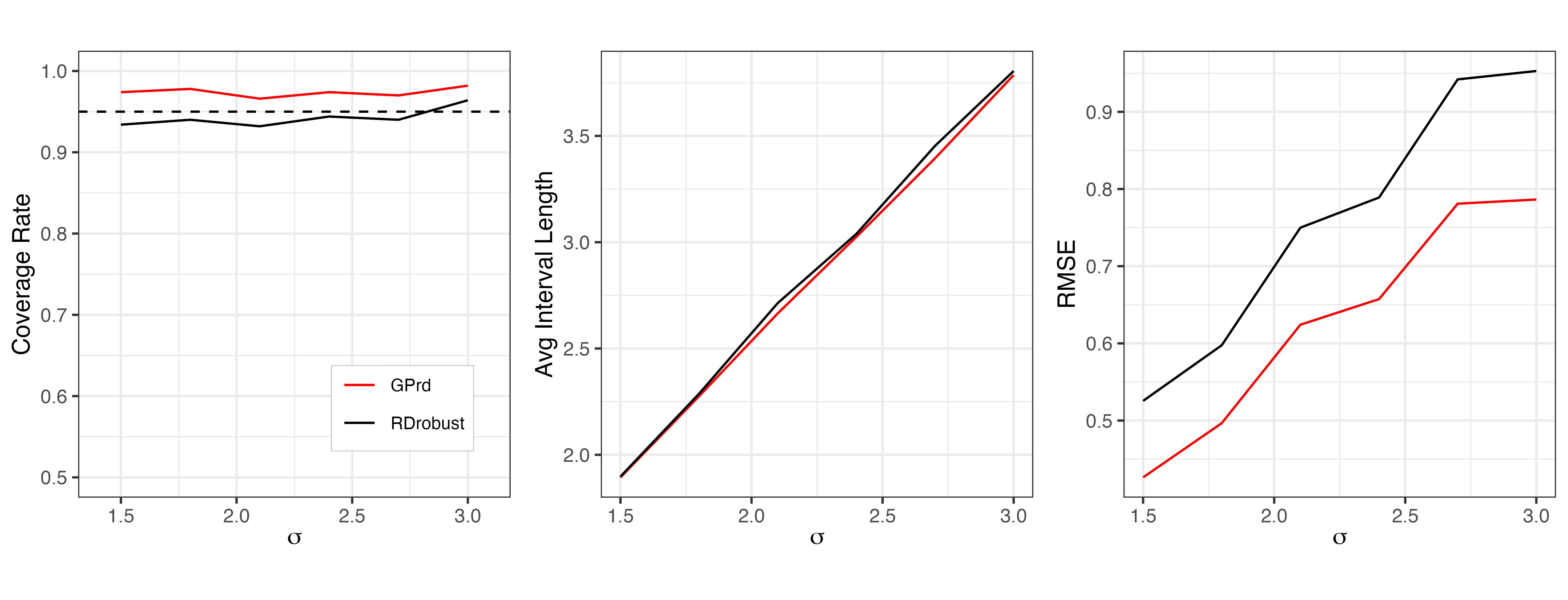}\\
\centering \textit{(b)} True effect of zero, $n=100$\\
\includegraphics[width=\linewidth]{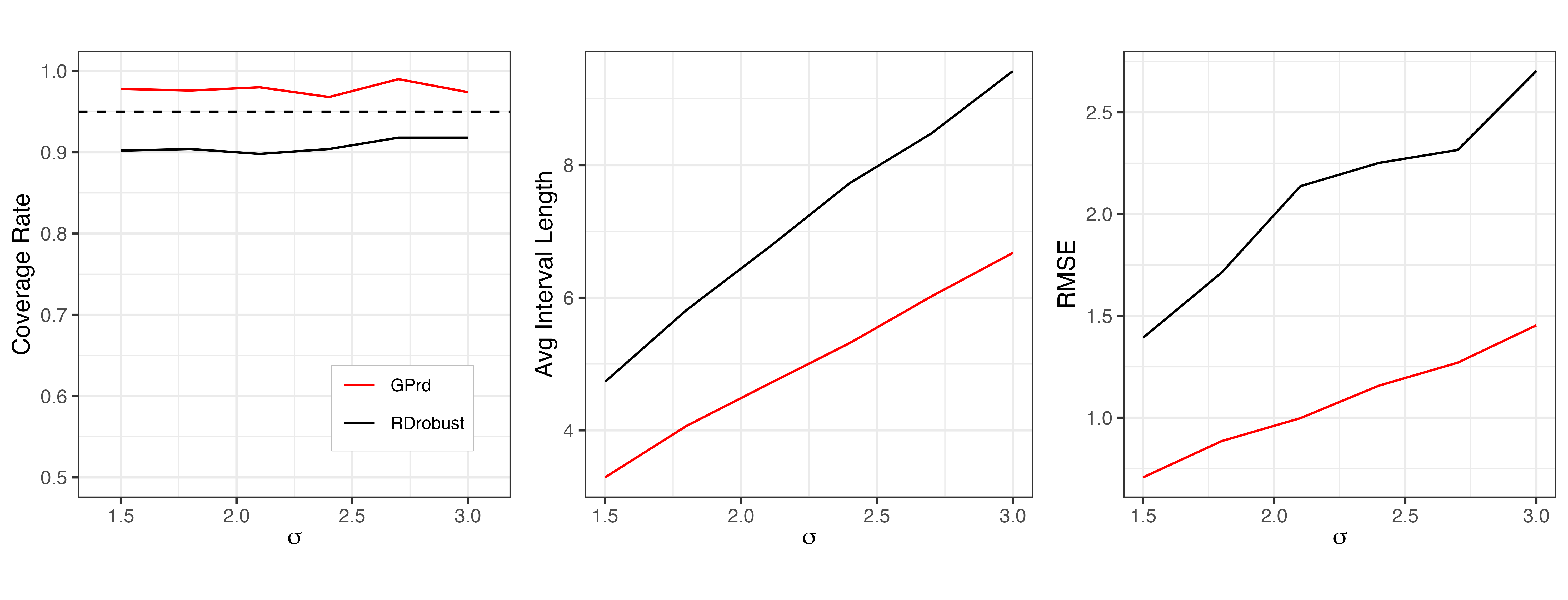}
\caption{The coverage rate, average length of 95\% confidence interval, and RMSE of \texttt{GPrd} (red) and \texttt{rdrobust} with ``robust'' options (black) in the total random setting. The ratio of the variance of $Y$ to $X$ is given by $\sigma$ on the horizontal axis. \textit{Top row:} True effect size is zero, 500 observations. \textit{Bottom row:} Effect size is zero, sample size is reduced to 100. Results with a true effect size of three look identical.}
\label{fig:rdd_totalrandom_summary}
\end{figure}

Figure~\ref{fig:rdd_totalrandom_summary} shows simulation results in this setting. Looking first at the top row, we see that \texttt{GPrd} has a higher coverage rate across all levels of $\sigma$, at 97-98\%, while \texttt{rdrobust} has a coverage rate between 93\% and 96\%. The two approaches have essentially identical interval sizes, which appropriately scale linearly with $\sigma$ (middle column). \texttt{GPrd} also shows much smaller RMSE across the 500 simulations. This RMSE benefit is amplified in smaller samples (bottom row). We examine and offer an explanation for this behavior in \ref{subsubsec:whyrmse}.

\subsubsection{RD case 2: Latent variable confounding simulation.}
\begin{figure}[hbt!]
\centering
\includegraphics[width=.8\linewidth]{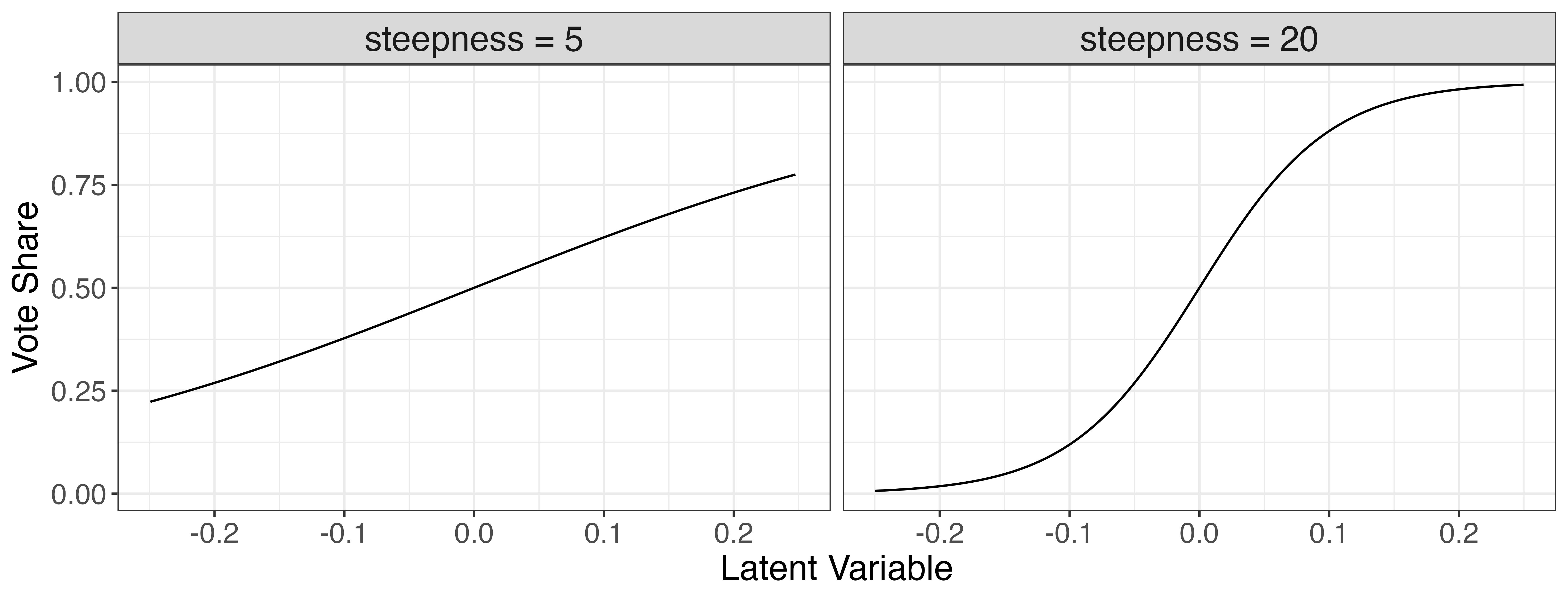}
\caption{Relationship between simulated latent variable $\mu$ (horizontal axis) and vote share $X$ (vertical axis) at two steepness ($s$) parameters used in the simulation.}
\label{fig:latentvar_shape}
\end{figure}

The total random case is a useful starting point, but may not produce sufficient risk of bias because there is no relationship between $X$ and $Y(d)$ on either side of the cutoff such that incorporating information farther from the cutoff risks adding bias to the estimate. To better simulate this concern we consider a latent confounding formulation. In each of 500 replications, we generate a random sample of size 200 according to the following:
\begin{enumerate}
\item Draw a latent variable: $\mu \sim \mathcal{U}(-0.25, 0.25)$.
\item Create our running variable, $X$, loosely resembling vote share in an electoral RD as a sigmoidal function of $\mu$, with $x_i = 1/(1 + exp(-s*\mu_i))$. The parameter $s$ represents a varying degree of steepness of this sigmoid, such that larger values produce a steeper sigmoid, increasing the risk of bias due to misspecification. See Figure~\ref{fig:latentvar_shape}.
\item The treatment $D$ represents ``winning the election'', coded as 1 if the vote share is greater than 0.5, and 0 otherwise. 
\item Generate the outcome $Y_i = 1.5 \mu_i + \tau D + \varepsilon$ where $\tau=3$ and $\varepsilon \sim \mathcal{N}(0, \sigma^2)$. The $\sigma$ will be varied by simulation setting to explore a range of realistic signal-to-noise ratios ($R^2$). 
\end{enumerate}

Next, purely data-driven RD specification risks using distant observations to estimate $\E[Y|X]$ at the cutoff, compromising the motivating logic for using RD. To address this concern, we consider options that set the bandwidth or trim the sample (or both) based on outside causal assumptions:
\begin{itemize}
\item \texttt{gp\_causal}: The investigator proposes a kernel bandwidth based on what they are willing to believe about how close units may covary in their outcomes. For example, it might be reasonable to propose that two observations whose vote shares are 2\% apart may be highly correlated, e.g., at 0.9. This implies a surprisingly small value for b at 0.005.\footnote{$k(X_i,X_j) = exp \left(-\frac{(X_i-X_j)^2}{b}\right) \rightarrow 0.9 \approx exp \left(-\frac{0.02^2}{0.005} \right)$. Correlations equal covariances here due to the rescaling of the outcome. The running variable is not rescaled, so that the 0.02 in the numerator remains correct.}
    
\item \texttt{gp\_causaltrim}: After choosing the \texttt{causaltrim} bandwidth, regions that are effectively irrelevant to estimation at the cutoff can be removed. This aids in transparency, but is also helpful because the value of $\sigma^2$ is a single global parameter, so trimming the data first ensures it is determined over only the range of the relevant range of the sample. In our simulation, we remove observations with vote shares below 0.4 or above 0.6, on the premise that users would likely argue that data outside this range are of no value to modeling what happens at the cutoff.
\end{itemize}

\begin{figure}[hbt!]
\centering
\includegraphics[width=\linewidth]{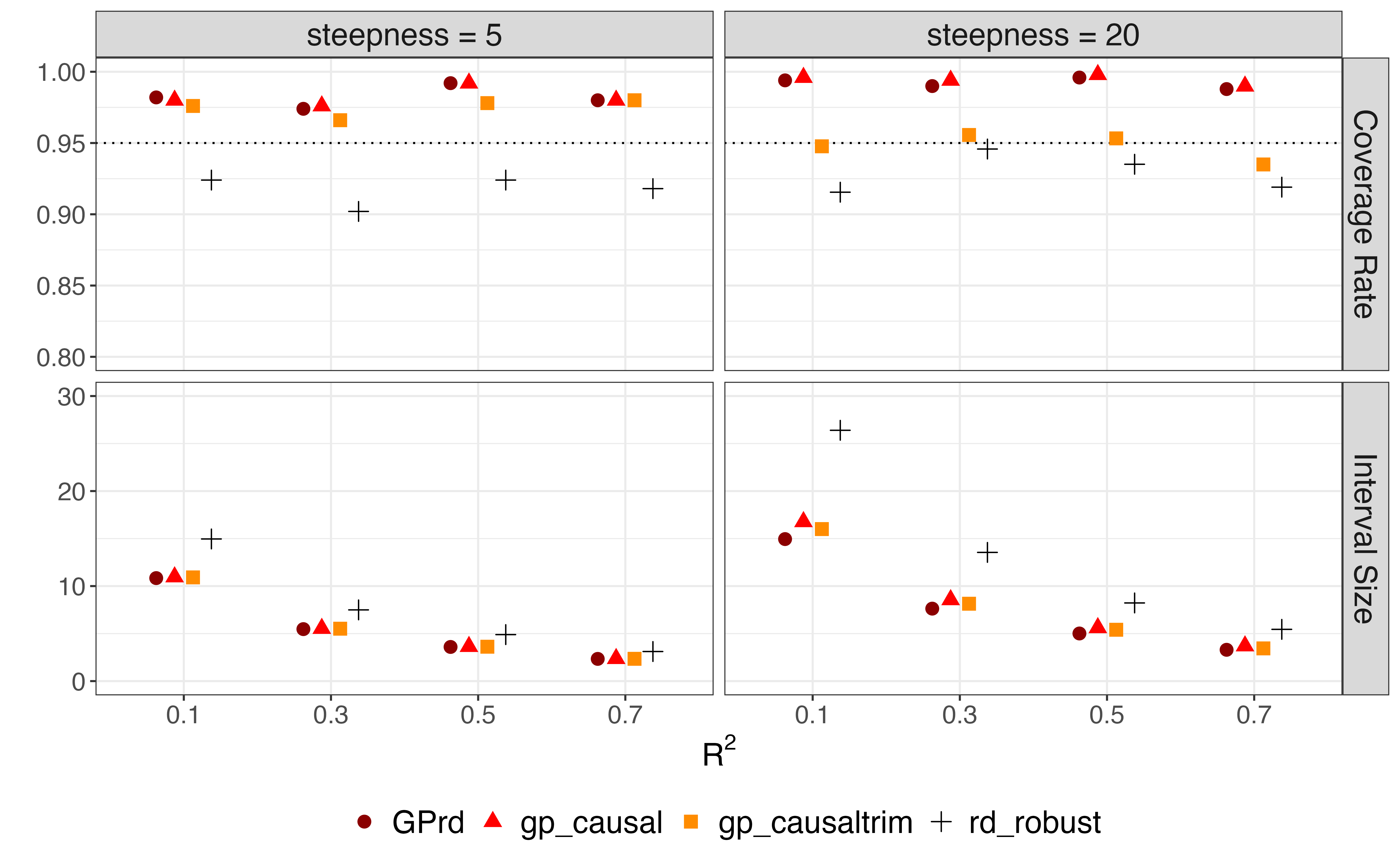}
\caption{RD simulation results with latent variable (effect size = 3). Each point represents the coverage rate, or average length of 95\% confidence interval of \texttt{GPrd, gp\_causal, gp\_causaltrim}, and \texttt{rdrobust} across 500 iterations of the latent variable simulations (n = 200). The horizontal axis represents the $R^2$.}
\label{fig:latentvar}
\end{figure}

Figure~\ref{fig:latentvar} summarizes the results over 500 iterations at each choice for the steepness ($s$) of the latent relationship between $\mu$ and $Y$ and at four different levels of $R^2$. All the approaches show similar interval sizes, with the exception of \texttt{rdrobust}, which yields wider intervals when steepness equals 20. While \texttt{rdrobust} tends to slightly under-cover the true effect, the GP-based approaches exhibit slight over-coverage, except for \texttt{gp\_causaltrim} at a steepness of 20. 

\subsubsection{RD case 3: Empirical application with benchmarking by placebo cutoffs}

Finally, we study performance of \texttt{GPrd} on real data, in a case where the ``true'' effect is knowable for benchmarking purposes. To do so, we consider the data from \citet{lee2004}, in which the forcing variable is Democrats' (two-party) vote share in US House Elections (1948 - 1990), and the outcome of interest is a measure of how liberal each representative's vote record is assessed to be, called the ADA score. While we cannot know the true effect size, it would be arguably zero when using any value other than 0.5 as a (placebo) cutoff.\footnote{In each placebo analysis, data are included only from below 0.5 or above it, but not from both, to avoid potential contamination due to a real treatment effect.} 

\begin{figure}[hbt!]
\centering
\includegraphics[width=0.8\linewidth]{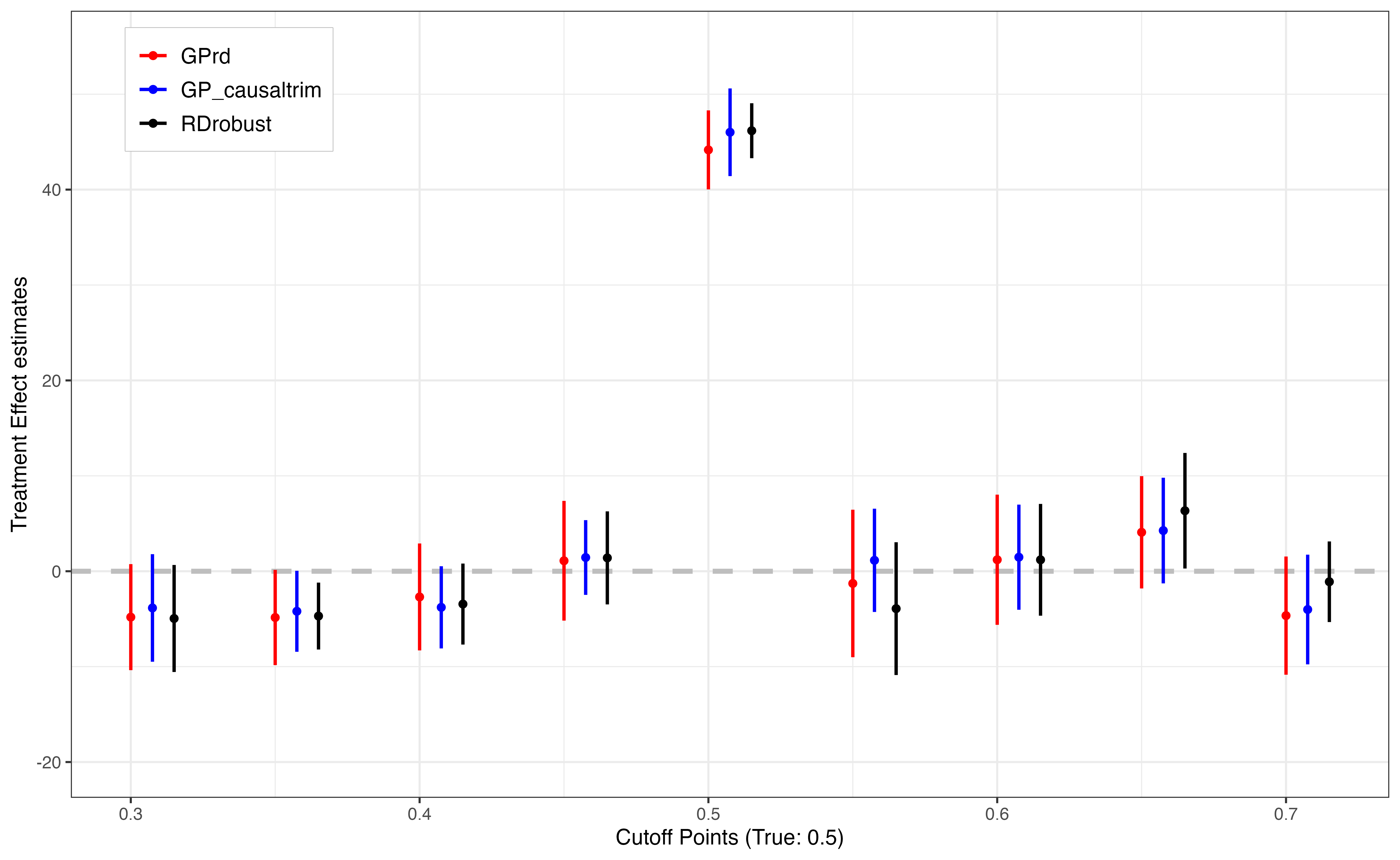}
\caption{RDD estimates using close election data with placebo cutoff points and different GP RD Specifications. The 95\% confidence interval of \texttt{GPrd} (\textit{red}), \texttt{gp\_causaltrim} (\textit{blue}), and \texttt{rdrobust} (\textit{black}) across eight different placebo cutoffs and the true cutoff point at 0.5 are presented.}
\label{fig:rdd_lee}
\end{figure}

We consider three estimators: \texttt{GPrd}, \texttt{gp\_causaltrim}, and \texttt{rdrobust}. Figure~\ref{fig:rdd_lee} shows results with the full data. The choice of estimator matters little to the point estimates at the true cutoff (0.5), though \texttt{GPrd} and \texttt{gp\_causaltrim} are more conservative in their uncertainty estimates. Looking at the eight placebo cutoffs, the differences are relatively small and while \texttt{rdrobust} suggests a statistically significant estimate at the cutoff of 0.35, \texttt{GPrd} and \texttt{gp\_causaltrim} very nearly do as well. At the cutoff of 0.65, the \texttt{rdrobust} interval again excludes zero, while the intervals for the GP approaches do not, but the actual difference is again relatively small.

However, the \citet{lee2004} data initially contains 13,577 observations without missingness on the running variable and outcome. While each of the placebo cutoff analyses uses only a subset of these points, there are still thousands of observations available to each of these estimates. The similarity of the GP and \texttt{rdrobust} approaches in this context fits the prediction that they should be similar with ample data. We also wish to know whether GP provides a suitable approach to RD for smaller sample sizes. We thus use the same data as the basis for a second analysis in which we examine much smaller sub-samples. Specifically, we
\begin{itemize}
\item Fix a (placebo) cutoff point: 0.3, 0.4, 0.5, 0.6, 0.7 (0.5 is the true cutoff).
\item Limit the data to just 0.1 above and below the cutoff in question (e.g., for cut=0.4, data with $0.3<X<0.5$ are used)  to maintain symmetry and comparability across cutoff estimates.
\item Sub-sample 200 observations (without replacement) from the remaining eligible observations.
\item Estimate the treatment effect and its uncertainty using various models.
\end{itemize}

\begin{figure}[hbt!]
\centering
\includegraphics[width=0.9\linewidth]{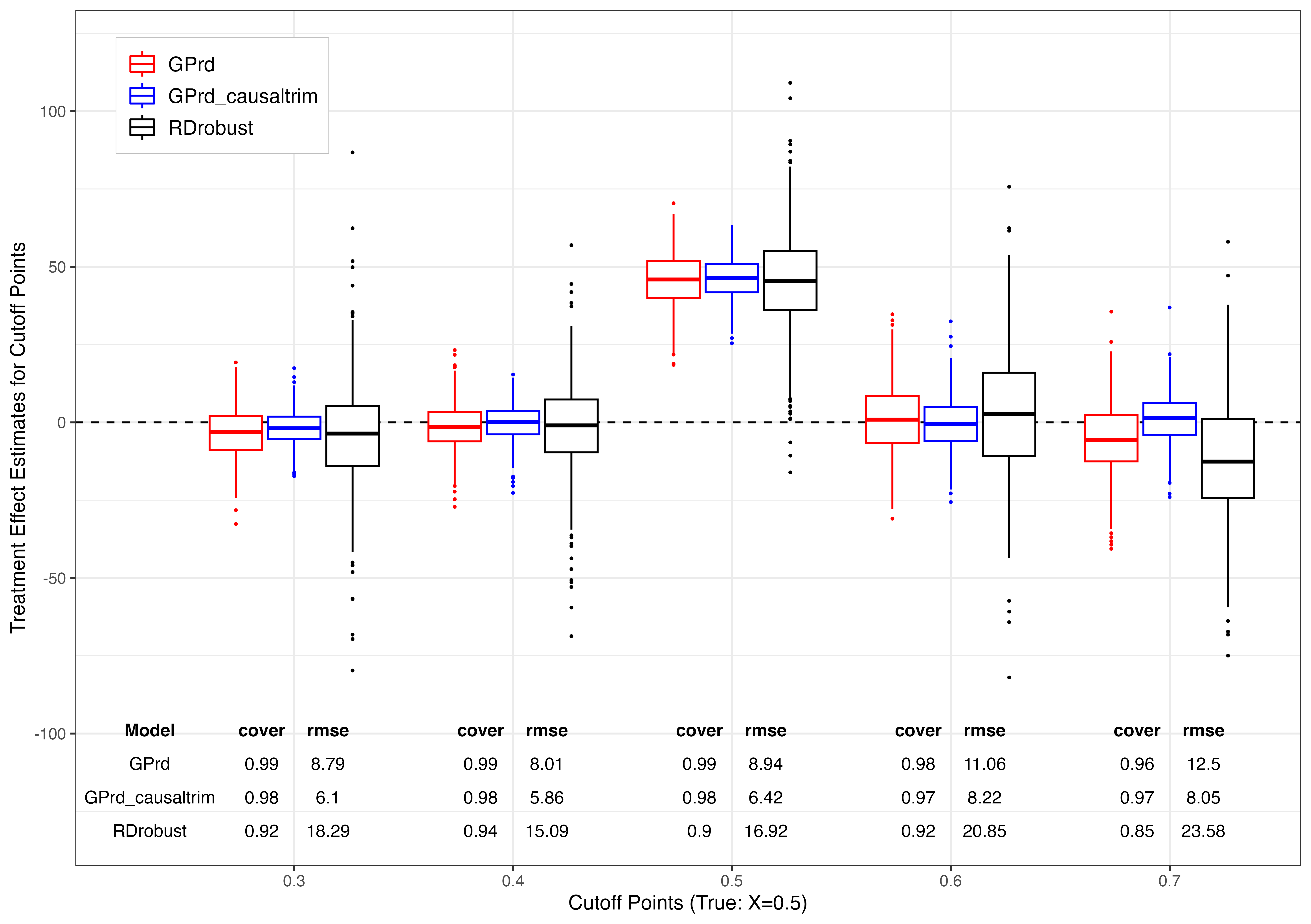}
\caption{RDD placebo cutoff subsampling simulation. Boxplots show the distribution of treatment effect estimates from \texttt{GPrd} (\textit{red}), \texttt{gp\_causaltrim} (\textit{blue}), and \texttt{rdrobust} (\textit{black}) across four different placebos and true cutoff point at 0.5.}
\label{fig:rdd_lee_subsample}
\end{figure}

We repeat this 500 times. Figure~\ref{fig:rdd_lee_subsample} shows the results. For coverage and RMSE calculations we use the ``full sample'' estimates above as if it were the long-run target to reveal the change in behavior due solely to sample size. In general, the GP approaches are similar and perform well. Coverage rates vary from 96\% to 99\% across the GP methods, while the \texttt{gp\_causaltrim} approach shows greater efficiency. As this approach makes stricter (and arguably transparency- and credibility-improving) choices, it is fortunate to also find it has desirable behavior. The \texttt{rdrobust} approach is not designed for the small sample setting, but we include it here to illustrate its sensitivity to sample size for those who might use it in smaller sample settings regardless. It shows occasional erratic estimates (omitted from Figure~\ref{fig:rdd_lee_subsample}) with RMSE values roughly twice those of \texttt{GPrd}.\footnote{Some extreme \texttt{rdrobust} estimates are truncated for graphical legibility. We set a truncation threshold (drop if $|\hat{\tau}| 
\geq 150$) which retains over 99\% of the estimates while maintaining the visual clarity of the figure. For more discussion on erratic estimates of \texttt{rdrobust} with small sample sizes in simulation settings, see \ref{app.fail}.} 

\paragraph{Heteroskedasticity.} As noted, the GP prior and the likelihood procedure for choosing $\sigma^2$ assume the DGP has normal errors with constant variance. In ~\ref{app.hetero}, we study three types of heteroskedastic noise, varying by degrees up to an extreme where the variance at some locations in $\mathcal{X}$ is four times that at others. Using the ``fully random'' simulation setting, across types, intensities of heteroskedasticity, and sample sizes (ranging from 100 to 500), \texttt{GPrd} performs very well on coverage rates, interval lengths, and RMSE in most circumstances. The exception occurs with one heteroskedasticity pattern (decreasing variance towards the cutoff, from both sides), with larger sample sizes and greater heteroskedasticity, where GPrd still has conservative coverage and preferable RMSE, but larger interval sizes.

\paragraph{Improved RMSE: Why and when?}\label{subsubsec:whyrmse} 
Though our initial motivation for employing GPs for RD was their inferential approach given the edge estimation problem, their most notable advantage is on RMSE. This improvement stems neither from employing more flexible functions, nor from the occasional extreme estimates produced by \texttt{rdrobust} in small samples. Even within the interquartile range, \texttt{rdrobust} exhibits considerably more variability than \texttt{GPrd}; see Figure~\ref{fig:rdd_lee_subsample} and ~\ref{app.fail}). As \citet{ornstein2022gaussian} emphasize, local polynomial models may be especially susceptible to extreme behaviors at the edges.  Complementing this, we argue that \texttt{GPrd} has especially low RMSE due to the role of $\sigma^2$ in regularizing the estimated CEFs on each side of the cutoff. When a small number of observations deviate substantially from others on a given side of the cutoff, GP treats them as noise rather than substantially deflecting the CEF to better fit them.\footnote{This regularization of CEFs on either side of the cutoff mutes only the influence of abrupt, localized changes near the boundary. Recall also that the mean/intercept of each CEF is not regularized. Consequently, this approach does not shrink the RD estimate itself towards zero. This is evidenced in our simulations, as they show identical RMSE regardless of the true effect size used.} 

To investigate this, in \ref{app.rdd.sims} we add seven simulation settings from prior RD literature, modified to control the signal-to-noise ratio (``true $R^2$''). In five of the cases, \texttt{GPrd} performs well, with nominal or better coverage and lower RMSE than \texttt{rdrobust}. One simulation (``Ludwig \& Miller'') includes a very severe deflection in the CEF at the cutoff, and on this case \texttt{GPrd} underperforms \texttt{rdrobust}. The ``Lee'' simulation also shows a substantial deflection. Here, \texttt{GPrd} outperforms \texttt{rdrobust} on RMSE when there is sufficient noise ($R^2 \leq 0.55$) but has higher RMSE at lower noise levels. These results fit with the reasoning that \texttt{GPrd} should excel where regularization of the CEF (especially near the cutoff) is a ``good bet''. We note, however, that GP can still handle relatively steep CEF changes---as demonstrated by the performance of all GP methods with the steep sigmoidal CEF in Figure~\ref{fig:latentvar}, across all noise levels.

\section{Discussion}

Our analysis shows that GPs provide a practical and principled tool for causal inference problems that hinge on predictions at the edge of the data. Their central advantage is the ability to incorporate extrapolation uncertainty, widening intervals as predictions rely more heavily on assumptions beyond the observed support. We demonstrate both the benefits and limitations of this approach in settings where poor overlap or  extrapolation pose the greatest risks, including treatment–control comparisons with limited covariate support, ITS, and RD, particularly in small to moderate samples.

A second contribution is to make GP methods more accessible for applied researchers. We (i) explain the approach in terms familiar to social scientists, with emphasis on uncertainty quantification; (ii) simplify and automate the fitting procedure, including hyperparameter management; and (iii) provide an implementation in the R package \texttt{gpss}, available at \href{https://doeunkim.org/gpss/}{http://doeunkim.org/gpss}.

\subsection{Limitations and future research}
Several limitations of the GP framework deserve emphasis. First, GP performance depends on its assumptions. The kernel function $k(X_i,X_j)$ is taken to approximate $cov(Y_i,Y_j)$. In practice, universal kernels such as the Gaussian perform well for predictions near observed data (within the kernel width), even if they do not match the true covariance structure exactly. For more extreme extrapolation, however, non-stationary kernels may be needed to encode plausible ways the CEF could evolve beyond the data, as we illustrated in the ITS setting. In addition, the GP prior assumes that $Y$ follows a multivariate normal distribution, and our procedure for choosing $\sigma^2$ relies on maximum likelihood under this assumption. Heteroskedastic or non-Gaussian residuals can violate these conditions. We briefly explored heteroskedasticity in the RD setting (~\ref{app.hetero}), but more work is needed to understand when such violations materially affect inference, and how to relax the assumptions. Finally, $\sigma^2$ acts as a regularization parameter, smoothing the fitted CEF. This shrinkage is often beneficial but can be costly when the true CEF features sharp deflections or discontinuities.

A second set of limitations concerns computation. Many GP implementations, including \texttt{gpss}, scale poorly beyond a few thousand observations because of the cost of constructing and inverting the kernel matrix. Promising approaches such as Nyström approximations, random feature methods, and kernel sketching (e.g., \citealp{chang2024generalized}) may help extend GPs to larger datasets.

Third, our empirical investigations are still limited. We have demonstrated performance in settings with one or a modest number of covariates (up to 10 in~\ref{app:lalonde}), but not yet in higher-dimensional problems. Likewise, while we have begun to examine how GP compares to other approaches in the various settings above, future work should further compare GP with other methods across varied tasks and DGPs.

Finally, the logic developed here suggests potential applications beyond the designs we studied. In problems of generalizability and transportability, for instance, GP provides a natural way to construct treatment effect estimates for populations with different covariate distributions, while reflecting the added uncertainty when such estimates rely on extrapolation. We leave this and other promising application areas to future research.

\paragraph{Funding Statement}
This research was not funded by any specific grant from any funding agency.

\paragraph{Competing Interests}
The authors declare no competing interests.

\paragraph{Data Availability Statement}
Replication materials are available at \url{https://doi.org/10.7910/DVN/7G092W} \citep{our_data}.

\paragraph{Supplementary Material}
Appendix materials are available at [TBD].

\begin{singlespacing}
\bibliography{bib.bib}    
\end{singlespacing}

\newpage
\pagenumbering{alph}
\appendix
\clearpage
\thispagestyle{empty} 

\begin{center}
{\Large\bf Online Appendix}\\[0.75em]
{\large\bf Inference at the data's edge: Gaussian processes for estimation and inference in the face of extrapolation uncertainty}\\[1em]
Soonhong Cho \quad Doeun Kim \quad Chad Hazlett\\
\end{center}




\vspace{1.5em}
\setcounter{page}{1}

\renewcommand{\thesection}{A.\arabic{section}}
\setcounter{section}{0}

\section{Interpretation and relationship to kernel ridge regression}\label{app.krlscompare}

This section clarifies the relationship between GP and kernel ridge regression. We show that while these methods share identical function spaces for CEF, they differ in their treatment of uncertainty and regularization.

Consider the posterior distribution for the outcomes at the observed points (``fitted values''), i.e., 
\begin{equation}
  \tilde{Y} \mid Y, X \sim \mathcal{N}(\K (\K+\sigma^2 I)^{-1}Y, \K+\sigma^2 I - \K (\K+\sigma^2 I)^{-1}\K^{\top}). 
\end{equation}
That is, though $Y$ for the observations in question are used for training, our posterior distribution after conditioning retains uncertainty even at those points, which is also the uncertainty we would have if we observed new observations at the same points. The CEF (or MAP) can be read from the mean function on the right-hand side:  $\E[Y^{\star} \mid X]=\K (\K+\sigma^2 I)^{-1}Y = \K c$ where $c=(\K+\sigma^2 I)^{-1}Y$. This is precisely the CEF employed by kernel ridge regression techniques such as kernel regularized least squares (KRLS, \citealp{hainmueller2014kernel}). 

Accordingly, we can understand the functional form of the CEF using the interpretations offered by \citet{hainmueller2014kernel}. Briefly, for example, we see that the conditional expectation of $Y$ is linear in the columns of $\K$, not in $\X$. This implies that each observation $i$ is effectively understood by its similarities to all other observations, $k(X_i,X_1),k(X_i,X_2)...,k(X_i,X_n)$, rather than by $X_i$ directly. The model thus predicts $Y_i$ as a linear combination of these similarity measures, i.e., ``[similarity of unit $i$ to unit 1, ..., similarity of unit $i$ to unit $N$]". As with KRLS, regularization is induced here through the $\sigma^2 I$.  

We can similarly view this function space as the set of functions built by placing a Gaussian kernel over every observation and summing these kernels with rescaled coefficients to form the CEF surface. 
The coefficients of the model linear in the kernel, $c$ in the model $\E[Y^{\star} \mid X]=\K c$, are equal to $c=(\K+\sigma^2 I)^{-1}Y$ here, but denoted as $(\K+\lambda I)^{-1}$ in \citet{hainmueller2014kernel}. The $\lambda$ in that setting arises strictly as a tuning parameter governing the degree of regularization, to be chosen by a cross-validation procedure. Here it is replaced by $\sigma^2$, which has a substantive meaning (in our rescaled version) as the fraction of variance in $Y$ that remains unexplained by the CEF. As described in ~\ref{subsec.additionaldetails}, we fit it using a marginal likelihood approach rather than cross-validation. 

\clearpage 

\section{Additional simulation results: Poor overlap case}\label{app:pooroverlap}

This section extends our findings from Section~\ref{sec:pooroverlap} on how GP handles uncertainty in treatment-control comparisons with poor overlap. We present simulation results across various one-covariate DGPs to more comprehensively evaluate GP performance.

\subsection{Simulation Design}
We compare three methods---Linear Model (LM), Gaussian Process (GP), and Bayesian Additive Regression Trees (BART)---across diverse DGPs, incuding both parametric and kernel-based ones. We maintain the same poor overlap setup as in Section~\ref{sec:pooroverlap}: treated units have values of $X$ between –3 and 1, while control units have values of $X$ between –2 and 3, creating two regions with no overlap (below –2 and above 1).

We vary the signal-to-noise ratio using two $R^2$ values (between $Y$ and the true CEF, not between $Y$ and $X$). In our main simulation with the quadratic DGP, we set $R^2=0.5$, representing a moderate signal-to-noise ratio where 50\% of outcome variance is explained by the model. We additionally test $R^2=0.3$ to examine the performance under noisier conditions of more challenging and realistic inference scenarios. The true treatment effect is set to 3 for all simulations.

We examine five distinct functions types:

\begin{enumerate}
\item \textbf{Quadratic function}: $f(x) = \alpha_0 + \alpha_1 x + \alpha_2 x^2$, with coefficients $\alpha_j \sim N(0,1)$. This parametric specification differs structurally from GP's kernel-based approach, producing global trends that challenge the Gaussian kernel's tendency to revert to the (pre-specified) mean beyond data support.

\item \textbf{Cubic functions with bounded derivatives}: $f(x) = \alpha_0 + \alpha_1 x + \alpha_2 x^2 + \alpha_3 x^3$, with coefficients constrained to ensure that the first derivative $f'(x)$ has a maximum absolute value bounded by $c$. We test two constraint magnitudes, $c=5$ and $c=10$, representing different levels of maximum steepness.

\item \textbf{Matérn kernel-based function}: Generated using 10 knots with values $x_j \sim \text{Unif}(-3,3)$ and coefficients $c_j \sim N(0,1)$. The Matérn kernel with smoothness parameter $\nu=1/2$ produces continuous but non-differentiable functions with sharper transitions than those generated by a Gaussian kernel:
\begin{equation}
k(x_i,x_j) = \frac{2^{1-\nu}}{\Gamma(\nu)} \left(\frac{\sqrt{2\nu}|x_i-x_j|}{l}\right)^{\nu} K_{\nu}\left(\frac{\sqrt{2\nu}|x_i-x_j|}{l}\right),
\end{equation}
where $K_{\nu}$ is the modified Bessel function, $l=\sqrt{b}$ is the length-scale parameter (in our notation), and $\nu=1/2$ controls the smoothness. The form of the kernel function for this $\nu=1/2$ value simplifies to the ``exponential'' kernel, i.e., $k(x_i,x_j) = \exp\left(-\frac{|x_i-x_j|}{l}\right)$.

\item \textbf{Gaussian kernel-based function}: Generated similarly to the Matérn functions but using a Gaussian kernel:
\begin{equation}
k(x_i,x_j) = \exp\left(-\frac{||x_i-x_j||)^2}{b}\right).
\end{equation}
\end{enumerate}

\begin{figure}[hbt!]
\centering
\includegraphics[width=1\linewidth]{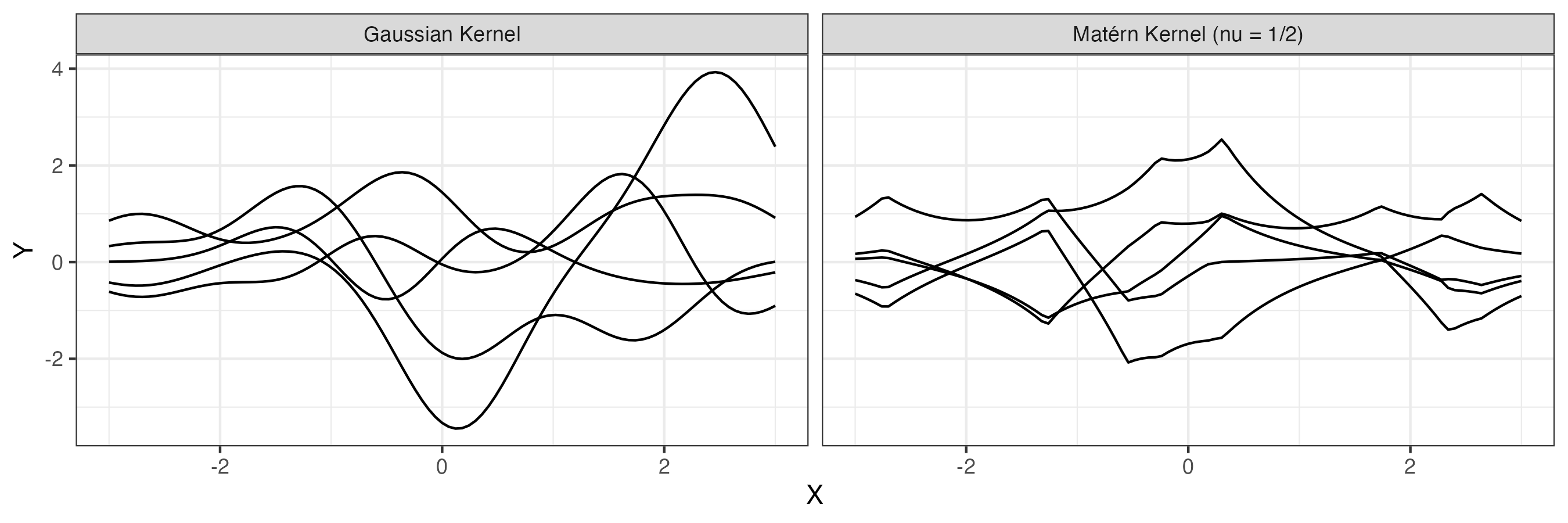}
\includegraphics[width=1\linewidth]{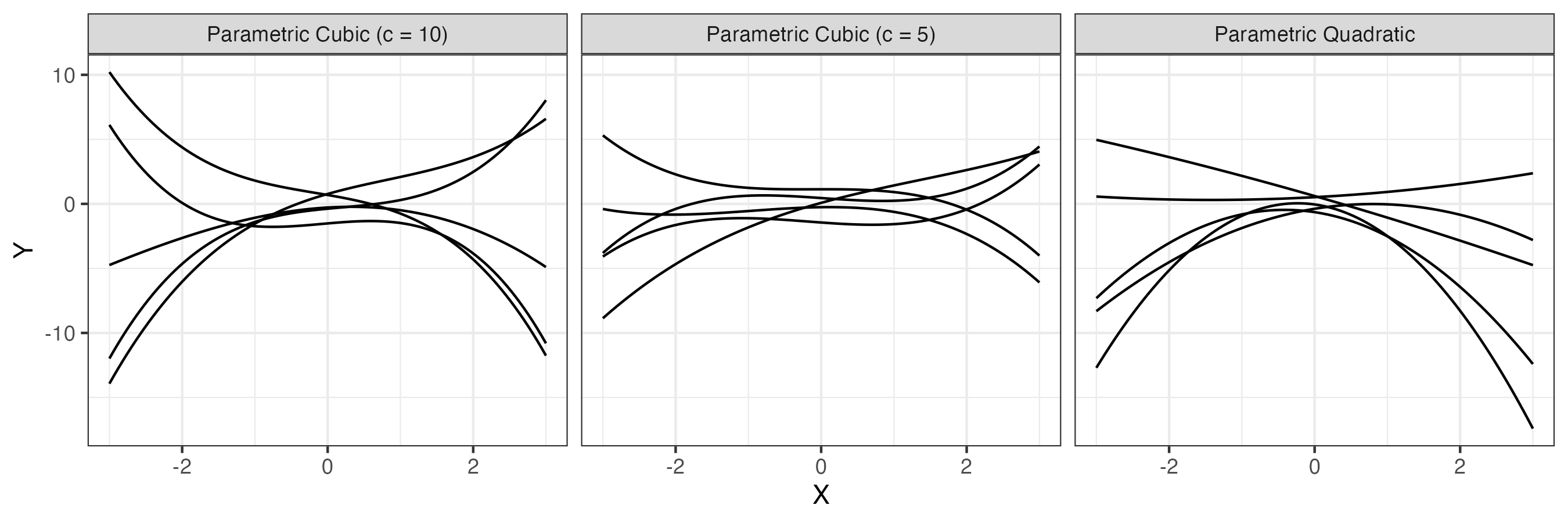}
\caption{5 samples of different random functions.}
\label{fig:sample_functions}
\end{figure}

Figure~\ref{fig:sample_functions} displays five random draws from each DGP, illustrating the diverse function classes being tested. The top row shows functions generated using kernel-based approaches (Gaussian and Matérn), which produce smoother, more localized variations. The bottom row depicts the parametric functions (quadratic and cubic), which exhibit stronger global trends and greater variability in steepness.

For each DGP and $R^2$ combination, we generate observed outcomes with noise to achieve the two signal-to-noise ratios described above. We conduct 500 Monte Carlo simulations, evaluating performance using three metrics: coverage rate, root mean squared error (RMSE), and interval length.

\subsection{Results}

\begin{table}[hbt!]
\centering
\footnotesize
\caption{Comparing the three models in terms of the coverage rate, the RMSE, the interval length of ATE estimation across different DGPs.}
\label{tab:pooroverlap_table}
\renewcommand{\arraystretch}{1.2}
\begin{tabular}{ccccccccccc}
\toprule
& & \multicolumn{3}{c}{LM} & \multicolumn{3}{c}{GP} & \multicolumn{3}{c}{BART} \\
\cmidrule(lr){3-5} \cmidrule(lr){6-8} \cmidrule(lr){9-11}
$R^2$ & DGP & \makecell{Coverage\\Rate} & RMSE & \makecell{Interval\\Length} & \makecell{Coverage\\Rate} & RMSE & \makecell{Interval\\Length} & \makecell{Coverage\\Rate} & RMSE & \makecell{Interval\\Length} \\
\midrule
  0.5 & Quadratic & 0.554 & 1.317 & 1.867 & 1 & 0.866 & 5.339 & 1 & 0.821 & 5.153 \\ 
  0.5 & Cubic, c=5 & 0.878 & 0.412 & 1.253 & 1 & 0.690 & 3.530 & 1 & 0.539 & 3.494 \\ 
  0.5 & Cubic, c=10 & 0.818 & 0.614 & 1.680 & 1 & 1.072 & 4.764 & 1 & 0.849 & 4.663 \\ 
  0.5 & Matérn & 0.870 & 0.330 & 1.015 & 1 & 0.281 & 2.711 & 1 & 0.287 & 2.745 \\ 
  0.5 & Gaussian & 0.786 & 0.458 & 1.095 & 1 & 0.306 & 2.895 & 1 & 0.344 & 2.912 \\ 
  0.3 & Quadratic & 0.830 & 1.217 & 2.715 & 1 & 0.904 & 7.489 & 1 & 0.840 & 7.498 \\ 
  0.3 & Cubic, c=5 & 0.950 & 0.500 & 1.898 & 1 & 0.765 & 5.209 & 1 & 0.626 & 5.253 \\ 
  0.3 & Cubic, c=10 & 0.934 & 0.680 & 2.609 & 1 & 1.216 & 7.229 & 1 & 0.967 & 7.233 \\ 
  0.3 & Matérn & 0.948 & 0.395 & 1.511 & 1 & 0.339 & 4.025 & 1 & 0.355 & 4.090 \\ 
  0.3 & Gaussian & 0.896 & 0.517 & 1.617 & 1 & 0.373 & 4.296 & 1 & 0.421 & 4.360 \\ 
\bottomrule
\end{tabular}
\end{table}

Table~\ref{tab:pooroverlap_table} presents results across all DGPs and $R^2$ values for ATE estimation. First, GP consistently achieves good coverage regardless of the DGP, demonstrating reliable uncertainty quantification even when the true functional form differs substantially from its encoded assumptions. In contrast, LM exhibits severe undercoverage in most scenarios. BART maintains overcoverage across all scenarios. The robust coverage of GP comes at the cost of wider interval lengths compared to LM, though these intervals are comparable to BART's.

RMSE varies across methods and DGPs. For kernel-based DGPs (Matérn and Gaussian), GP typically outperforms LM in RMSE, achieving both better point estimates and better coverage. For parametric DGPs (quadratic and cubic), LM sometimes achieves lower RMSE despite poor coverage. These finding highlight a trade-off: methods with narrower intervals may appear more precise when their assumptions are correct but risk severe undercoverage when these assumptions are violated.

The performance difference between $R^2=0.5$ and $R^2=0.3$ settings further illuminates these patterns. As the signal-to-noise ratio decreases (moving from $R^2=0.5$ to $R^2=0.3$), LM's coverage improves somewhat but requires wider intervals, while GP's RMSEE penalty remains minimal. This improvement occurs because increased noise masks some model misspecification issues. GP and BART maintain good coverage regardless of noise level, adapting their interval widths accordingly.

\begin{figure}[hbt!]
\centering
\includegraphics[width=.9\linewidth]{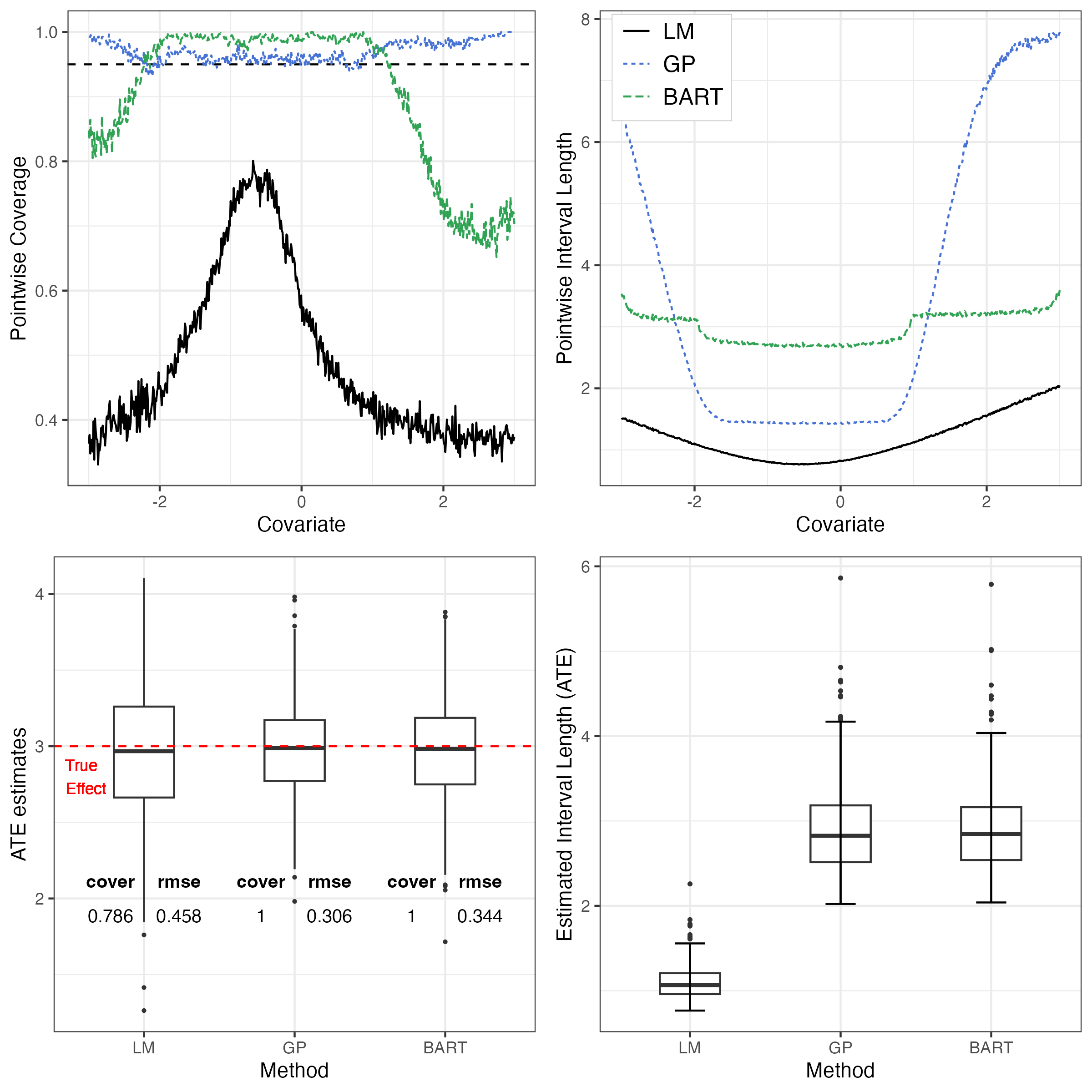}
\caption{Pointwise (top) and average (bottom) performance of GP model compared to LM and BART. The same plots as in Figure~\ref{fig:pooroverlap_summary} except that the DGP is a Gaussian kernel-based random function.}
\label{fig:pooroverlap_summary_gaussian}
\end{figure}

Finally, we note that using a smooth function DGP, which is more favorable to the Gaussian kernel than steep polynomials, enables GP to achieve better pointwise coverage behavior. Unlike the other methods, GP produces nearly nominal coverage in areas of good common support and then transitions to over-coverage in areas of poor overlap (Figure~\ref{fig:pooroverlap_summary_gaussian}). This conservative behavior is perhaps desirable, for many purposes.

\clearpage

\section{Illustration with a multivariate benchmark under varying overlap} \label{app:lalonde}

We briefly demonstrate the use of GP for covariate-adjusted comparisons in a setting with multiple covariates and a known experimental benchmark. We use the well-known National Supported Work Demonstration (NSW) \citep{LaLonde1986}, a job training program designed to help disadvantaged workers build employable skills from 1975 to 1979. Participants were randomly assigned to either a treatment group (job training) or a control group (no job training). The data also include a separate control group from outside the experiment, constructed using the Panel Study of Income Dynamics (PSID). We work with the \citet{dehejia1999causal} dataset, which contains data for males with two pre-training years (1975, 1976) of earnings data, along with 1978 (post-training) earnings data as the outcome. 

Using the experimental treatment and control groups, a difference-in-means estimate of the treatment effect is 1,794 USD. A linear regression controlling for observed covariates estimates a treatment effect of 1,671 USD.\footnote{Here and below, we use the following covariates: age, education (years of formal schooling), earnings in 1974, earnings in 1975, unemployment in 1974 and 1975 (indicators for earnings of zero), race/ethnicity indicators (Black, Hispanic, and (omitted) White); marital status, and an indicator for no high school degree.} In the non-experimental sample, as has been well-known since \citet{LaLonde1986}, linear regression produces a very poor estimate (4 USD). This is largely due to the large number of observational control units that differ greatly from the treated units, visible in pre-treatment earnings and employment status.

Our GP estimator produces an estimated ATT of 1,907 USD, close to the experimental benchmarks (1,671 and 1,794 USD). While the reasonable point estimate is reassuring, our central concern regards how counterfactual uncertainty should reflect the degree of overlap---that is, how uncertainty in predicted counterfactual outcomes should increase as treated units become more dissimilar from control units. For the ATT, this example actually offers very good overlap: while many control units are unlike treated units, most treated units have numerous nearby control observations available. Nevertheless, treated units vary in how densely they are surrounded by control observations, and our counterfactual beliefs for a given treated unit (i.e. $p(Y_i(0))$ for a unit $i$ that was treated) should reflect this.

\begin{figure}[hbt!]
\centering
\includegraphics[width=.49\linewidth]{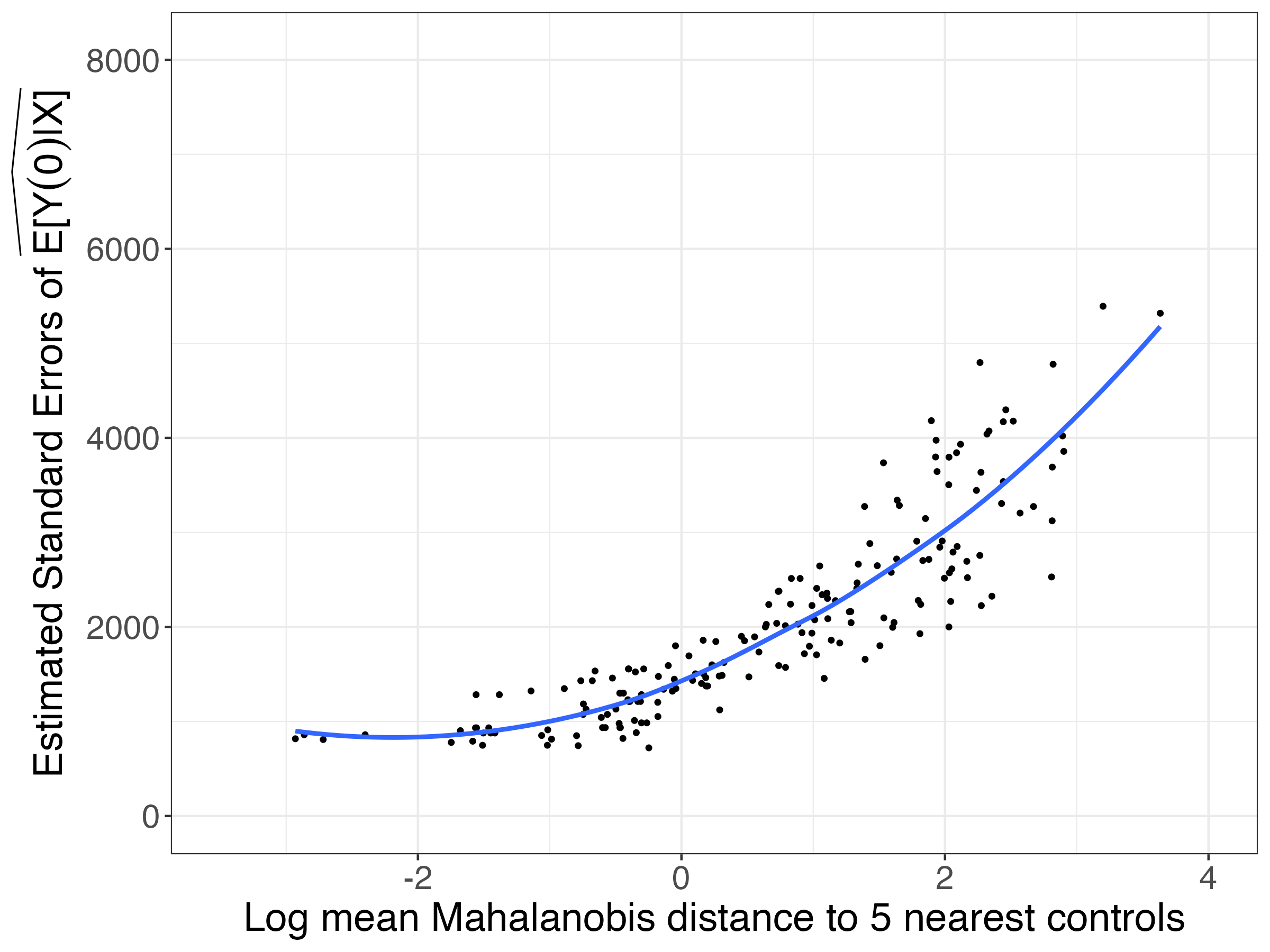} 
\includegraphics[width=.49\linewidth]{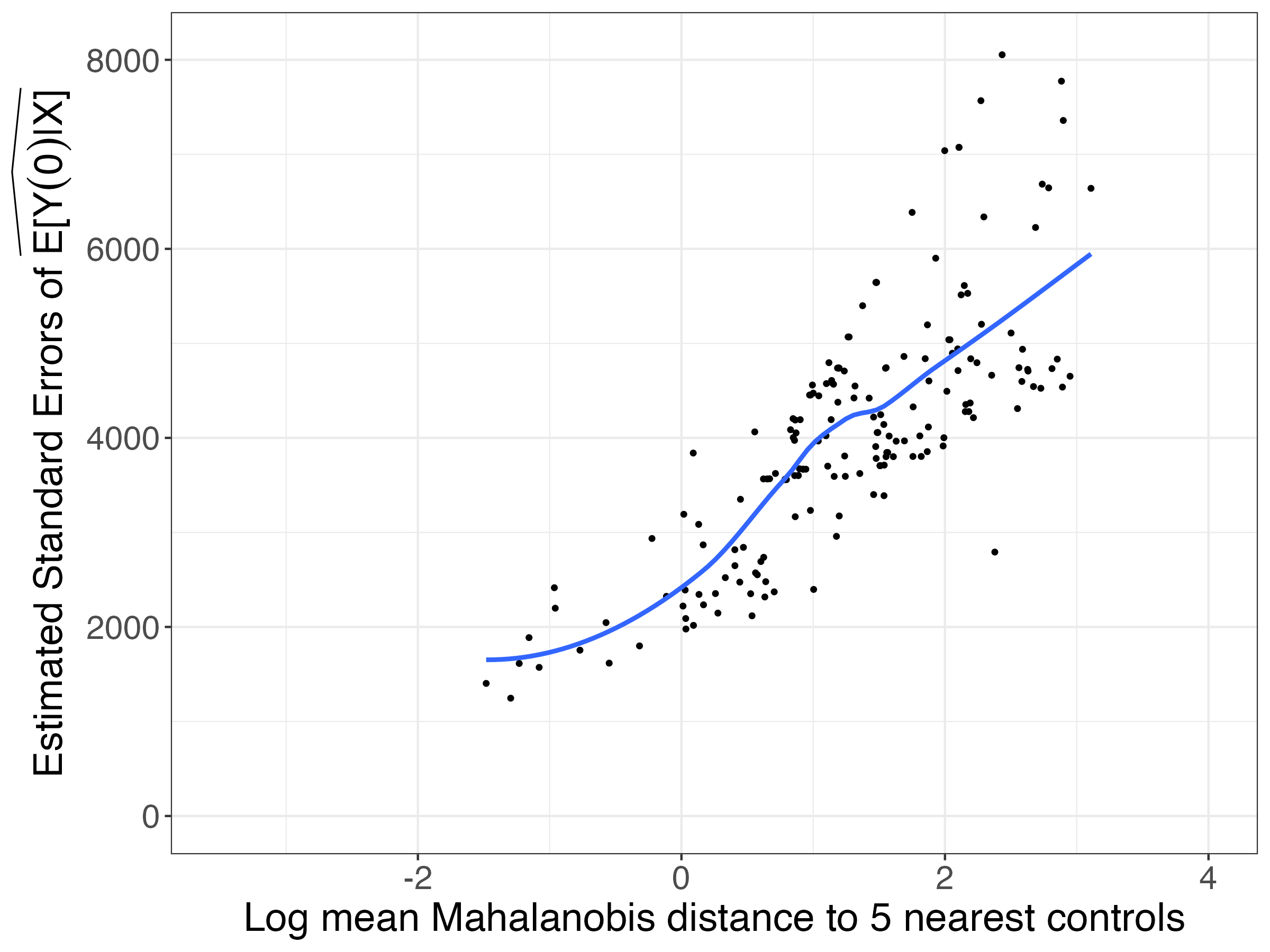} 
\caption{Estimated standard errors of $Y(0)$ for treated units as a function of the log of average Mahalanobis distance to 5 nearest control units. The plots illustrate how the uncertainty estimates (of the CEF) for treated units change with log of average Mahalanobis distance to 5 nearest control units. The non-experimental data (right) exhibits overall larger standard errors compared to the experimental data (left), indicating less confidence over estimates with poor overlap between treated and control groups. Each point represents the estimated standard error of CEF of untreated potential outcome for an individual treated unit. The blue solid line is a locally smoothed fit (LOESS) to these points. In both plots, as treated units become more dissimilar from control units in covariate space (i.e., poor overlap), the standard errors of counterfactual predictions (and those of corresponding estimated treatment effect) increase.}
\label{fig:lalonde}
\end{figure}

GP provides a reasoned basis for estimating this uncertainty, illustrated in Figure~\ref{fig:lalonde}. Counterfactual uncertainty for each treated unit is quantified by $\widehat{SE}(\reallywidehat{\E[Y(0) \mid X]})$ on the vertical axes. The degree of overlap is quantified here as the average Mahalanobis distance of the observation to the nearest five control observations, which we log for visualization. The left panel shows this when using controls from the experimental group, while the right panel shows this using controls from the observational (PSID) sample.

We can see, first, that the experimental group offered more comparable control comparisons to more of the treated observations, though both the experimental and observational studies have treated units for whom the nearest five controls are farther away. Second, the observational data shows greater overall counterfactual uncertainty, indicating less confidence in estimates due to poorer overlap. In both studies, counterfactual uncertainty increases as treated units become more dissimilar from control units (i.e., as overlap decreases). That said, we recall that this uncertainty is not solely a function of distance to the nearest comparisons, to the mean, or to any other point or group of points. Rather, this computation is made based on Expression~\ref{eq.main}, and depends on the density of nearby data, the precise shape of the kernel function, and the overall unexplained variation, $\sigma^2$.

\clearpage 

\section{Causal assumptions for ITS}\label{app.causalITS}

We outline the key causal identification assumptions required for ITS. While we focus on estimation concerns in the main text (~\ref{sec:its}), estimation results in this context are typically of interest insofar as they gauge the causal effect of the event on the outcome. The difference between observed post-treatment outcomes and extrapolated non-treatment outcomes at a given time $t \geq T$ can only be interpreted as the causal effect of that event insofar as the predicted $Y(0)$ values closely represent what would have truly happened absent treatment.

We break this into two constituent assumptions. First, the (possibly high-ordered) trend information required to effectively model $Y(0)$ in the post-treatment period must be (i) available in the pre-treatment period and (ii) sufficiently detected and modeled by the estimation approach. Second, no event that influences the outcome, other than the event of interest or its knock-on effects, occurred within the comparison period after treatment. That is, if some event not considered the event of interest occurs and influences the outcome, it would affect $Y(0)$. A proper causal contrast should account for this by utilizing the affected $Y(0)$. However, since this information was not available in the pre-treatment outcome, it cannot be incorporated into the estimate of $Y(0)$. Although it is a substantive and untestable question, we limit our attention to short-term post-treatment periods to mitigate the concern. We emphasize, however, that our focus here is on estimation challenges inherent to ITS approaches, which must address these causal identification concerns regardless of the estimator employed. 

\clearpage

\section{Additional simulation results: Regression discontinuity case}\label{app.rd}

We provide additional simulation results to further evaluate GP performance under various challenging scenarios in the RD setting: (1) heteroskedastic error structures, (2) latent variable models with varying sample sizes, (3) small-sample conditions under which conventional (and GP-based) estimators fail or produce extreme uncertainty intervals, and (4) selected ``worst-case'' functional forms to challenge GP's regularization properties. These simulations complement the main text by stress-testing GP under conditions that probe its limitations in the RD setting.

\subsection{Total random simulation with heteroskedasticity}\label{app.hetero}

For each of 500 simulations, we generate the data using the following process:
\begin{itemize}
    \item $x_i \sim \mathcal{U}(-2, 2)$
    \item $y_{\text{true},i} \sim \tau \cdot I(x>0)$
    \item $y_i = y_{\text{true},i} + \varepsilon$
    \item $\varepsilon \sim \mathcal{N}(0, \sigma(x))$,
\end{itemize}
where $\sigma(x)$ represents one of three different error forms: 
\begin{itemize}
    \item decreasing toward the middle: $\sigma_1^2(x) = v_0 + v_0(\kappa-1)|x/2|$
    \item increasing toward the middle: $\sigma_2^2(x) = v_0 + v_0(k - 1)(2-|x|)/2$
    \item increasing to the right: $\sigma_3^2(x) = v_1 + v_1(k-1)((x+2)/2)$.
\end{itemize}

Note that to maintain our $R^2 \approx 0.33$, i.e. $cor(y_{\text{true}}, y)^2 \approx 0.33$, we set $v_0 = 4Var(Y_{\text{true}})/(\kappa + 1)$ for $\sigma_1(x)$ and $\sigma_2(x)$ and $v_1 = 2Var(Y_{\text{true}})/(\kappa + 1)$ for $\sigma_3(x)$. We parameterized the noise level such that $\kappa = 0$ indicates homoskedasticity. When $\kappa = 2$, the maximum variance doubles for $\sigma_1$ and $\sigma_2$, or triples for $\sigma_3$. Similarly, $\kappa  = 4$ corresponds to a fourfold increase in maximum variance for $\sigma_1$ and $\sigma_2$, or a sevenfold increase for $\sigma_3$ compared to $\kappa = 0$.

\begin{figure}[hbt!]
\centering \textit{(a)} $\kappa = 1$\\
\includegraphics[width=.9\linewidth]{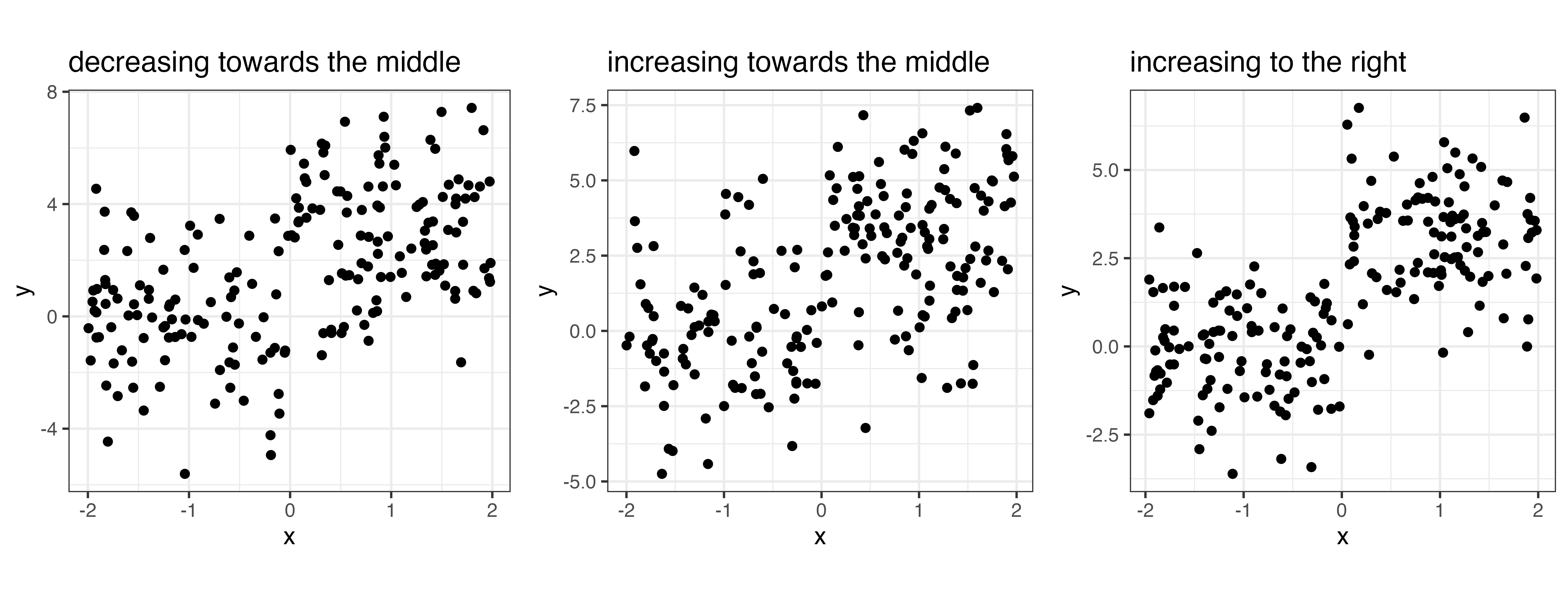}\\
\centering \textit{(b)} $\kappa = 2$\\
\includegraphics[width=.9\linewidth]{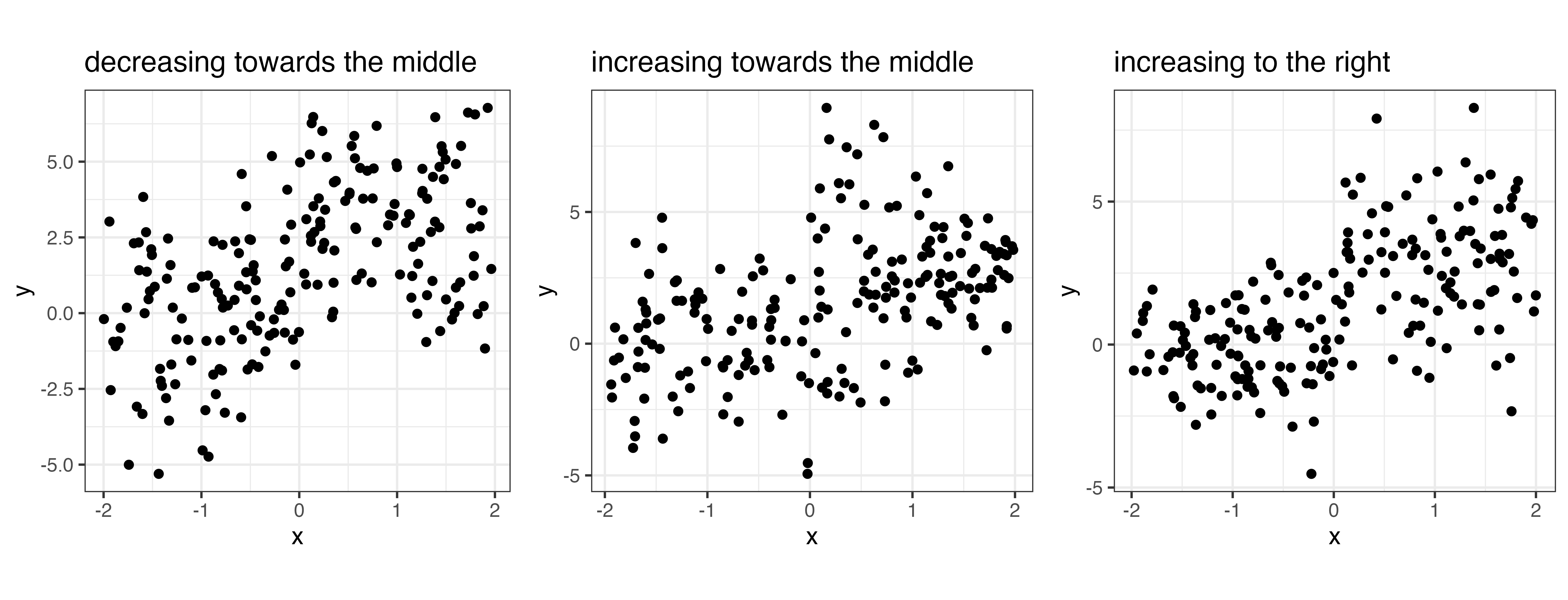}\\
\centering \textit{(c)} $\kappa = 4$\\
\includegraphics[width=.9\linewidth]{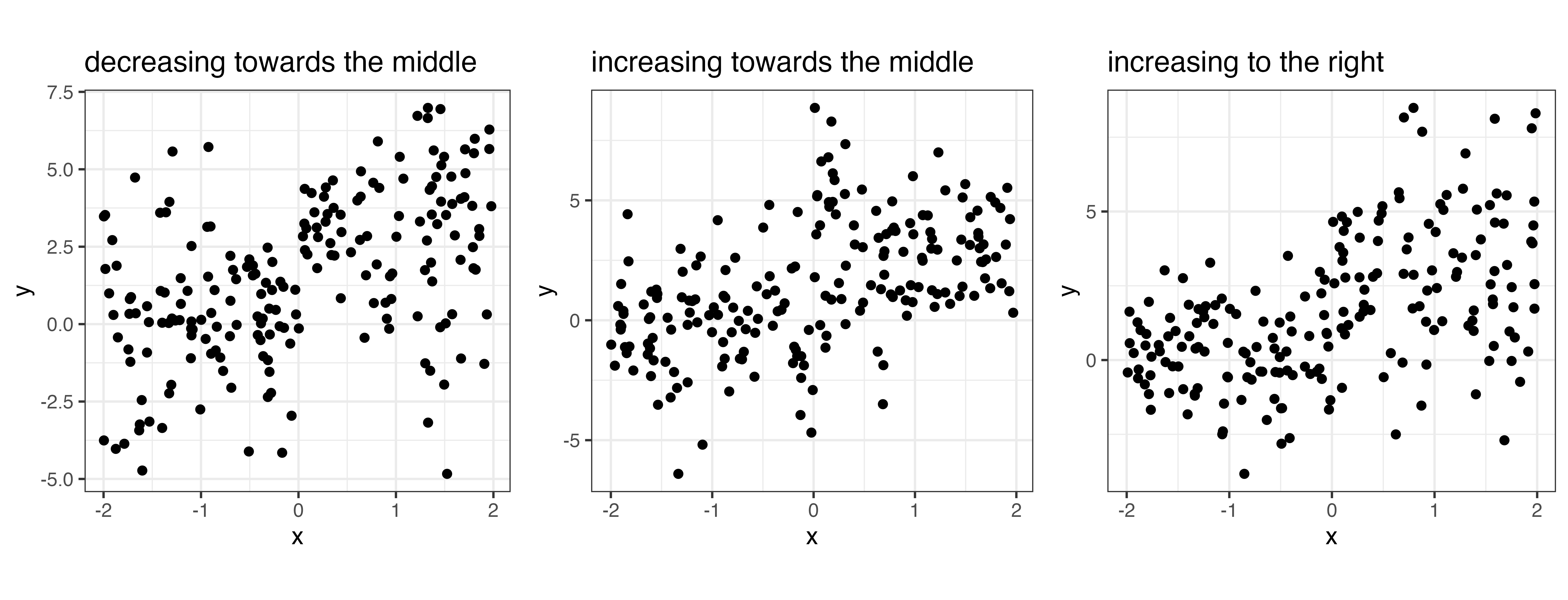}\\
\caption{Generated $X$ and $Y$ from one simulation for each error term at a different level of $\kappa$.}
\label{fig:rdd_totalrandom_hetero_desc}
\end{figure}

\begin{figure}[hbt!]
\centering \textit{(a)} decreasing towards the middle,$\sigma_1^2(x) = v_0 + v_0(\kappa-1)|x/2|$, with true effect of three, $n=100$\\
\includegraphics[width=.9\linewidth]{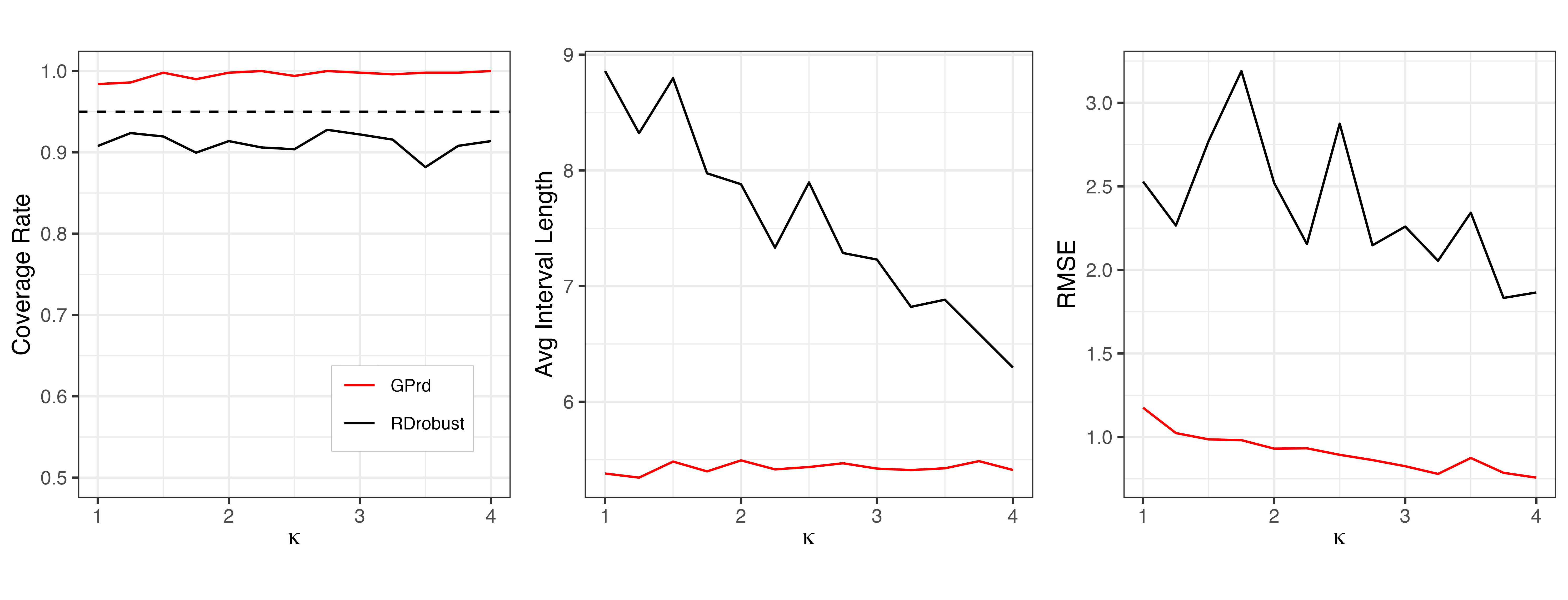}\\
\centering \textit{(b)} increasing towards the middle, $\sigma_2^2(x) = v_0 + v_0(\kappa - 1)(2-|x|)/2$, with true effect of three, $n=100$\\
\includegraphics[width=.9\linewidth]{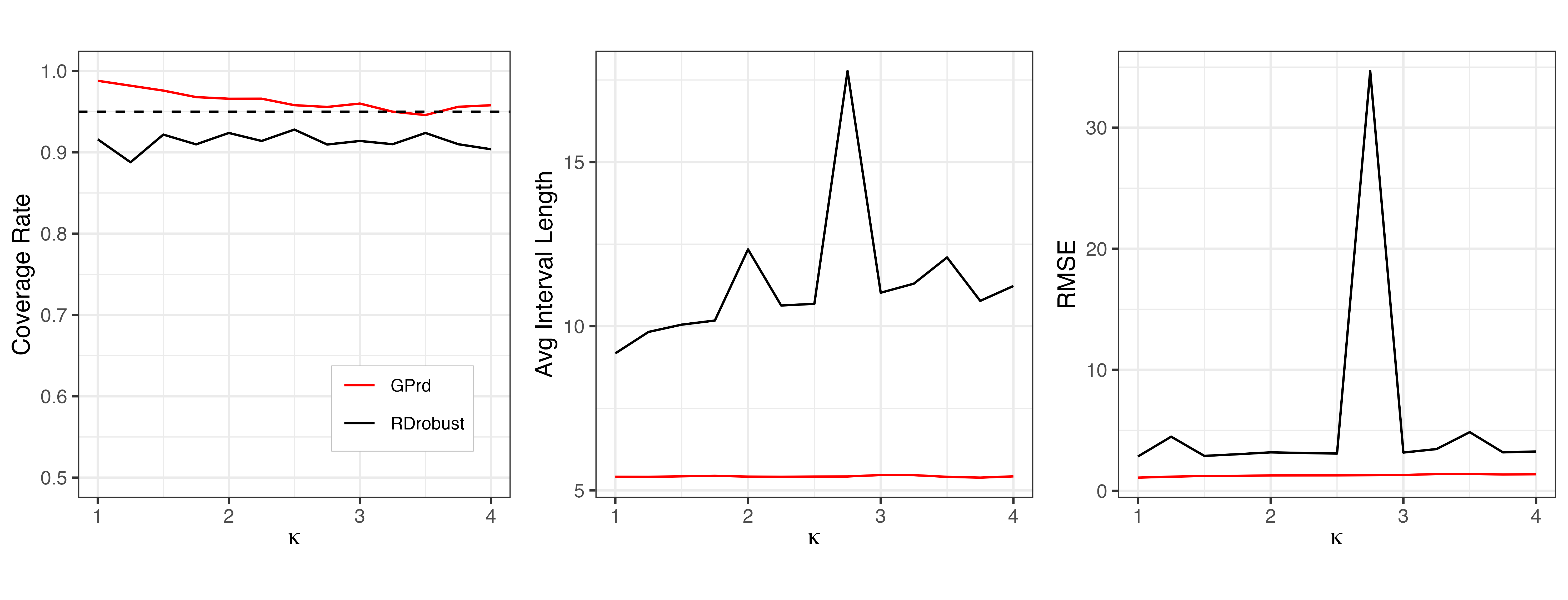}\\
\centering \textit{(c)} increasing to the right, $\sigma_3^2(x) = v_1 + v_1(\kappa-1)((x+2)/2)$, with true effect of three, $n=100$\\
\includegraphics[width=.9\linewidth]{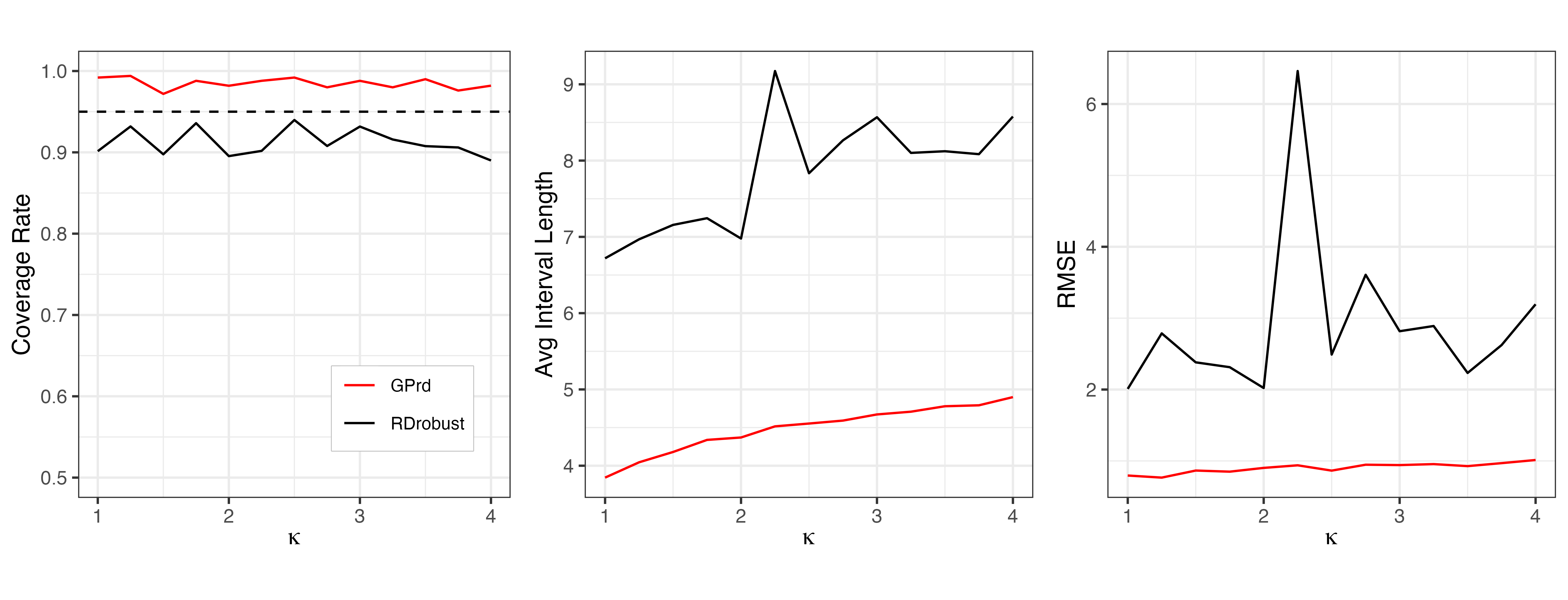}
\caption{\textbf{n = 100:} The coverage rate, average length of 95\% confidence interval, and RMSE of \texttt{GPrd} (red) and \texttt{rdrobust} with ``robust'' options (black) in the total random setting. The horizontal axis indicates the heteroskedasticity parameter, $\kappa$.To improve visualization, extreme values with interval lengths greater than 200 were omitted.}
\label{fig:rdd_totalrandom_hetero_n100}
\end{figure}

\begin{figure}[hbt!]
\centering \textit{(a)} decreasing towards the middle,$\sigma_1^2(x) = v_0 + v_0(\kappa-1)|x/2|$, with true effect of three, $n=200$\\
\includegraphics[width=.9\linewidth]{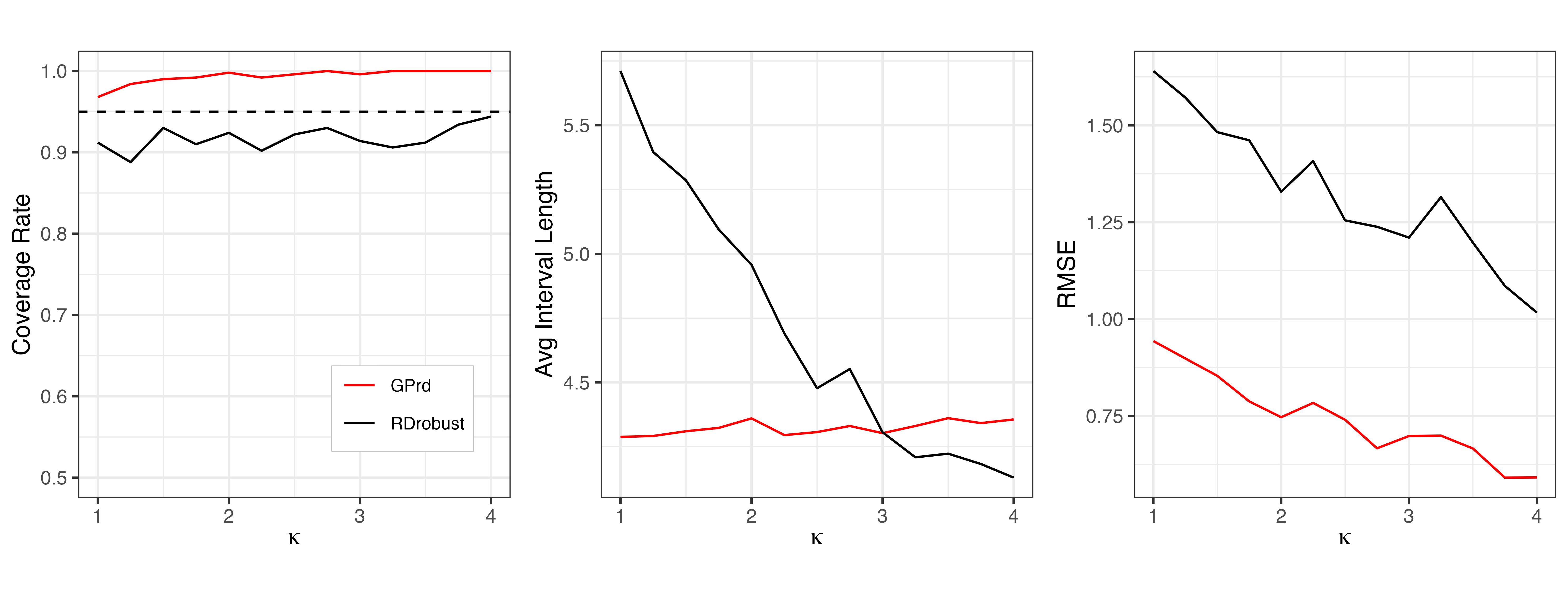}\\
\centering \textit{(b)} increasing towards the middle, $\sigma_2^2(x) = v_0 + v_0(\kappa - 1)(2-|x|)/2$, with true effect of three, $n=200$\\
\includegraphics[width=.9\linewidth]{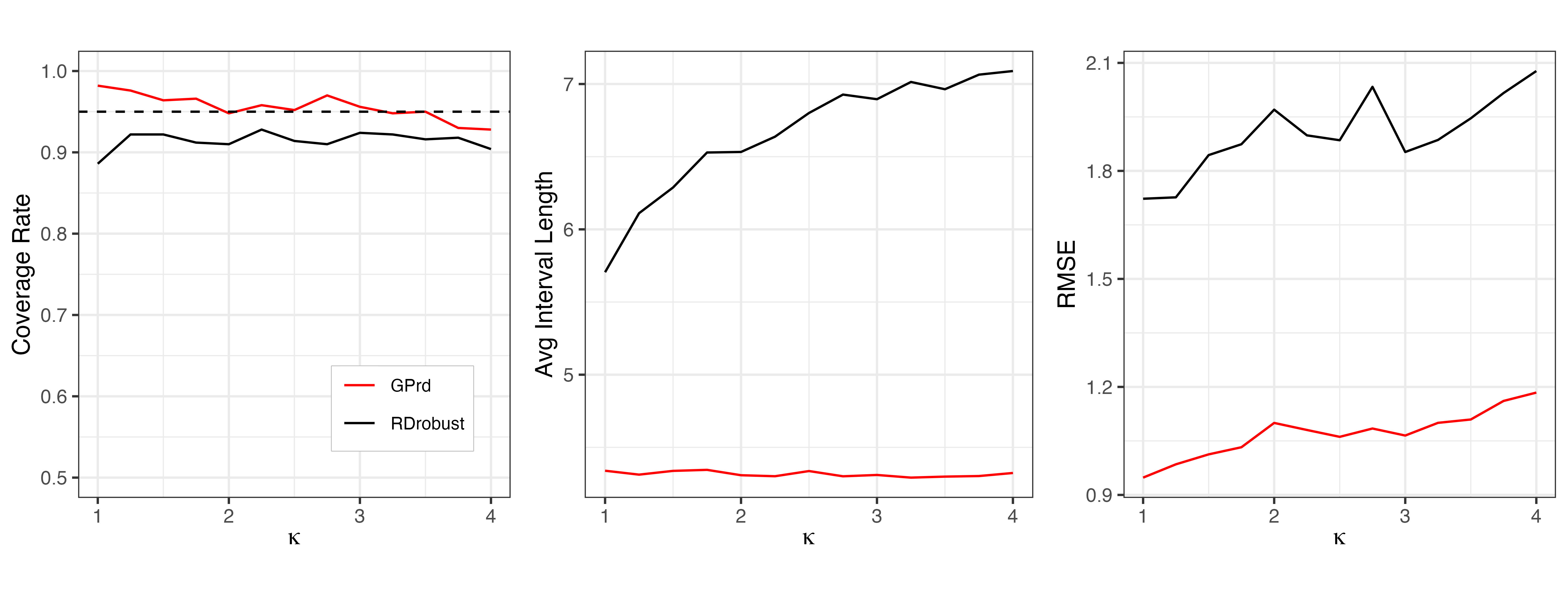}\\
\centering \textit{(c)} increasing to the right, $\sigma_3^2(x) = v_1 + v_1(\kappa-1)((x+2)/2)$, with true effect of three, $n=200$\\
\includegraphics[width=.9\linewidth]{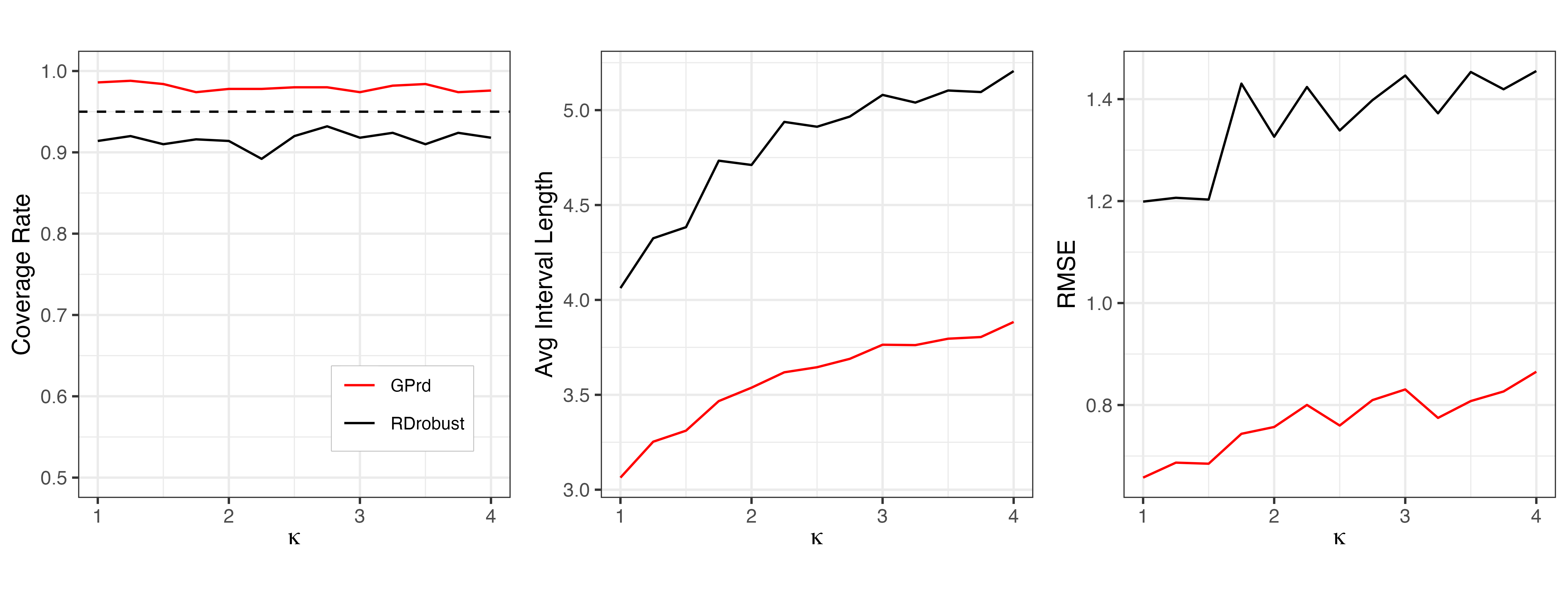}
\caption{\textbf{n = 200:} The coverage rate, average length of 95\% confidence interval, and RMSE of \texttt{GPrd} (red) and \texttt{rdrobust} with ``robust'' options (black) in the total random setting. The horizontal axis indicates the heteroskedasticity parameter, $\kappa$.}
\label{fig:rdd_totalrandom_hetero_n200}
\end{figure}

\begin{figure}[hbt!]
\centering \textit{(a)} decreasing towards the middle,$\sigma_1^2(x) = v_0 + v_0(\kappa-1)|x/2|$, with true effect of three, $n=500$\\
\includegraphics[width=.9\linewidth]{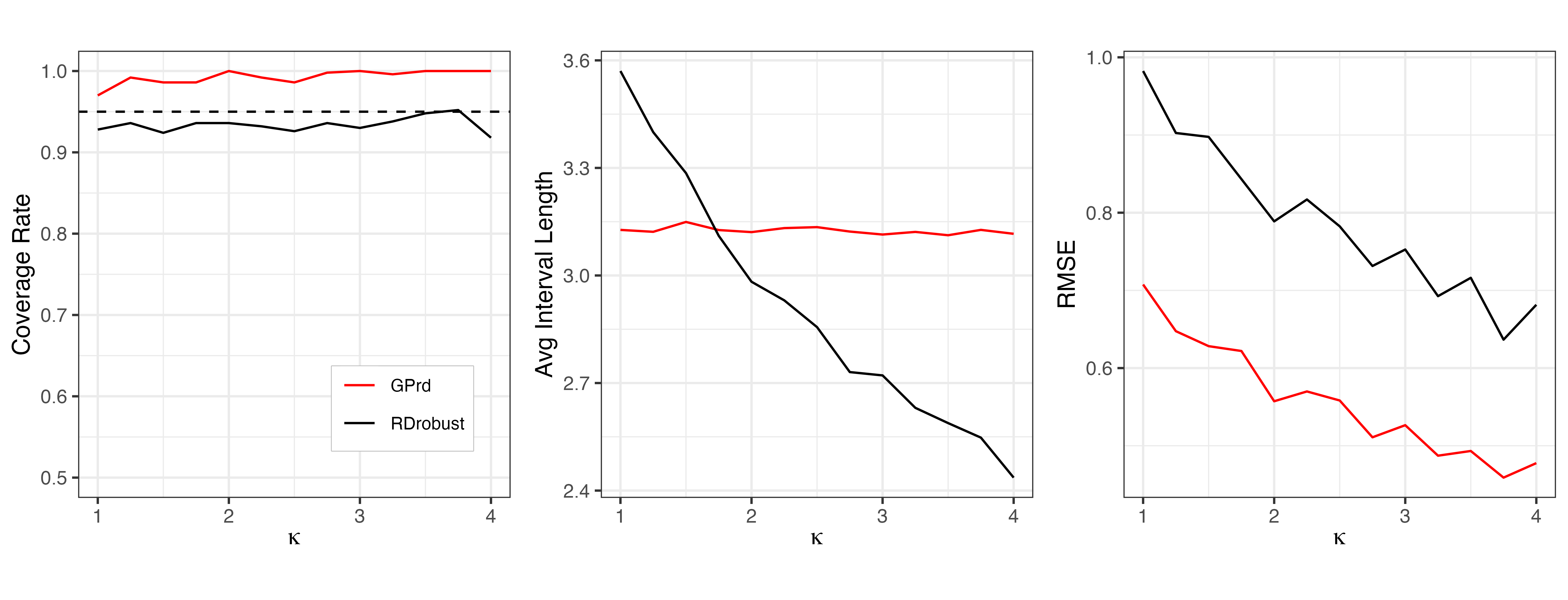}\\
\centering \textit{(b)} increasing towards the middle, $\sigma_2^2(x) = v_0 + v_0(\kappa - 1)(2-|x|)/2$, with true effect of three, $n=500$\\
\includegraphics[width=.9\linewidth]{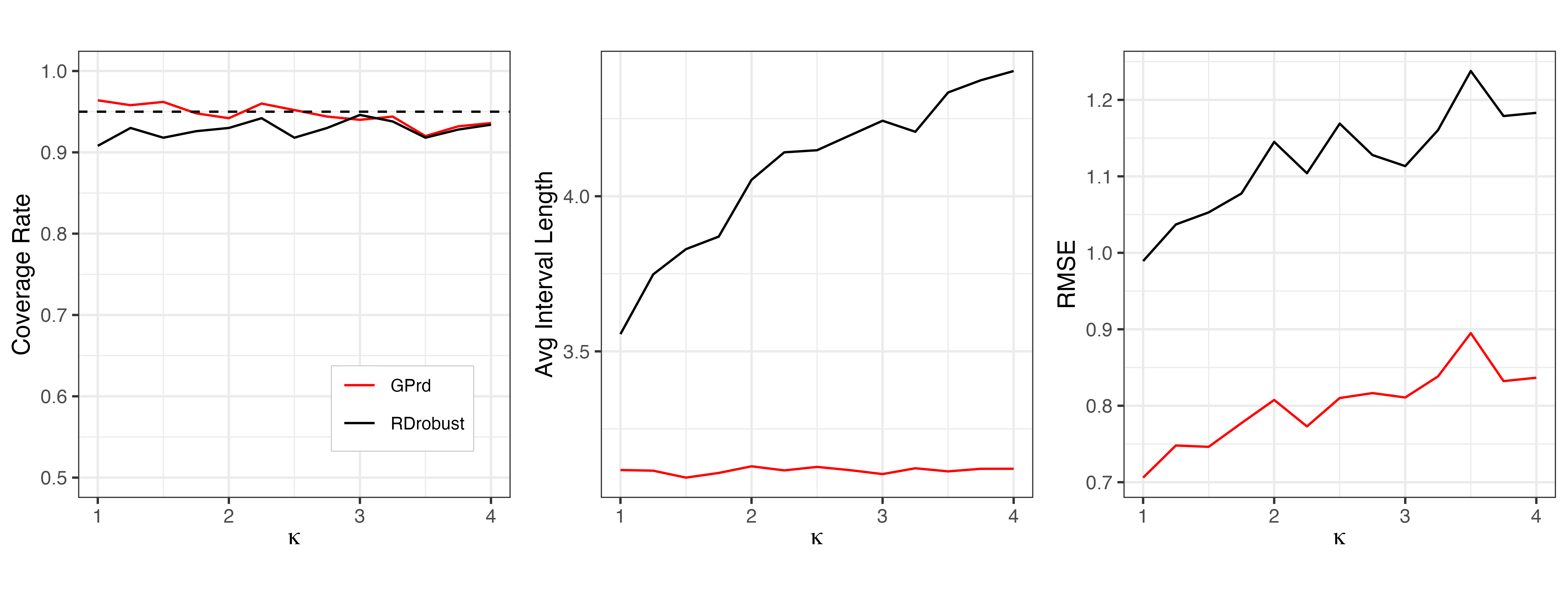}\\
\centering \textit{(c)} increasing to the right, $\sigma_3^2(x) = v_1 + v_1(\kappa-1)((x+2)/2)$, with true effect of three, $n=500$\\
\includegraphics[width=.9\linewidth]{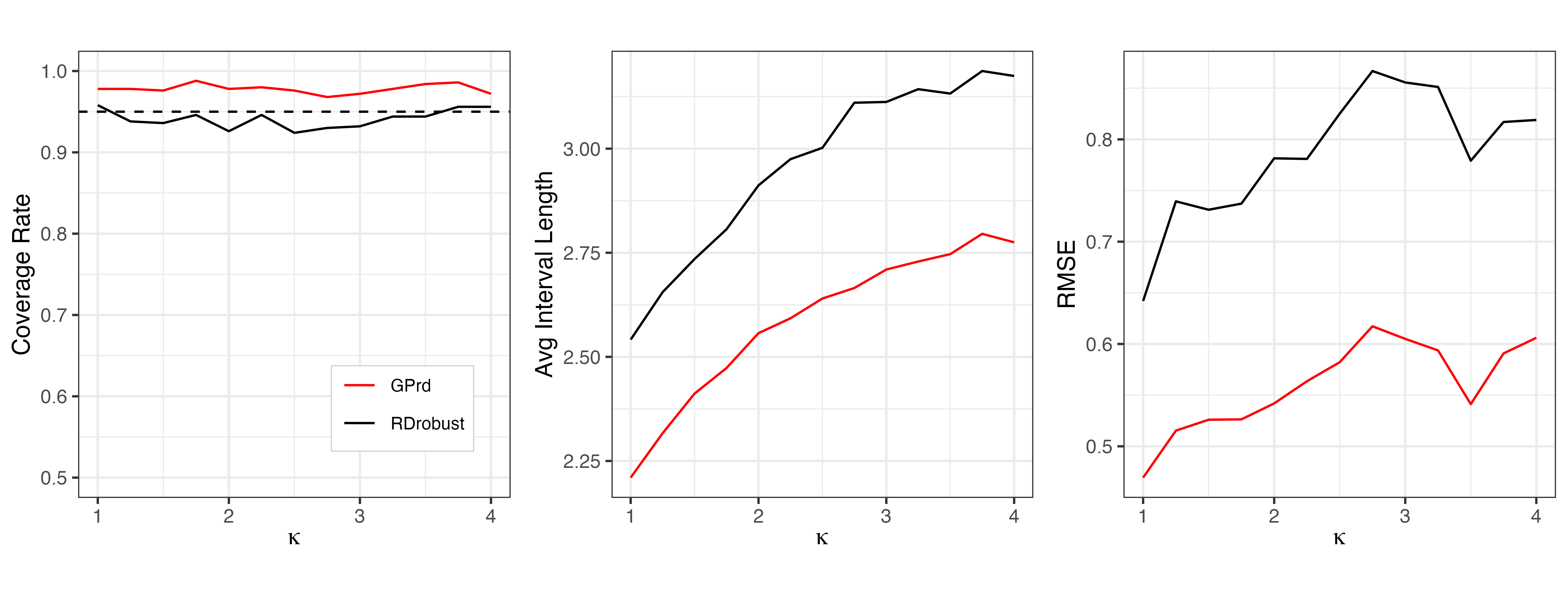}
\caption{\textbf{n = 500:} The coverage rate, average length of 95\% confidence interval, and RMSE of \texttt{GPrd} (red) and \texttt{rdrobust} with ``robust'' options (black) in the total random setting. The horizontal axis indicates the heteroskedasticity parameter, $\kappa$.}
\label{fig:rdd_totalrandom_hetero_n500}
\end{figure}

We draw a sample of size 100, 200, 500 with $\kappa$ varying from 1 to 4. We use a cutoff of 0 for $X$ to determine treatment eligibility status, and the true effect is 3. 
To provide some visual description of heteroskedasticity in each error form, Figure \ref{fig:rdd_totalrandom_hetero_desc} illustrates the scatter plots of a simulated $X$ and $Y$ (n = 200) at $\kappa =$ 1 and 2. Figure \ref{fig:rdd_totalrandom_hetero_n100}, \ref{fig:rdd_totalrandom_hetero_n200}, and \ref{fig:rdd_totalrandom_hetero_n500} display the simulation results.

\clearpage

\subsection{Latent variable confounding simulation with different number of observations}\label{app.latentdiffn}

\begin{figure}[hbt!]
\centering
\includegraphics[width=\linewidth]{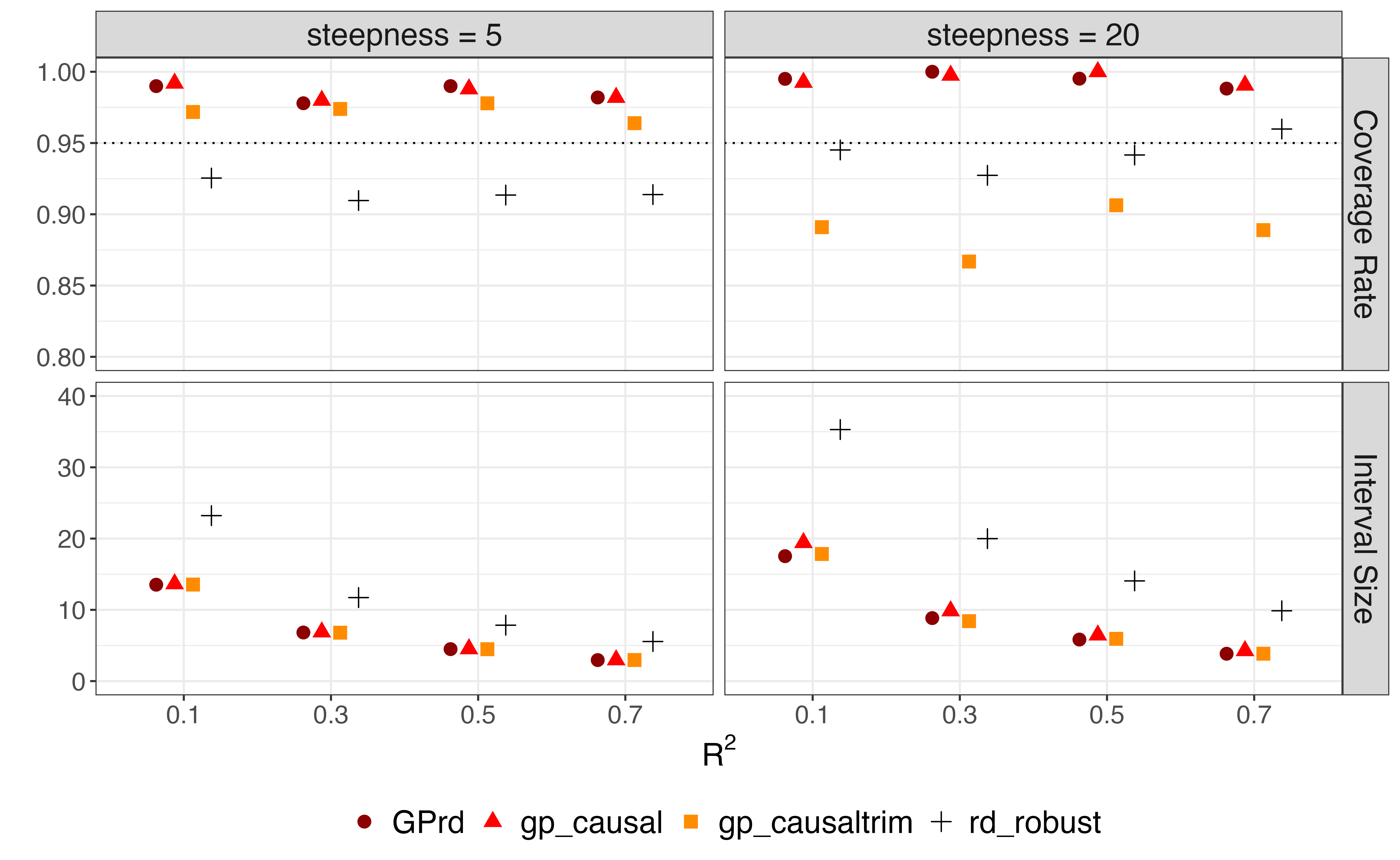}
\caption{RD 500 simulation results with latent variable (effect size = 3, n = 100). Simulation results from \texttt{rdrobust} with confidence interval lengths exceeding 100 have been excluded for legibility.}
\end{figure}

\begin{figure}[hbt!]
\centering
\includegraphics[width=\linewidth]{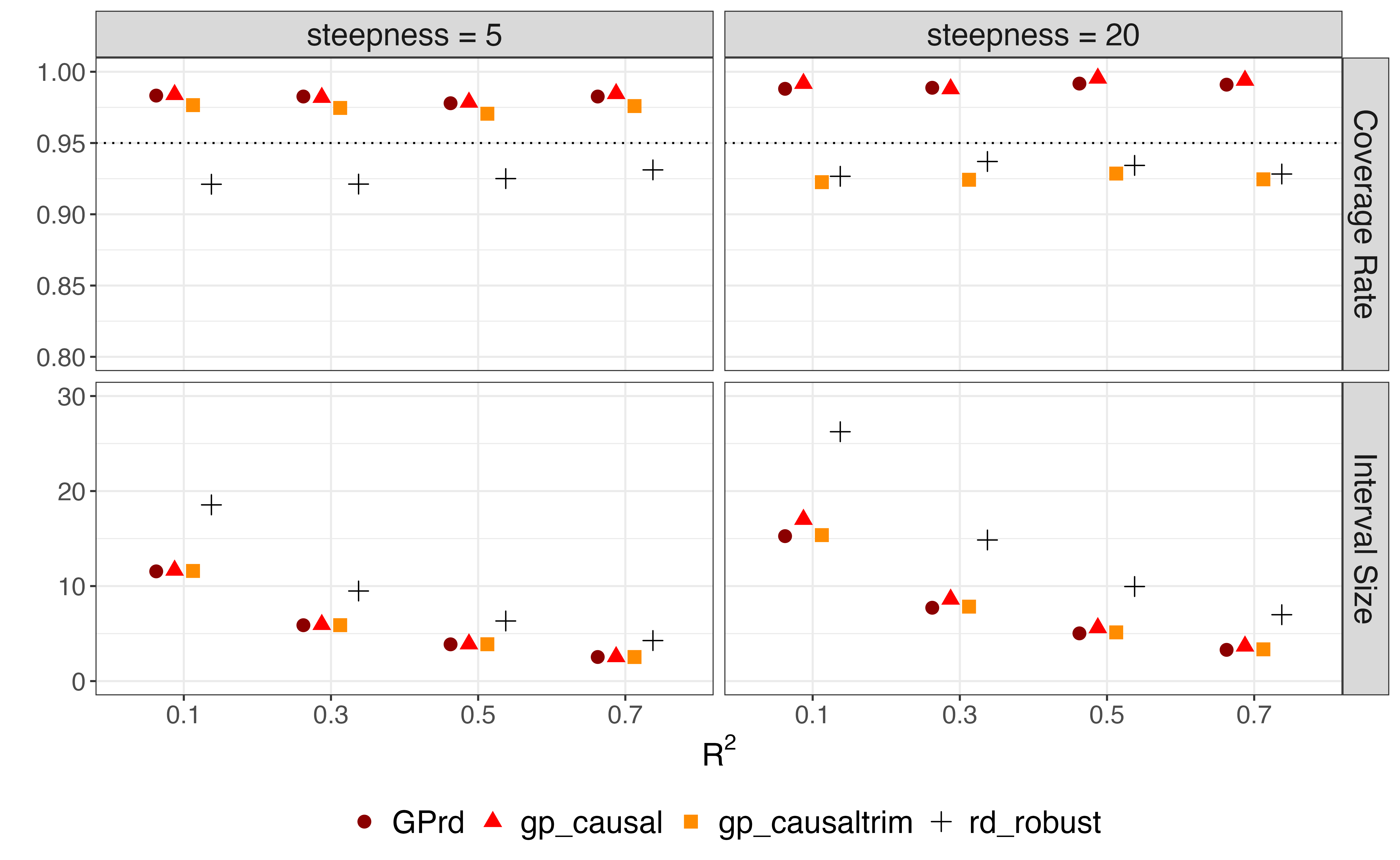}
\caption{RD 500 simulation results with latent variable (effect size = 3, n = 500). Simulation results from \texttt{rdrobust} with confidence interval lengths exceeding 100 have been excluded for legibility.}
\end{figure}

\clearpage

\subsection{Estimator failures and interval outliers at small sample size in latent variable confounding simulations}\label{app.fail}

\paragraph{Error rate: 500 simulations with n = 100} 

\begin{figure}[hbt!]
\centering
\includegraphics[width=\linewidth]{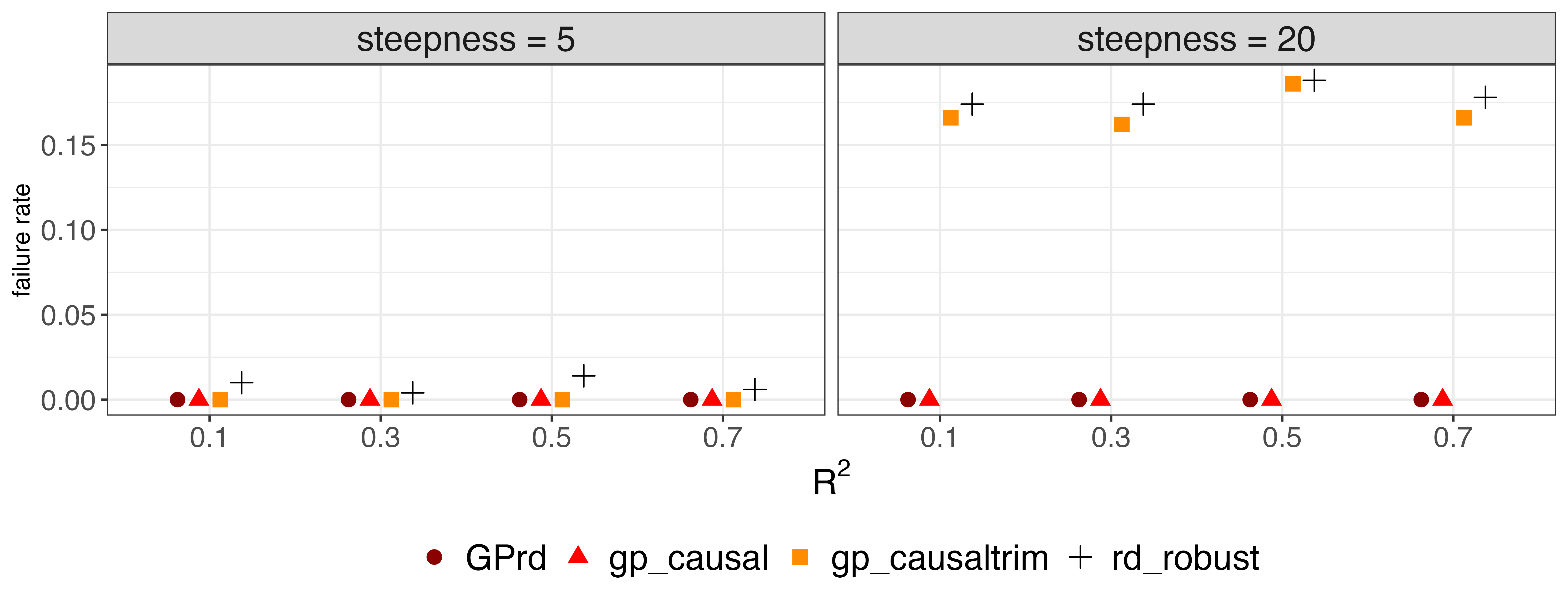}
\caption{Error rate in 500 simulation results with latent variable confounding (effect size = 3, n = 100)}
\label{fig:latent_errorraten100}
\end{figure}

The figure below displays how many simulations failed to run successfully for each model across different R2 values and the interval lengths we obtained from each simulation (Figure \ref{fig:latent_errorraten100}). While the estimators are stable when steepness = 5, when steepness = 20, \texttt{rdrobust} and \texttt{gp\_causaltrim} do not run at all sometimes. The graphs show the failure rate in 500 simulations.

\paragraph{Interval length: 500 simulations with n = 100 for steepness = 20} 

\begin{figure}[hbt!]
\centering
\includegraphics[width=\linewidth]{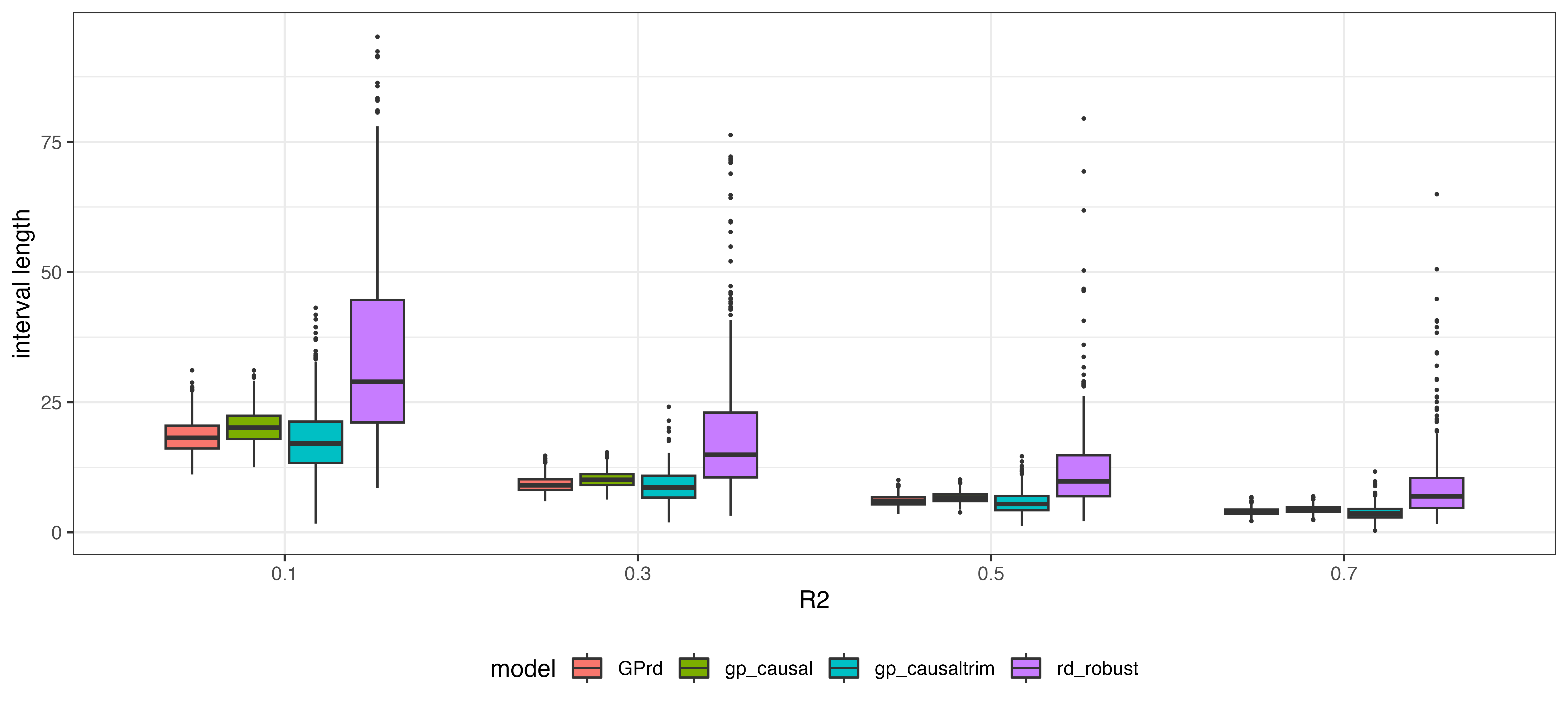}
\caption{Interval length in 500 simulation results with latent variable confounding (effect size = 3, n = 100)}
\label{fig:latent_lengthn100}
\end{figure}

\texttt{rdrobust} shows some extreme length with its maximum value in a simulation of 1,292,623,605 when steepness = 20. Out of 4,000 simulations (500 for each R2 and steepness), 68 results from rdrobust have the interval length that goes over 100. For GP, the maximum length is 43.1 for \texttt{gp\_causaltrim}. Even after removing observations with the interval length over 100, we can still see that \texttt{rdrobust} shows extremely large lengths (Figure \ref{fig:latent_lengthn100}).

\clearpage

\subsection{Examining potential worst-case scenarios for \texttt{GPrd}}\label{app.rdd.sims}
Following simulations from \citet{branson2019nonparametric}, \citet{calonico2014robust} and \citet{imbens2012}, we use the seven different functional forms: CATE1, CATE2, cubic, curvature, a functional form from \citet{lee2008randomized}, a functional form from \citet{ludwig2007}, and Quadratic.

\begin{itemize}
    \item CATE1: $0.42 + 0.1 * \mathcal{I}(x>0) + 0.84 - 3*x^2 + 7.99*x^3 - 9.01 * x^4 + 3.56 * x^5$
    \item CATE2:$0.42 + 0.1 * \mathcal{I}(x>0) + 0.84 + 7.99*x^3 - 9.01 * x^4 + 3.56 * x^5$
    \item Cubic: $4x^3$ when $x<0$, $3x^3$ when $x \geq 0$
    \item Curvature:
    \begin{itemize}
        \item $.48+1.27x -0.5*7.18x^2+0.7*20.21x^3+1.1*21.54x^4+1.5*7.33x^5$ when $x<0$,
        \item $3.52+.84x-0.1*3*x^2-0.3*7.99x^3-0.1*9.01x^4+3.56x^5$ when $x \geq 0$
    \end{itemize}
    \item Lee:
    \begin{itemize}
        \item $.48+1.27*x+7.18*x^2+20.21*x^3+21.54*x^4+7.33*x^5$ when $x<0$,
        \item $.52+.84*x-3*x^2+7.99*x^3-9.01*x^4+3.56*x^5$ when $x \geq 0$
    \end{itemize}
    \item Ludwig \& Milner:
    \begin{itemize}
        \item $3.71+2.3*x+3.28*x^2+1.45*x^3+0.23*x^4+0.03*x^5$ when $x<0$,
        \item $0.26+18.49*x-54.81*x^2+74.3*x^3-45.02*x^4+9.83*x^5$ when $x \geq 0$
    \end{itemize}
    \item Quad: $4*x^2$ when $x<0$, $3*x^2$ when $x \geq 0$
\end{itemize}

\begin{figure}[hbt!]
\centering
\includegraphics[width=\linewidth]{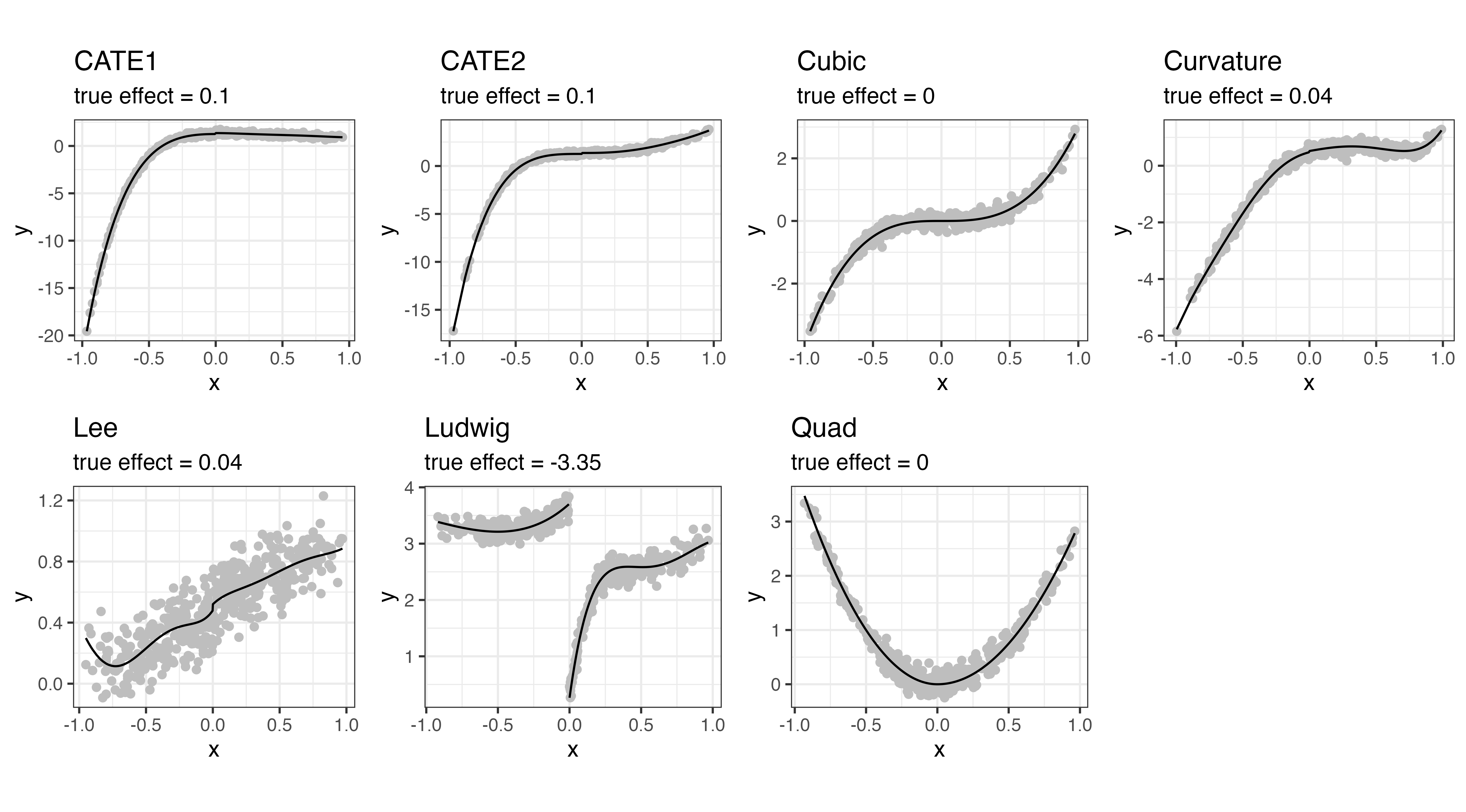}
\caption{Functional forms for the potential worst-case scenario simulations}
\label{fig:functionshape}
\end{figure}

\begin{figure}[hbt!]
\centering
\includegraphics[width=\linewidth]{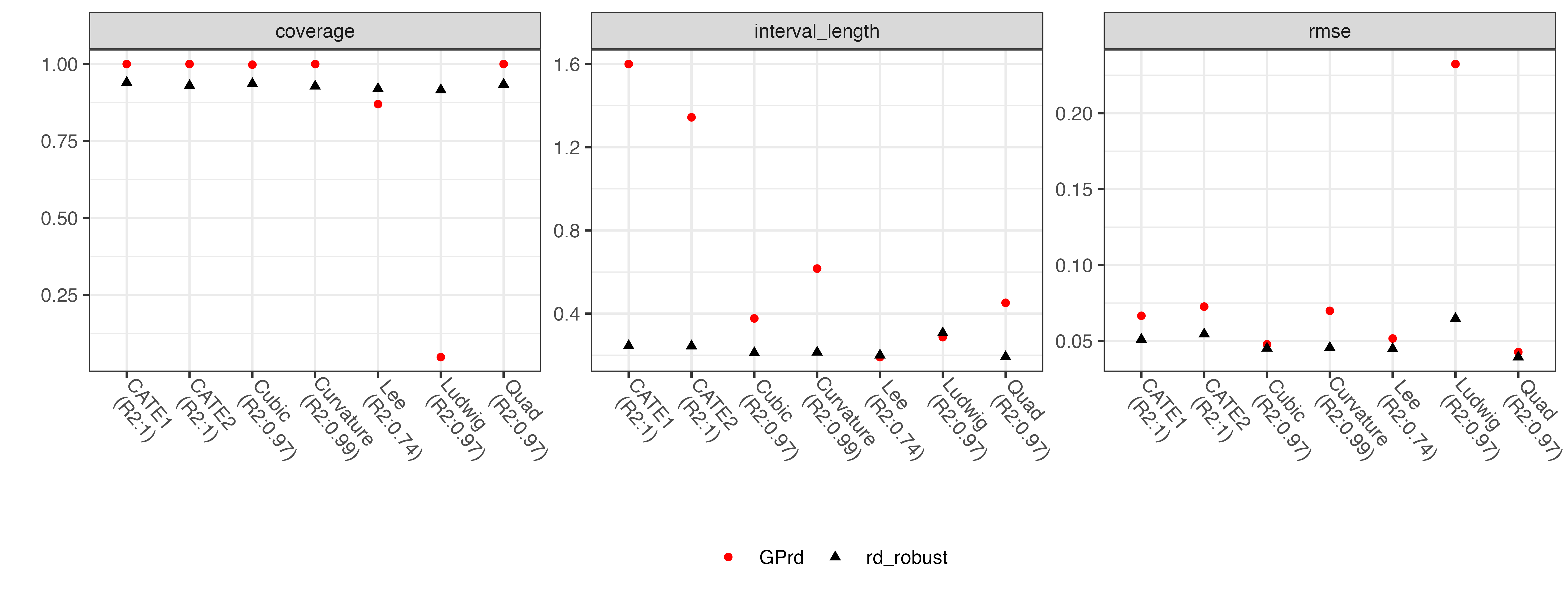}
\caption{RD 500 simulation results}
\label{fig:rdd_worst}
\end{figure}

Figure \ref{fig:rdd_worst} shows the results from the 500 simulations for each functional form. However, the signal-to-noise ($R^2$) values for these simulations were not typically well calibrated to those expected from real data, especially in the social sciences, which we expect to more often be around .33-0.5. As shown in the horizontal axis, $R^2$  values are generally very high (between 0.96 and 1), with the Lee case showing the lowest value at 0.64. We replicated the same simulations across different $R^2$ (Figure \ref{fig:worstsim1} and Figure \ref{fig:worstsim2}).

\begin{figure}[hbt!]
\centering \textit{(a)} CATE1\\
\includegraphics[width=.9\linewidth]{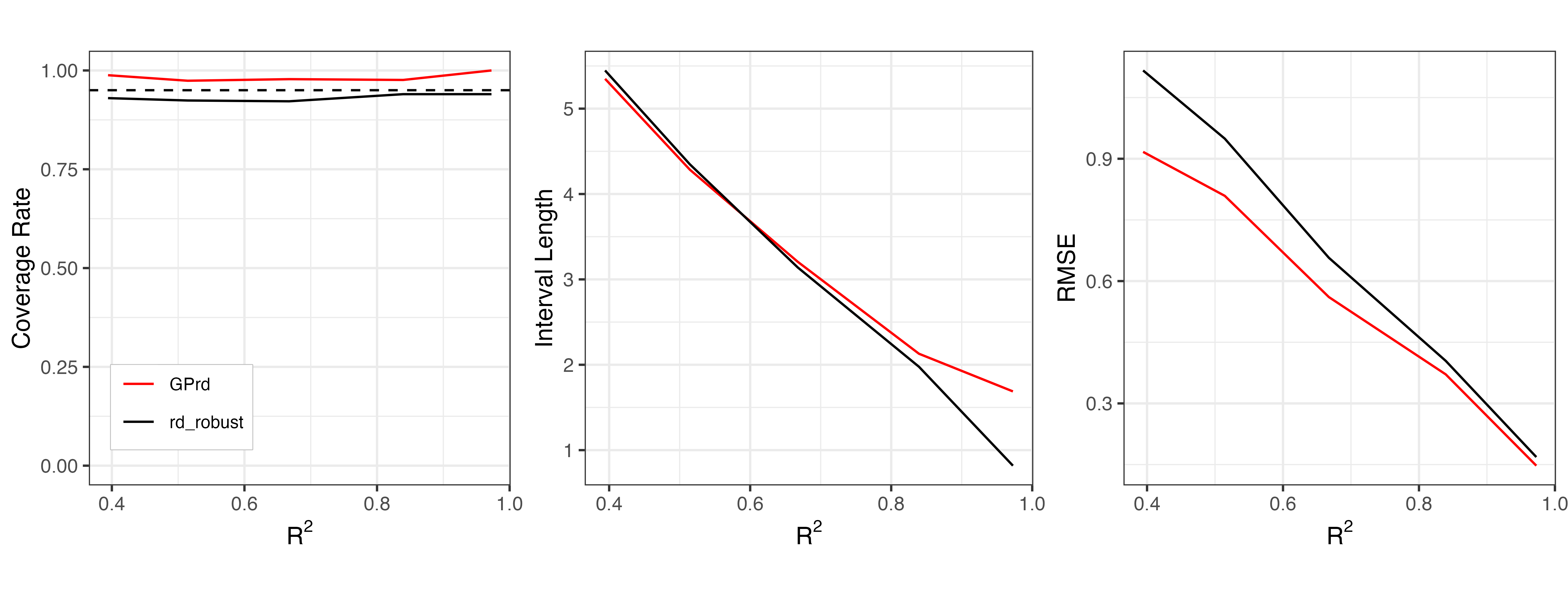}\\
\centering \textit{(b)} CATE2\\
\includegraphics[width=.9\linewidth]{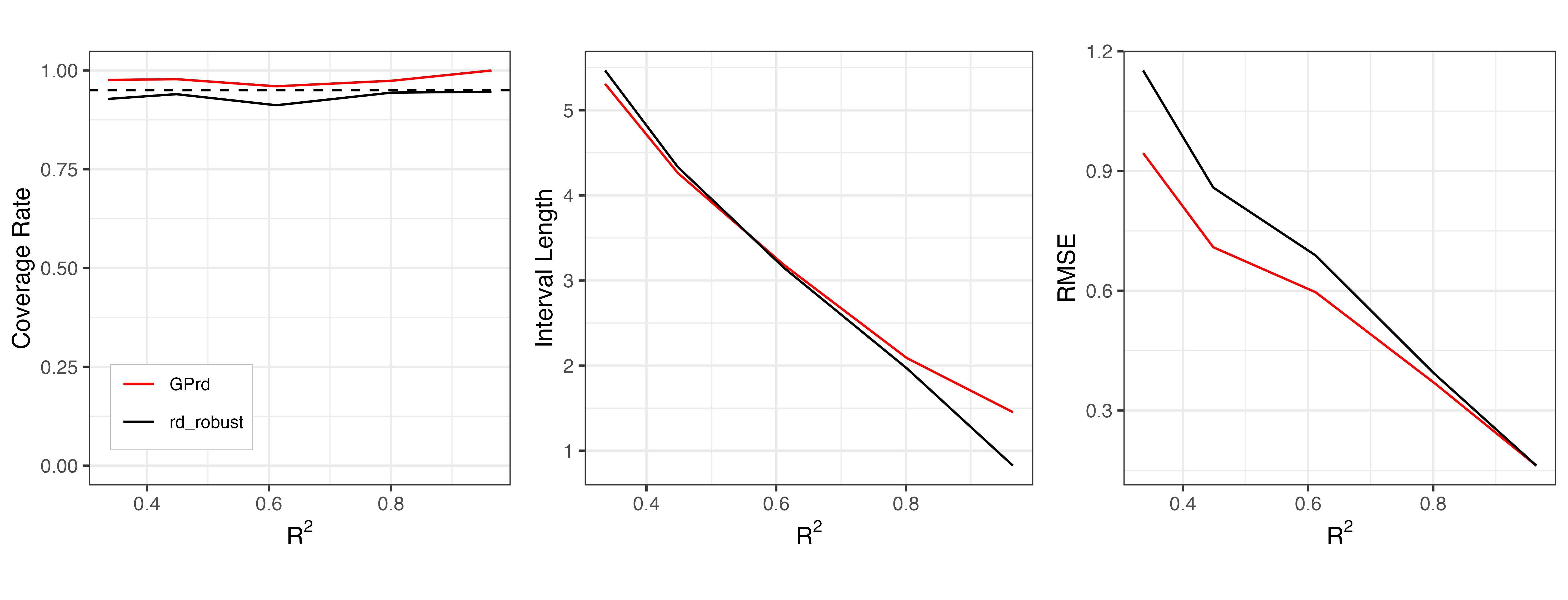}\\
\centering \textit{(c)} Cubic\\
\includegraphics[width=.9\linewidth]{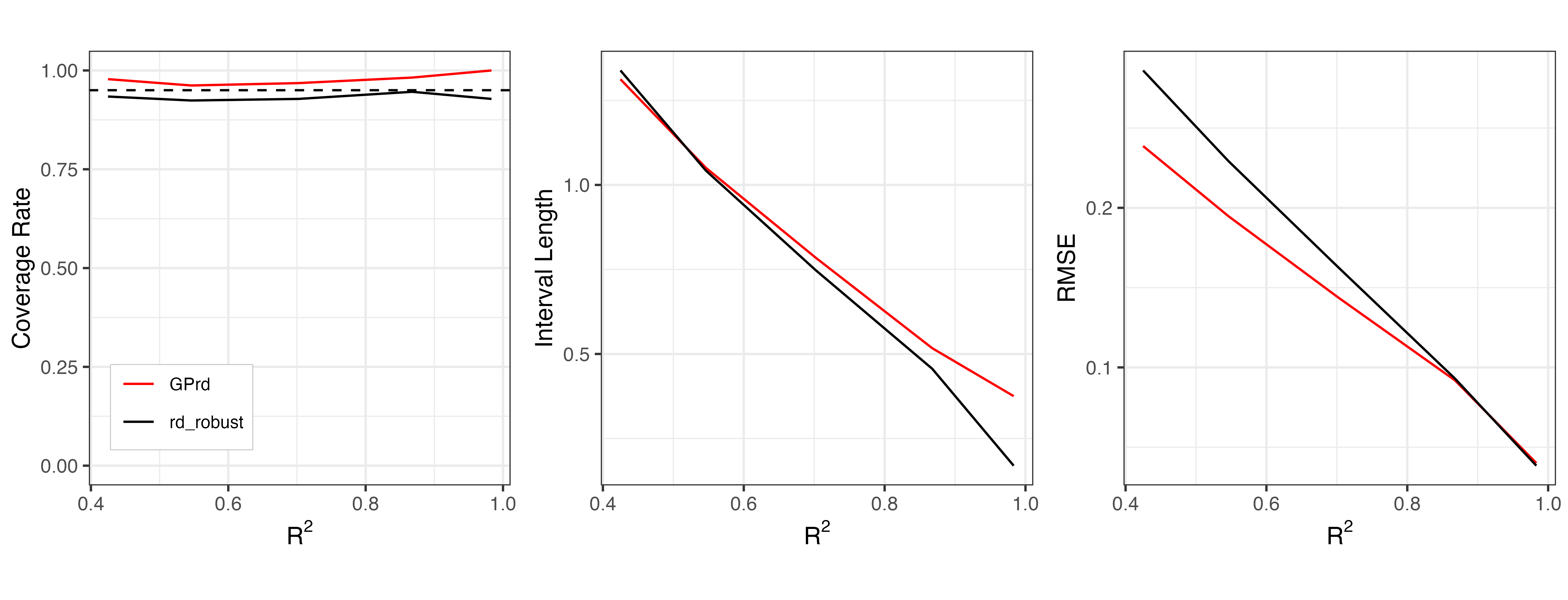}
\caption{RD 500 simulation results for seven different functional forms across different $R^2$}
\label{fig:worstsim1}
\end{figure}

\begin{figure}[hbt!]
\centering \textit{(d)} Curvature\\
\includegraphics[width=.9\linewidth]{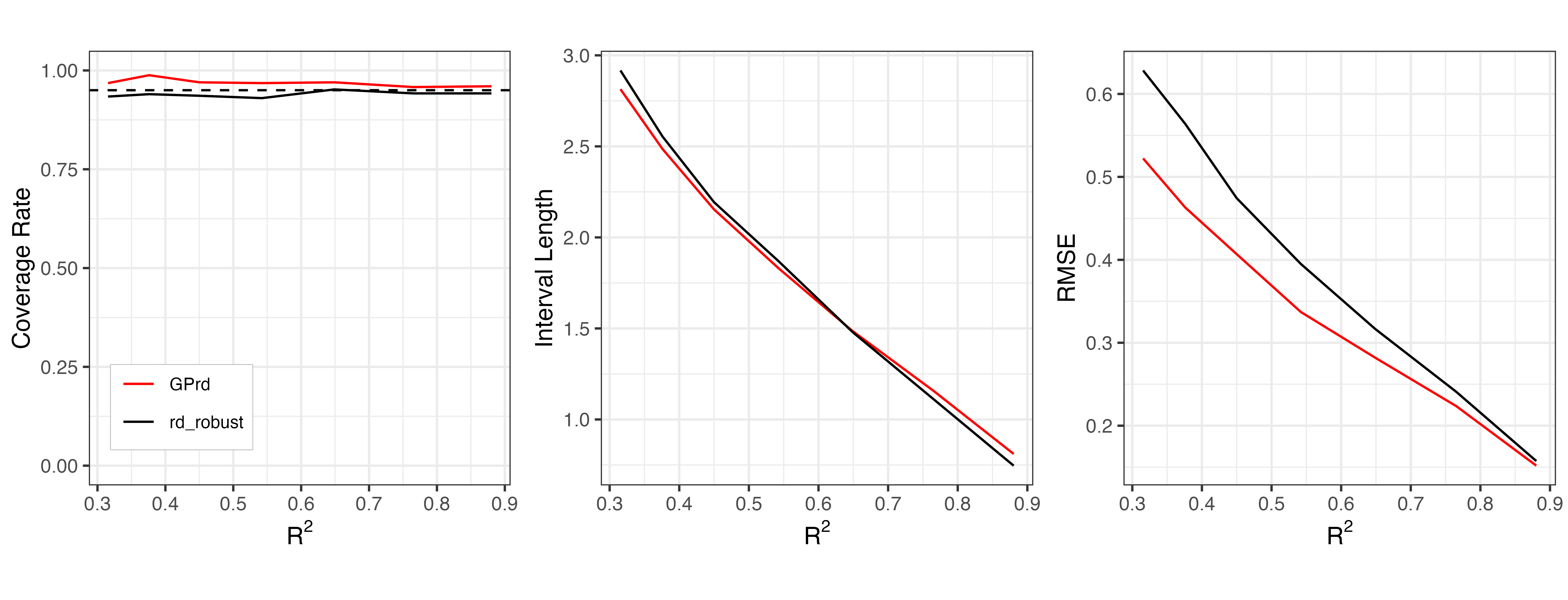}\\
\centering \textit{(e)}  Lee(2008) \\
\includegraphics[width=.9\linewidth]{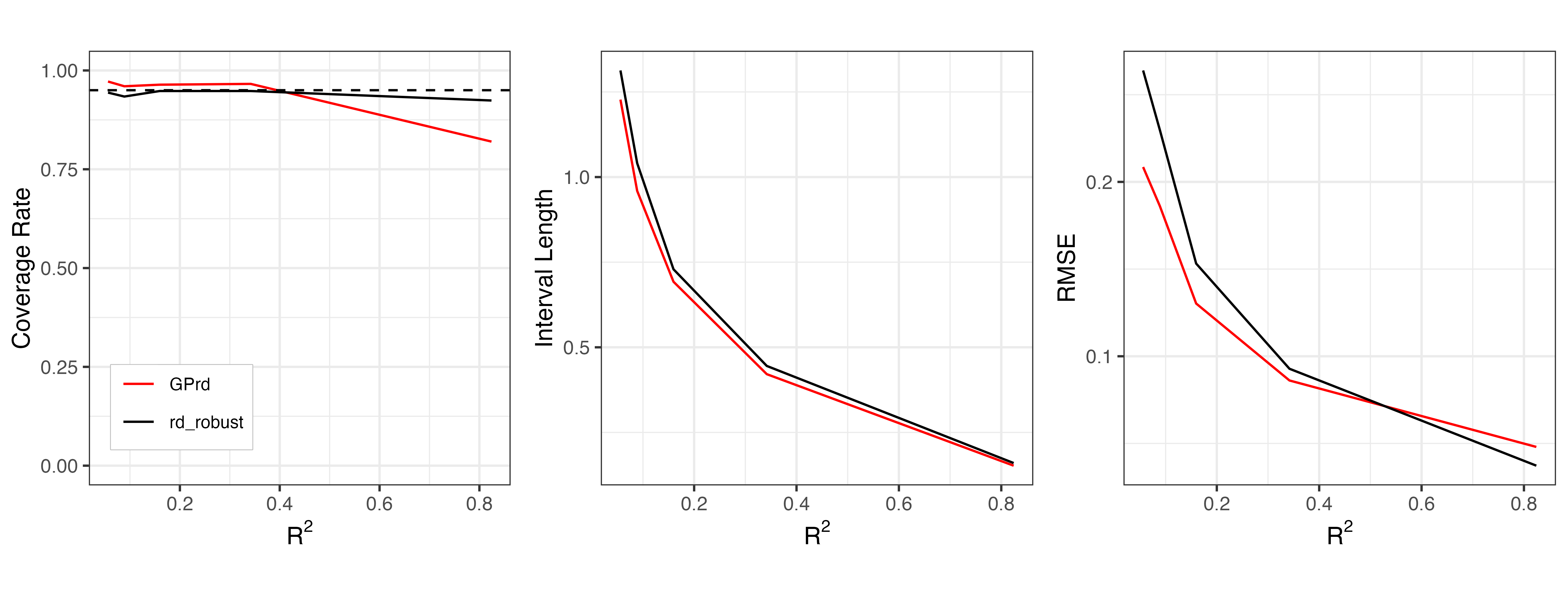}\\
\centering \textit{(f)} Ludwig \& Milner (2007)\\
\includegraphics[width=.9\linewidth]{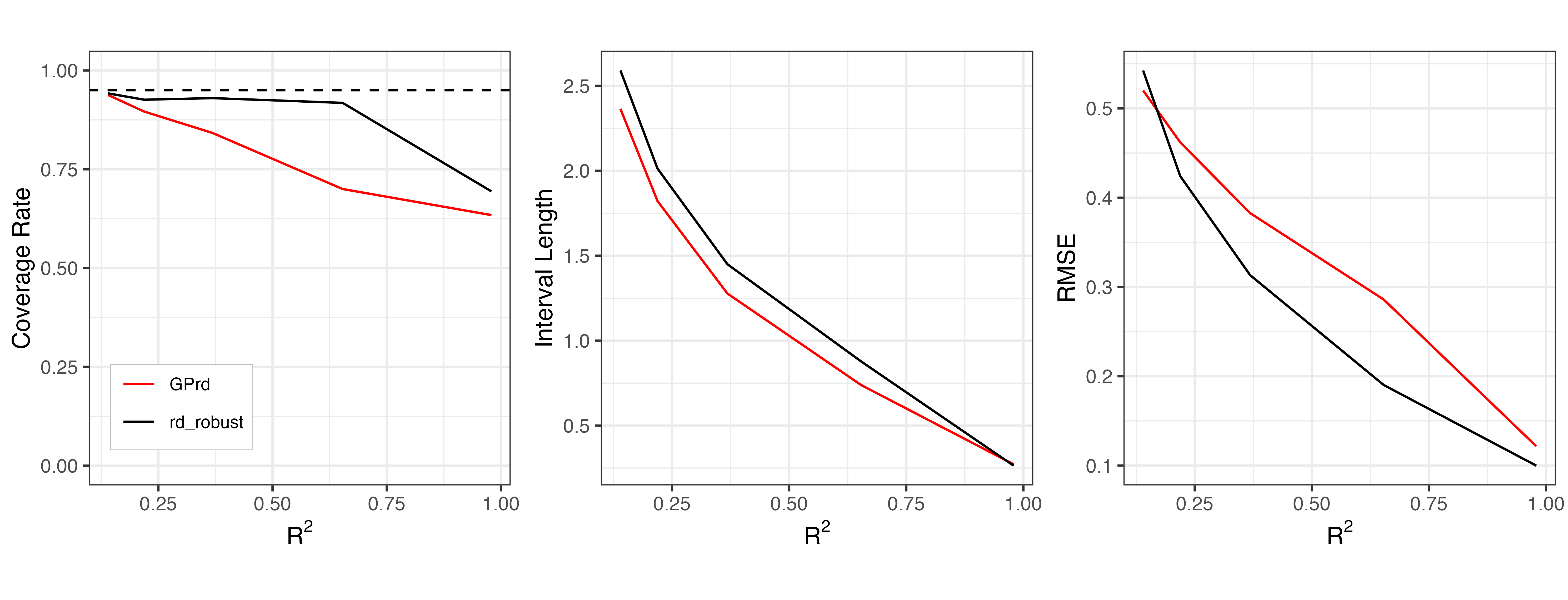}\\
\centering \textit{(g)} Quad\\
\includegraphics[width=.9\linewidth]{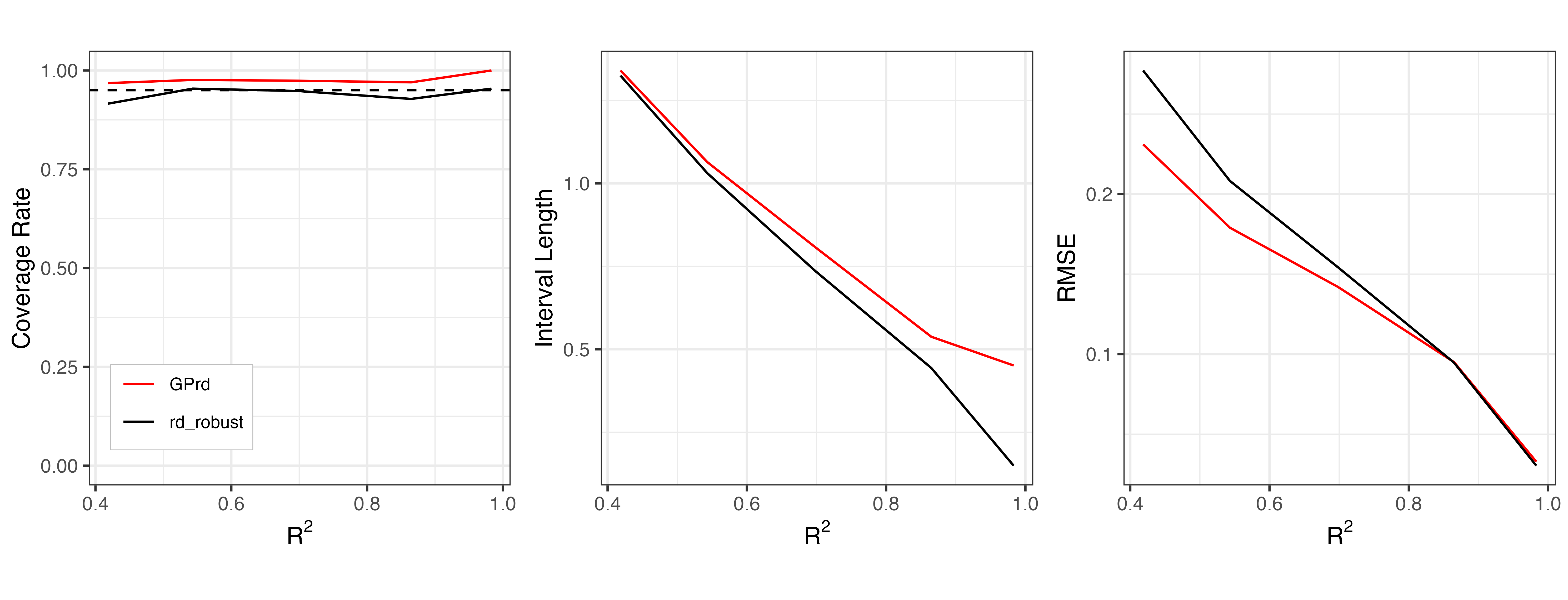}
\caption{(Continued) RD 500 simulation results for seven different functional forms across different $R^2$}
\label{fig:worstsim2}
\end{figure}

\end{document}